\documentclass[fleqn,preprint,review,12pt]{elsarticle}

\usepackage{physics}
\usepackage{amsmath}
\usepackage{amssymb}
\usepackage{esint}
\usepackage{siunitx}
\usepackage{multicol}
\usepackage{pdflscape}
\usepackage{multirow}
\usepackage{multicol}
\usepackage{nicefrac}
\usepackage{booktabs}
\usepackage{float}
\usepackage{svg}
\usepackage{epstopdf}   
\usepackage{cuted}

\usepackage{lineno}
\usepackage{caption}
\usepackage{subcaption}
\makeatletter
\let\LN@align\align
\let\LN@endalign\endalign
\renewcommand{\align}{\linenomath\LN@align}
\renewcommand{\endalign}{\LN@endalign\endlinenomath}
\let\LN@gather\gather
\let\LN@endgather\endgather
\renewcommand{\gather}{\linenomath\LN@gather}
\renewcommand{\endgather}{\LN@endgather\endlinenomath}
\makeatother

\newcolumntype{P}[1]{>{\centering\arraybackslash}p{#1}}
\newcolumntype{M}[1]{>{\centering\arraybackslash}m{#1}}

\newcommand{\ensmean}[1]{\left\langle #1 \right\rangle}
\newcommand{\tmean}[1]{\overline{#1}}
\newcommand{\enstmean}[1]{\left\langle \overline{#1} \right\rangle}
\let\oldfrac\frac
\renewcommand{\frac}[2]{%
  \mathchoice
    {\oldfrac{#1}{#2}}
    {#1/#2}
    {#1/#2}
    {#1/#2}
}

\DeclareSIUnit\density{\kilogram\per\meter\tothe{3}}
\DeclareSIUnit\velocity{\meter\per\second}

\newcommand{\noah}[1]{\noindent \color{black} #1\normalcolor}
\newcommand{\revision}[1]{\noindent \color{black} #1\normalcolor}
\journal{}

\begin{document}
\begin{frontmatter}

\title{Understanding Flow Dynamics in Membrane Distillation: Effects of Reactor Design on Polarization}  

\author[1]{Yinuo Yao}
\cortext[cor1]{Corresponding author}
\author[1]{Siqin Yu}
\author[1]{Ilenia Battiato\corref{cor1}}

\address[1]{Department of Energy Science and Engineering, Stanford University, Stanford, CA 94305, USA}

\begin{abstract}
\noah{Optimization and design of full-scale  membrane distillation (MD) systems usually require Sherwood and Nusselt correlations that are developed from lab-scale systems. However, entrance effects in lab-scale systems can significantly impact heat, mass and momentum transfer in the reactor, therefore affect the accuracy of the developed experimental Sherwood and Nusselt} correlations. Here, Computational Fluid Dynamics (CFD) simulations using OpenFOAM are performed to understand the effects of right-angled bends and inlet design on flow dynamics, temperature and concentration polarization in MD systems. Simulation results show that the presence of right-angled bends and inlets with sudden expansions lead to the formation of Dean vortices. Dean vortices enhance perpendicular mixing in MD systems and reduce  both temperature and concentration polarization. Temperature and concentration polarization coefficients in MD systems with right-angled bends and inlets with sudden expansions vary significantly for the same volumetric flow rate. \noah{Our studies show that lab-scale systems with the same volumetric flow rate but different designs lead to significantly different Nusselt and Sherwood correlations. This study demonstrates the importance of CFD-informed design of lab-scale systems to minimize entrance effects and suppress Dean vortices for consistent model development and calibration across multiple scales.}

\medskip

\noindent\textbf{Graphic abstract}

\bigskip

\noindent\includegraphics[width=10cm]{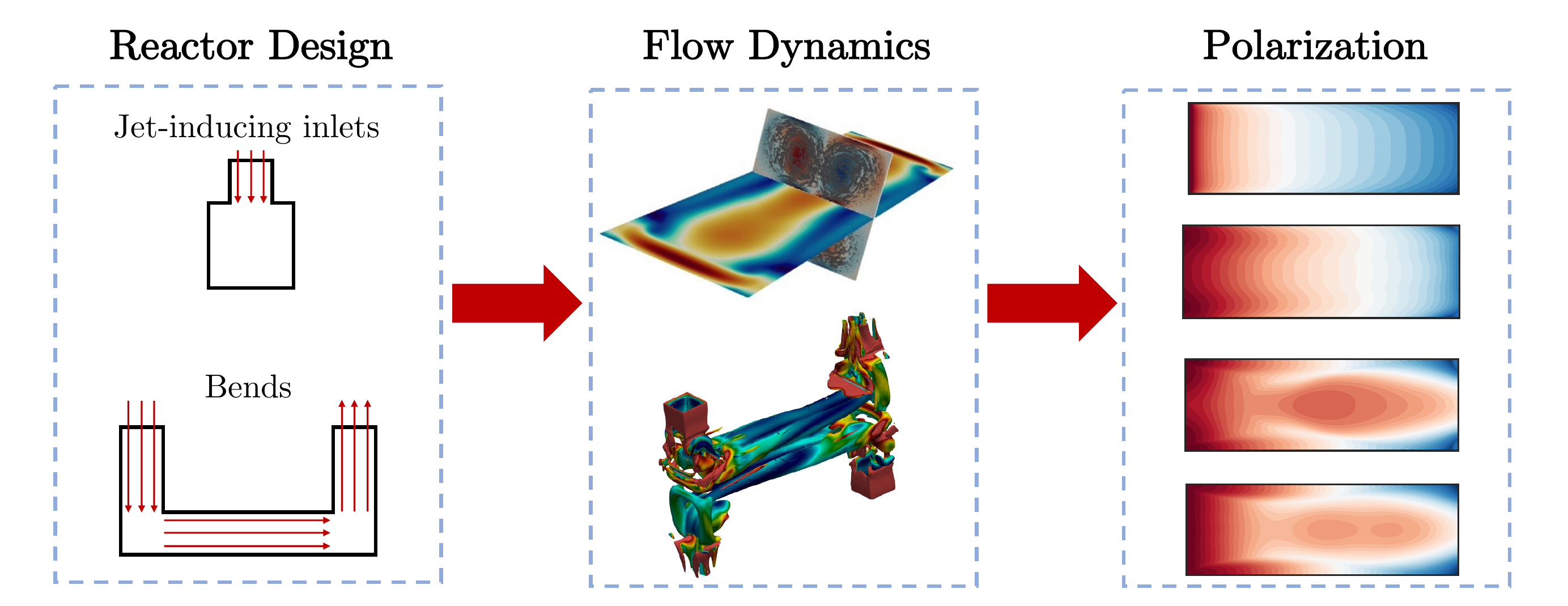}
\end{abstract}


\end{frontmatter}

\section{Introduction}
Over the past few decades, desalination processes of unconventional water sources (i.e., seawater, brackish water, treated used-water etc.) have received considerable attention to improve local water supply in the era of climate change~\cite{Khawaji2008-oi, Miller2003-wh,Subramani2015-qi,Shin2021-ug}. Typical membrane-based desalination processes are reverse osmosis (RO), forward osmosis (FO) and membrane distillation (MD). Nowadays, RO has been widely adopted to separate salt and impurities from water, producing water quality that meets the drinking standard. Although promising, the cost of treating unconventional water sources is still significantly higher than that of treating conventional sources. One challenge associated with desalination is the production of brine~\cite{Greenlee2009-fh}. Membrane distillation is proposed to treat the high concentration brine produced. 

One of the challenges in operating MD systems is related to membrane fouling. Membrane fouling is a complex phenomenon in which foulant accumulates on the membrane surface due to deposition or adsorption arising from temperature and concentration polarization as a result of the coupling between heat transfer, mass transport and flow dynamics~\cite{Tijing2015-gz, Curcio2010-of,Gryta2008-lz, Hickenbottom2014-er}. Severe membrane fouling will reduce permeate flux and increase the maintenance cost in long-term operation. One method of minimizing membrane fouling is to develop models, based on the predicted temperature and concentration, that can be used to optimize MD systems ~\cite{Dudchenko2020-mf,Dudchenko2022-pz,Hitsov2015-au}. To establish a relationship between design, operating parameters and temperature polarization, multiple models have been developed based on the Nusselt number correlation~\cite{Ali2013-gr,Phattaranawik2003-he,Curcio2005-pd} which establishes a relationship between the Nusselt number, the Reynolds number and the Prandtl number~\cite{Gryta2008-lz,Phattaranawik2003-he}. Similarly to temperature polarization, concentration polarization can be predicted by models based on the Sherwood number correlation which correlates the Sherwood number with  Reynolds number and Schmidt number~\cite{Ve2021-rv,Lokare2019-xe}. In addition to developing theoretical models, numerous experimental studies have focused on understanding heat and mass transfer in MD systems~\cite{Ohta1990-mx,Ali2013-gr,Chen2009-ro,Qtaishat2008-bt}.

However, one challenge associated with model development and experiments is the ``tyranny of scales'' where models and experiments are developed and performed on the laboratory scale (on the order of centimeters)~\cite{Lee2016-ar} while the industrial scale is usually on the order of meters~\cite{Chafidz2019-lo}, resulting in significantly different polarization and flow dynamics. Song~\cite{Song2007-aw}  investigated a novel reactor design on a larger scale (on the order of decimeter), but still smaller than the typical industrial scale. Due to the limited size of  experimental laboratory bench modules, the inlet and module designs may have strong influences on the measured temperature and concentration polarization as well as permeate fluxes~\cite{Salem2019-yv, Dudchenko2020-mf}. Dudchenko \emph{et al.}~\cite{Dudchenko2022-pz} compared different existing Nusselt number correlations with experimental results. They  concluded that the best correlation still had significant errors in predicting permeate fluxes and its performance was also highly dependent on module designs. Yet, the underlying physical mechanisms leading to such discrepancy between bench experiments and theoretical correlations are not fully understood.

In general, two typical module designs  influence  measured quantities in  MD systems: right-angled bends and inlets~\cite{Salem2019-yv,Lee2016-ar}. Over the past few decades, both experiments and simulations have demonstrated the impact of right-angled bends on the flow dynamics~\cite{Nivedita2017-dv,Bottaro1991-yc,Finlay1988-pr,Helin2009-zm,Naphon2006-pz, Altay2021-rv,Wang2022-rf,Bayat2017-zh,Ligrani1988-do}. Chen and Zhang~\cite{Chen2003-ha} investigated the coupling of thermal distribution and flow characteristics. Lira~\cite{De_Brito_Lira2022-dq} studied the difference in flow dynamics and mass transport between  straight and curved channels in a chemical system. In a typical MD system, the inlet is designed with a surface area  smaller than the main channel such that impinging jet is generated. Previous studies have shown that impinging jets would result in a high local heat and mass transfer coefficient. Gardon and Akfirat~\cite{Gardon1965-ay} found that the maximum heat transfer would occur at the stagnation point of the jet when the inlet spacing is greater than four times of the inlet diameter. Lytle and Webb~\cite{Lytle1994-uj} observed that the heat transfer coefficient exhibits two local maxima away from the stagnation point. Sparrow and Wong~\cite{Sparrow1975-iy} measured the dependence of the mass transport coefficient at the stagnation point on the Reynolds number. Feroz~\cite{Feroz2006-dk} observed that the mass transfer coefficient
increases with increasing the inlet diameter. The impinging jet has also been reported to enhance vortex formation and transition to turbulence~\cite{Fox1993-jv,Tesar2002-uq,Le_Song2007-lo}. To date, the combined effects of right-angled bends and inlets on flow dynamics, temperature and concentration polarization on the membrane in MD systems are still not fully understood. With increasing computational power, high-fidelity CFD simulations have been used to understand the impact of flow dynamics on system operation and optimization~\cite{Hitsov2015-au,Salem2019-yv,Bottaro1991-yc,Wang2022-rf,De_Brito_Lira2022-dq,Le_Song2007-lo,Yao2021-ex, Lou2019-xa, Lou2021-ix, Yao2022-du, Yao2021-ky,Yao2021-rg, Ling2019-nm, Ling2021-ba, Zhou2021-ry}. A thorough understanding of the effects will provide quantitative guidelines for model development.

In this work, we use of computational fluid dynamics (CFD) simulations to study the effects of right-angled bends and inlets on direct-contact membrane distillation (DCMD) system performance. \noah{The objective of this work is to quantify entrance effects and understand the underlying governing physics that ultimately result in performance discrepancies between lab-scale and full-scale MD systems. Such a fundamental understanding can guide the design of lab-scale MD systems to minimize entrance effects and enable the development of more accurate models and correlations for full-scale systems.} Four designs with different volumetric flow rates will be simulated to evaluate the flow dynamics, and its coupling to temperature and concentration polarization. The paper is organized as follows. In Section~\ref{sec:goven_eqs}, we describe the governing equations. The verification of the numerical implementation and the setup of the simulations are presented in Sections~\ref{sec:val} and~\ref{sec:sim_setup}, respectively. In Section~\ref{sec:results}, the effects of right-angled bends and inlets on vortex formation, as well as temperature and concentration polarization are discussed. The relationship between vortex formation and modeling errors is presented in Section~\ref{sec:disscussion}. We present concluding remarks in Section~\ref{sec:conclusions}.

\section{Methodology}
\subsection{Governing equations}
\label{sec:goven_eqs}
We consider membrane-desalinating a stream of sodium chloride solution that flows through the feed side of the DCMD systems.  The permeate is collected on the draw side while sodium chloride is rejected and retained on the feed side of the DCMD systems. In this study, we incorporate the effects of varying density and temperature by solving the compressible Navier–Stokes equations in a three-dimensional membrane distillation (MD) system
\begin{align}
    \label{eq:ns_eq}
    \pdv{\left(\rho \boldsymbol{u}\right)}{t} + \div{\left(\rho\boldsymbol{u}\boldsymbol{u}\right)} = -\grad{p} + \div{\boldsymbol{\sigma}} + \rho\boldsymbol{g},
\end{align}
with $\boldsymbol{\sigma}$ the viscous stress tensor, defined as
\begin{align}
  \label{eq:stress}
    \boldsymbol{\sigma} = \mu \left( \grad\boldsymbol{u} + \left(\grad\boldsymbol{u}\right)^T - \frac{2}{3}\left(\div{\boldsymbol{u}}\right)\mathbf{I}\right),
\end{align}
subject to continuity,
\begin{align}
  \label{eq:continuity}
    \pdv{\rho}{t} + \div{\left(\rho\boldsymbol{u}\right)} =0.
\end{align}
In Equations~\eqref{eq:ns_eq}-\eqref{eq:continuity}, $\boldsymbol{u}$~[\si{\velocity}]$=\begin{pmatrix} u & v & w\end{pmatrix}^T$ is the velocity vector, $\rho$~[\si{\density}] is the fluid density, $p$~[\si{\pascal}] is the total pressure, $\mu$~[\si{\kilogram\per\meter\per\second}] is the dynamic viscosity of the fluid, and $g$~[\si{\meter\per\second\tothe{2}}]$=\begin{pmatrix} 0 & 0 & -9.81 \end{pmatrix}$ is the gravitational acceleration vector. The temperature equation is given by the energy equation as 
\begin{align}
    \pdv{\left(\rho h\right)}{t} + \div{\left(\rho \boldsymbol{u} h\right)} = \div{\left(k\grad T\right)} + \pdv{p}{t} + \boldsymbol{u} \cdot \grad{p},
\end{align}
where $h$~[\si{\kilo\joule\per\kilogram}] is the specific enthalpy, $k$~[\si{\watt\per\meter\per\kelvin}] is the thermal conductivity and $T$~[\si{\kelvin}] is the temperature. The relationship between $h$ and $T$ is given by
\begin{align}
    \int_{T_{stp}}^T c_p(T, C) \ \dd T = h,
\end{align}
where $c_p$~[\si{\joule\per\kilogram\per\kelvin}] is the specific heat capacity of the fluid, $T_{stp} = 273.15$~\si{\kelvin} is the standard temperature and $C$~[\si{\density}] is the concentration of sodium chloride. The transport of sodium chloride is governed by the scalar transport equation such that
\begin{align}
    \label{eq:scalar_eq}
    \pdv{C}{t} + \div{\left(\boldsymbol{u} C \right)} = \div{\left(D \grad{C}\right)},
\end{align}
where $D$~[\si{\meter\tothe{2}\per\second}] is the molecular diffusion coefficient.

At the inlets for both the feed and draw sides, Dirichlet boundary conditions for velocity, temperature and concentration, and Neummann boundary condition for pressure are applied as
\begin{subequations}
\begin{align}
    &\boldsymbol{u}(\boldsymbol{x}, t) = u_{in,i}, \quad \boldsymbol{x} \in \Gamma_{in,i}, \\
    &T(\boldsymbol{x}, t) = T_{in,i}, \quad \boldsymbol{x} \in \Gamma_{in,i}, \\    
    &C(\boldsymbol{x}, t) = C_{in,i}, \quad \boldsymbol{x} \in \Gamma_{in,i},  \\
    &\grad{p}(\boldsymbol{x}, t)\cdot \boldsymbol{n}(\boldsymbol{x}) = 0, \quad \boldsymbol{x} \in \Gamma_{in,i},
\end{align}
\end{subequations}
where $i = f$ or $d$ refers to the feed and draw sides of the MD system, respectively, $u_{in,i}$, $T_{in,i}$ and $C_{in,i}$ are the inlet velocity, temperature and concentration, $\boldsymbol{n}$ is the normal vector that points outwards of the surface and $\Gamma_{in,i}$ is the surface of the inlet for subdomain $i$. At the outlets of the feed and draw channels, we apply  a Dirichlet boundary condition to pressure and Neummann boundary conditions to velocity, temperature and concentration as
\begin{subequations}
\begin{align}
    &\grad{\boldsymbol{u}}(\boldsymbol{x}, t)\cdot \boldsymbol{n}(\boldsymbol{x})  = 0, \quad \boldsymbol{x} \in \Gamma_{out,i}, \\
    &\grad{T}(\boldsymbol{x}, t)\cdot \boldsymbol{n}(\boldsymbol{x}) = 0, \quad \boldsymbol{x} \in \Gamma_{out,i}, \\    
    &\grad{C}(\boldsymbol{x}, t)\cdot \boldsymbol{n}(\boldsymbol{x}) = 0, \quad \boldsymbol{x} \in \Gamma_{out,i},  \\
    &p(\boldsymbol{x}, t) = 0, \quad \boldsymbol{x} \in \Gamma_{out,i},
\end{align}
\end{subequations}
where $\Gamma_{out,i}$ is the surface of the outlet for the subdomain $i$. For the walls, the no-slip boundary condition is applied to velocity and zero-gradient boundary conditions are applied to pressure, temperature and concentration such that
\begin{subequations}
\begin{align}
    &\boldsymbol{u}(\boldsymbol{x}, t)  = 0, \quad \boldsymbol{x} \in \Gamma_{wall}, \\
    &\grad{T}(\boldsymbol{x}, t)\cdot \boldsymbol{n}(\boldsymbol{x}) = 0, \quad \boldsymbol{x} \in \Gamma_{wall}, \\    
    &\grad{C}(\boldsymbol{x}, t)\cdot \boldsymbol{n}(\boldsymbol{x}) = 0, \quad \boldsymbol{x} \in \Gamma_{wall},  \\
    &\grad{p}(\boldsymbol{x}, t)\cdot \boldsymbol{n}(\boldsymbol{x}) = 0, \quad \boldsymbol{x} \in \Gamma_{wall},
\end{align}
\end{subequations}
where $\Gamma_{wall}$ are the surfaces of the channels.

To simulate flow through the membrane and its effects on temperature and concentration polarization, we adopt the boundary condition formulation by Lou \emph{et al.}~\cite{Lou2019-xa,Lou2021-ix} and extend it to three-dimensions. Transmembrane permeate flux $J_w$~[\si{\kilogram\per\meter\tothe{2}\per\second}] is modeled as 
\begin{align}
    J_w = A\left(p_{m,f} - p_{m,d}\right),
\end{align}
where $A$~[\si{\meter\tothe{3}\per\second}] is the vapor permeability and $p_{m,f}$~[\si{\pascal}] and $p_{m,d}$~[\si{\pascal}] are the vapor pressure on the feed and draw sides, respectively. Without resolving the transport in the membrane pores~\cite{Lou2019-xa,Lou2021-ix,Hitsov2015-au}, we determine the local vapor pressure $p_{m,i}$~[\si{\pascal}] with respect to temperature and concentration as 
\begin{align}
    p_{m,i} (C, T) = a_w (C) p_{sat} (T),
\end{align}
with $a_w$~[-] the water activity 
\begin{align}
    a_w = 1 - 0.03122m + 0.001482m^2,
\end{align}
and $m$~[\si{\mole\per\kilogram}]  the sodium chloride molality. The  vapor saturation pressure  $p_{sat}$~[\si{\pascal}] is calculated with the Antoine equation~\cite{Smith1950-ra} as 
\begin{align}
    p_{sat} = \exp\left(23.238 - \frac{3841}{T_{m} - 45}\right),
\end{align}
where  $T_{m}$~[\si{\kelvin}] is the temperature on the membrane surface.

Heat transport on the membrane is modeled as a balance between convective and conductive heat transfer. Transmembrane heat conduction $q_m$~[\si{\watt\per\meter\tothe{2}}] is modeled as ~\cite{Sarti1985-yq, Lou2019-xa} 
\begin{align}
    q_m = h_m \left(T_{m,f} - T_{m,d}\right),
\end{align}
where $T_{m,f}$~[\si{\kelvin}] and $T_{m,d}$~[\si{\kelvin}] are the membrane temperature of the feed and draw sides, respectively, and $h_m$~[\si{\watt\per\meter\tothe{2}\per\kelvin}] is the conductive heat
transfer coefficient of the membrane. 
 A balance between diffusive and convective fluxes provides the boundary condition for mass transport on the membrane~\cite{Martin2010-be, Ling2019-nm, Ling2021-ba}. In general, the boundary conditions for the membrane on the feed side are given by
\begin{subequations}
\begin{align}
    &\boldsymbol{u}(\boldsymbol{x}, t) = \frac{J_w}{\rho_f} \boldsymbol{n}_f, \quad \boldsymbol{x} \in \Gamma_{m,f}\\
    &k\grad\boldsymbol{T}(\boldsymbol{x}, t) \cdot \boldsymbol{n}_f(\boldsymbol{x}) = -\lambda J_w - h_m \Delta T_m, \quad \boldsymbol{x} \in \Gamma_{m,f} \\
    &D\grad\boldsymbol{C}(\boldsymbol{x}, t) \cdot \boldsymbol{n}_f(\boldsymbol{x}) = - \frac{J_w}{\rho_f} C_{m,f}, \quad \boldsymbol{x} \in \Gamma_{m,f}, \\
    &\grad{p}(\boldsymbol{x}, t)\cdot \boldsymbol{n}_f(\boldsymbol{x}) = 0, \quad \boldsymbol{x} \in \Gamma_{m,f},
\end{align}
\end{subequations}
where $\Delta T_m = \left(T_{m,f} - T_{m,d}\right)$ is the membrane temperature difference between feed and draw sides, $\boldsymbol{n}_f$ is the normal vector that points outward of the feed-side membrane, $\rho_f$~[\si{\density}] is the feed-side fluid density, $C_{m,f}$~[\si{\density}] is the sodium chloride concentration on the feed side, $\lambda$~[\si{\joule\per\kilogram}] is the latent heat of water and $\Gamma_{m,f}$ is the surface of feed-side membrane. For the draw side, the boundary conditions are given by
\begin{subequations}
\begin{align}
    &\boldsymbol{u}(\boldsymbol{x}, t) = -\frac{J_w}{\rho_d} \boldsymbol{n}_d, \quad \boldsymbol{x} \in \Gamma_{m,d}\\
    &k\grad\boldsymbol{T}(\boldsymbol{x}, t) \cdot \boldsymbol{n}_d(\boldsymbol{x}) = \lambda J_w + h_m \Delta T_m, \quad \boldsymbol{x} \in \Gamma_{m,d} \\
    &D\grad\boldsymbol{C}(\boldsymbol{x}, t) \cdot \boldsymbol{n}_d(\boldsymbol{x}) =  \frac{J_w}{\rho_d} C_{m,d}, \quad \boldsymbol{x} \in \Gamma_{m,d}, \\
    &\grad{p}(\boldsymbol{x}, t)\cdot \boldsymbol{n}_d(\boldsymbol{x}) = 0, \quad \boldsymbol{x} \in \Gamma_{m,d},
\end{align}
\end{subequations}
where $\boldsymbol{n}_d$ is the normal vector that points outward of the draw-side membrane, $\rho_d$~[\si{\density}] is the draw-side fluid density, $C_{m,d}$~[\si{\density}] is the sodium chloride concentration on the draw side and $\Gamma_{m,d}$ is the surface of draw-side membrane. To incorporate the effect of temperature and concentration in the simulations, we calculate the thermophysical properties of the fluid as a function of temperature and concentration. Details can be found in~\ref{sec:thermo_prop}.

\begin{figure*}
    \centering
    \begin{subfigure}[ht]{0.49\textwidth}
    \caption{}
    \centerline{
     {\includegraphics[width=\textwidth]{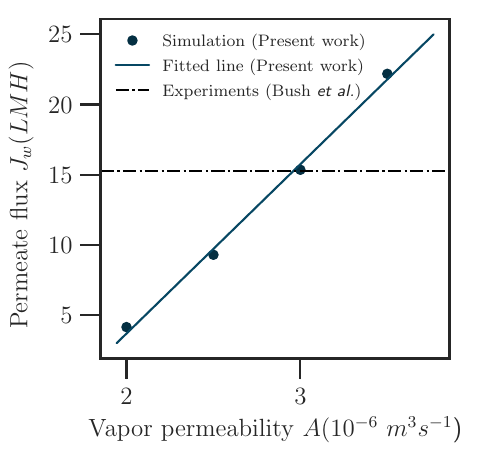}}}
    
    \label{fig:A_Calibration}
    \end{subfigure}
    \begin{subfigure}[ht]{0.49\textwidth}
     \caption{}
    \centerline{
     {\includegraphics[width=\textwidth]{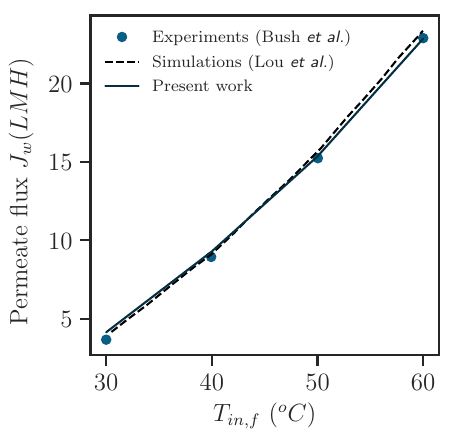}}}
    \label{fig:deltaT_flux}
     \end{subfigure}
     \caption{(a) Calibration of vapor permeability $A$ based on experimental results and (b) validation and verification of present work with results by Lou \emph{et al.}~\cite{Lou2019-xa}.}
     \label{fig:validation}
\end{figure*}

\subsection{Numerical method implementation and validation}
\label{sec:val}
In this study, we solve the Navier-Stokes equation (equation~\ref{eq:ns_eq}) by implementing a solver in OpenFOAM~\cite{Weller1998-ho} based on the low-Mach number flow approximation~\cite{Lessani2006-dk}. For the pressure-velocity coupling, the noniterative method, pressure-implicit algorithm with operators splitting (PISO) is adopted to satisfy mass conservation using predictor-corrector steps~\cite{Issa1986-ld}. The governing equations are discretized with first-order backward Euler in time to avoid time-step constraints. Divergence terms are discretized with second-order accurate upwind schemes~\cite{Warming1976-pw} while Laplacian terms are discretized with a second-order central difference scheme. The numerical solver used in this study is  validated and verified through direct comparison with experiments~\cite{Vanneste2018-vu, Bush2016-cw} and two-dimensional simulations by Lou \emph{et al.}~\cite{Lou2019-xa}. The inlet concentration of sodium chloride, $C_{in,f}$, is set to \SI{1}{\density} and the inlet velocities in both feed and draw sides, $u_{in,f}$ and $u_{in,d}$, are set as laminar parabolic profiles such that
\begin{align}
    u_{in,i} (y) = \frac{3u_{\max}}{2}\left(1 - \frac{y^2}{H^2} \right),
\end{align}
where $u_{\max}= \SI{0.124}{\meter\per\second}$ is the maximum velocity, and $H$~[\si{\meter}] is the height of the channel. 

To determine vapor permeability $A$, we perform four simulations with $T_{in,d} = \SI{20}{\celsius}$, $T_{in,f} = \SI{50}{\celsius}$ and $A=\num{2e-6}$, $\num{2.5e-6}$, $\num{3e-6}$, $\num{3.5e-6}$~\si{\meter\tothe{3}\per\second}. Figure~\ref{fig:validation}(\subref{fig:A_Calibration}) shows the permeate flux $J_w$ as a function of vapor permeability $A$. The optimal vapor permeability is given by the intersection between the experimental result (black) and the best-fit line (blue), which corresponds to $A = \num{2.96e-06}$~\si{\meter\tothe{3}\per\second}. The  remaining simulations use such value for $A$.

To validate and verify our numerical results with the experimental and simulation results by Lou \emph{et al.}~\cite{Lou2019-xa}, we set the inlet temperature on the draw side $T_{in,d}$ to  \SI{20}{\celsius} \revision{while the inlet temperature on the feed side $T_{in,f}$ varies between \SI{30}{\celsius} and \SI{60}{\celsius} with an increment of \SI{10}{\celsius}.} Figure~\ref{fig:validation}(\subref{fig:deltaT_flux}) shows the permeate flux as a function of the inlet temperature on the feed side, $T_{in,f}$. As demonstrated, the difference between our results and the experimental and simulation results by Lou \emph{et al.}~\cite{Lou2019-xa} is negligible, demonstrating the accuracy of the developed solver. Differently from \cite{Lou2019-xa, Lou2021-ix}, the solver implemented in OpenFOAM can efficiently utilize high-performance computing clusters to perform three-dimensional simulations, as shown in the following analyses. 

\subsection{Three-dimensional simulations setup}
\label{sec:sim_setup}
\begin{figure*}
    \centering
    \begin{subfigure}[ht]{0.583\textwidth}
    \caption{}
    \centerline{
     {\includegraphics[width=\textwidth]{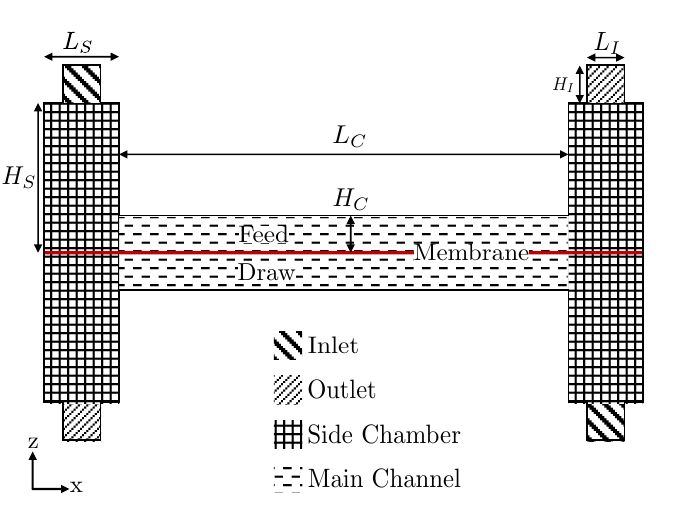}}}
    \label{fig:5x5_dim_xz}
    \end{subfigure}
    \begin{subfigure}[ht]{0.25\textwidth}
    \caption{}
    \centerline{
     {\includegraphics[width=\textwidth]{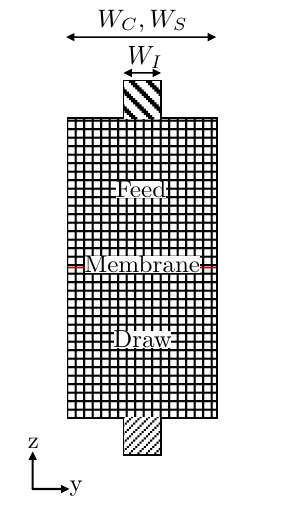}}}
    \label{fig:5x5_dim_yz}
    \end{subfigure}

     \caption{Illustration of geometry schematic diagram and dimensions.}
     \label{fig:schema_diag_dim}
\end{figure*}

\begin{figure*}
    \centering
    \begin{subfigure}[ht]{0.24\textwidth}
    \caption{\quad MD-5x10S}
    \centerline{
     {\includegraphics[width=\textwidth]{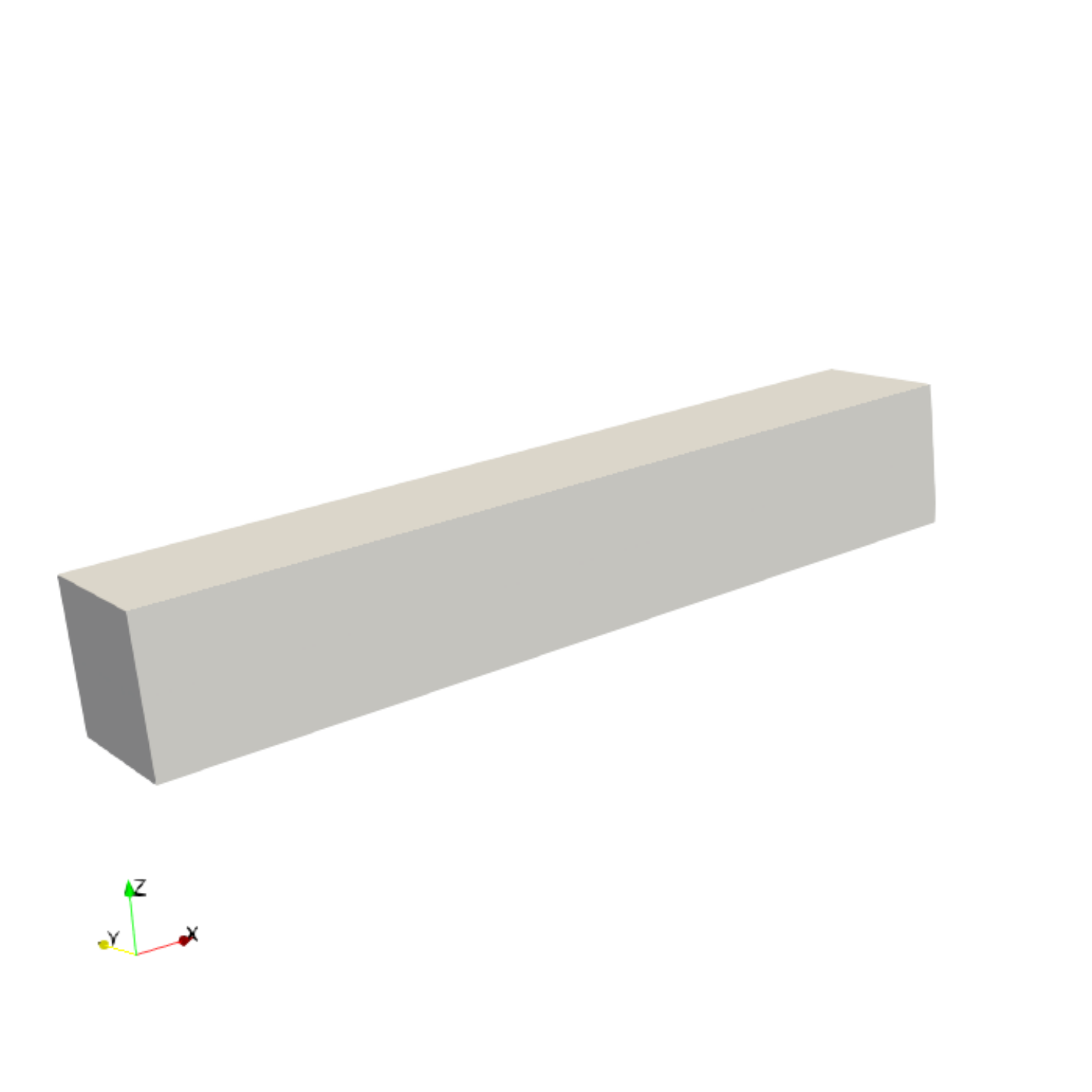}}}
    \label{fig:5x10s_diag}
    \end{subfigure}
    \hfill
    \begin{subfigure}[ht]{0.24\textwidth}
    \caption{\quad MD-5x10}
    \centerline{
     {\includegraphics[width=\textwidth]{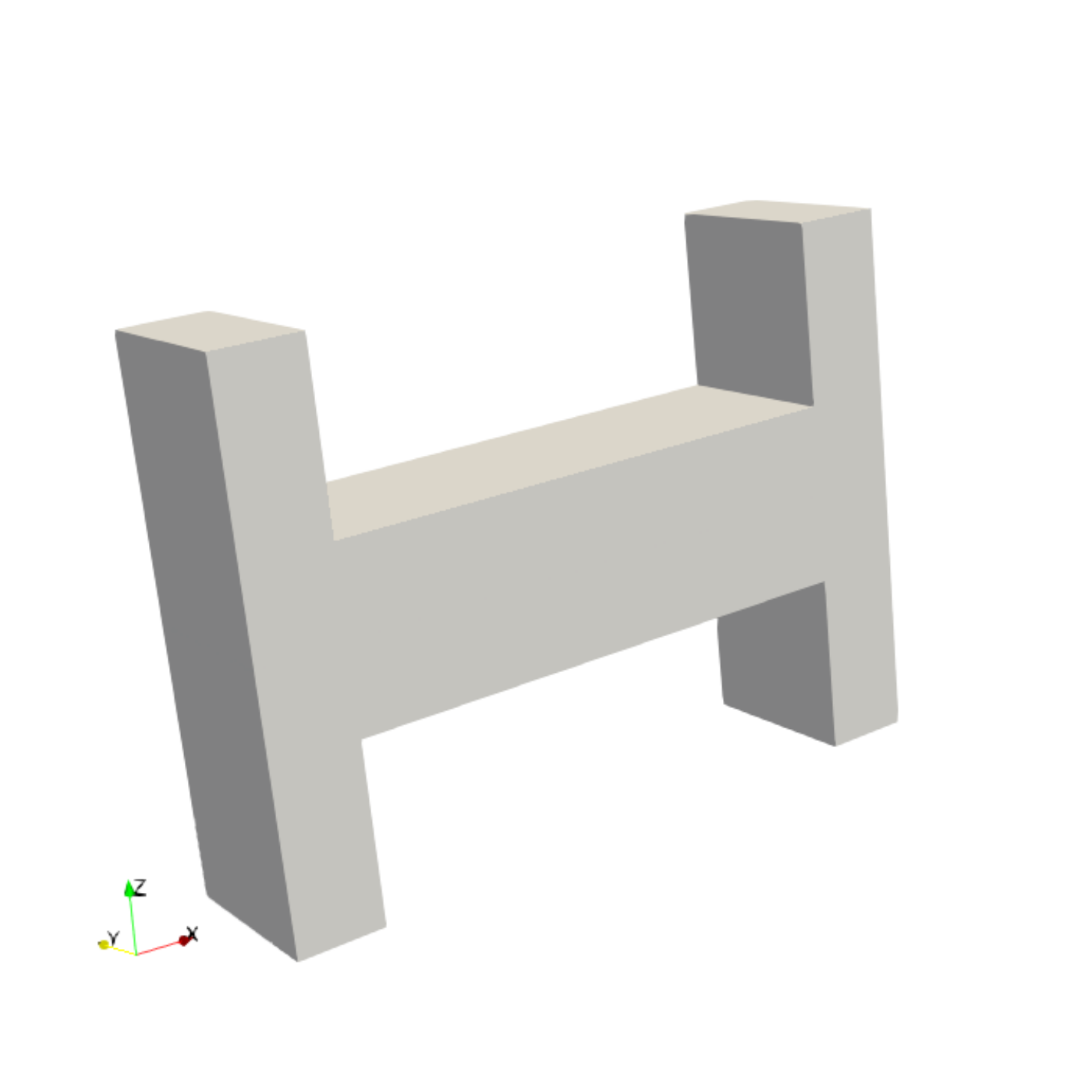}}}
    \label{fig:5x10_diag}
    \end{subfigure}
    \hfill
    \begin{subfigure}[ht]{0.24\textwidth}
    \caption{\quad MD-5x5}
    \centerline{
     {\includegraphics[width=\textwidth]{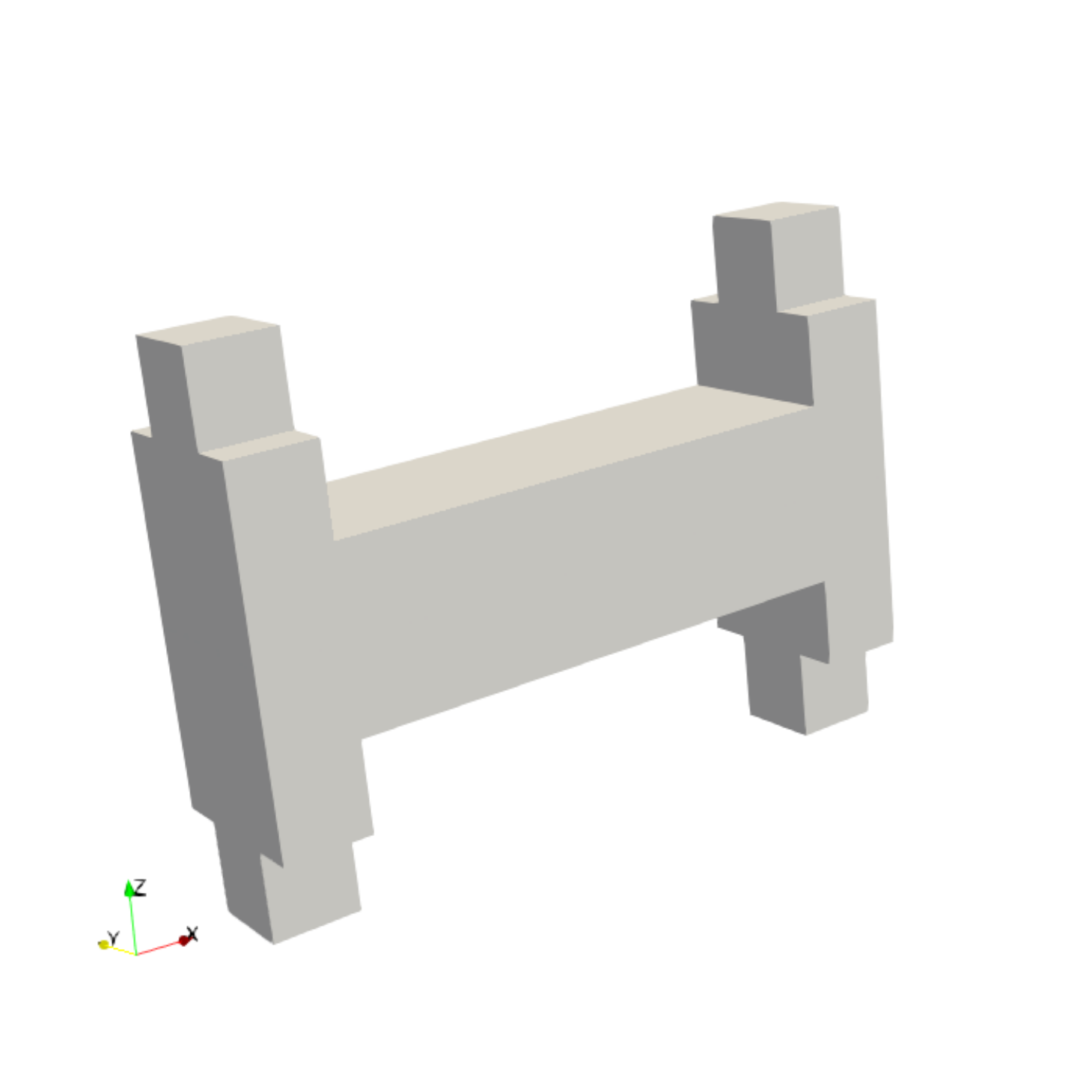}}}
    \label{fig:5x5_diag}
    \end{subfigure}
    \hfill
    \begin{subfigure}[ht]{0.24\textwidth}
     \caption{\quad MD-3x3}
    \centerline{
     {\includegraphics[width=\textwidth]{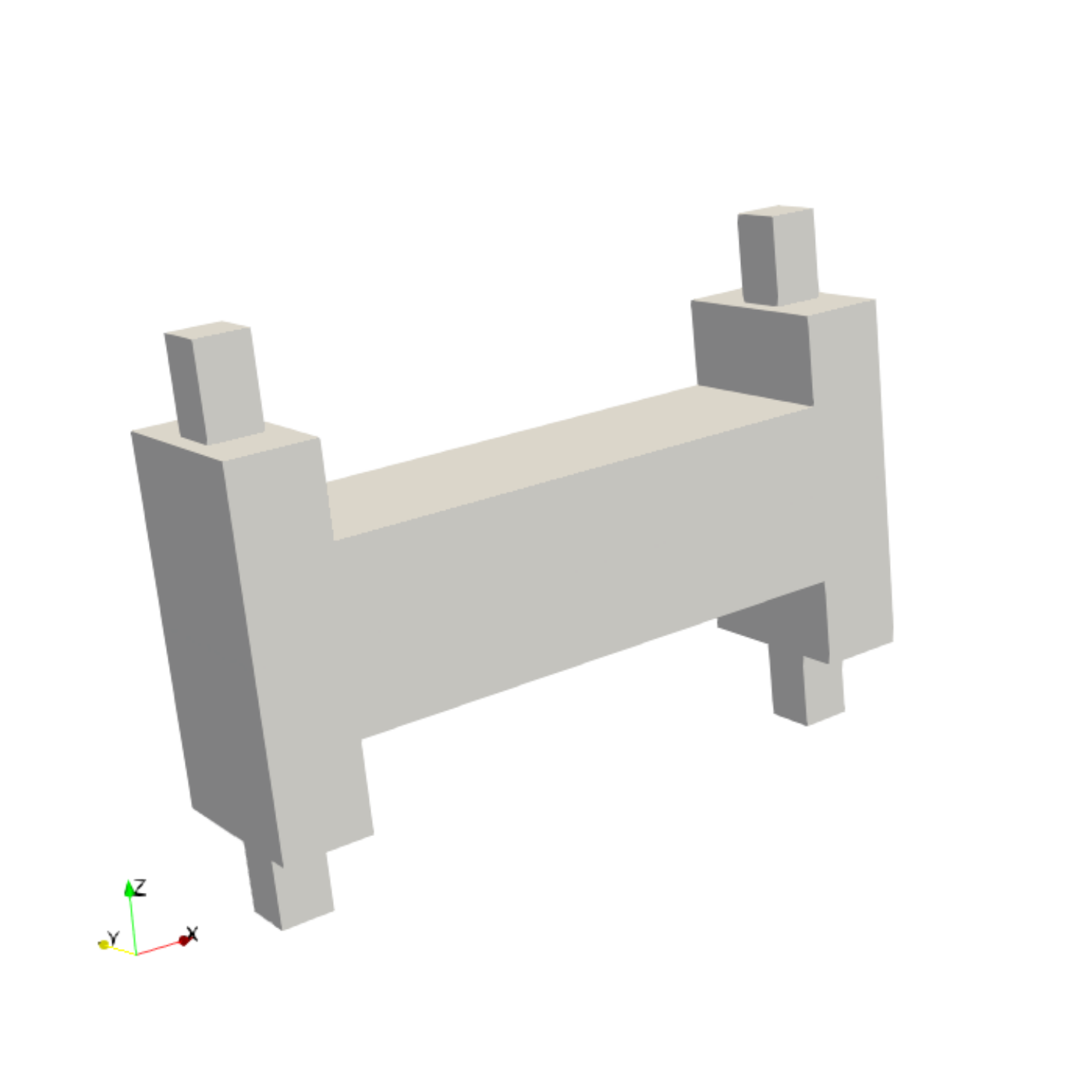}}}
    \label{fig:3x3_diag}
     \end{subfigure}
     \caption{Schematic diagrams of different geometries}
     \label{fig:schema_diag}
\end{figure*}

Three-dimensional simulations are performed in rectangular channels with four representative designs to elucidate the effects of right-angled bends and inlet. Dimensions and sketches of each design are shown in Figure~\ref{fig:schema_diag_dim} and in Figure~\ref{fig:schema_diag}, respectively. The thermophysical properties of the fluid, such as viscosity, density, etc., are calculated with the models described in ~\ref{sec:thermo_prop}. The grid spacing is uniform in the $x$-, $y$- and $z$-directions ($\Delta{x} = \Delta{y} = \Delta{z}_{\max} = \num{1.25e-4}$~\si{\meter}) except in proximity of the membrane  (details about grid convergence studies can be found in Section~\ref{sec:grid_conv}). To fully resolve the boundary layers near the membrane, a non-uniform grid spacing is used such that $\Delta{x} = \Delta{y} = \num{1.25e-4}$~\si{\meter} and $\Delta{z}_{\min} = \num{6.25e-6}$~\si{\meter}. In these simulations, we used adaptive time-step size such that the maximum Courant number
\begin{align}
    CFL_{\max} = \max \left( \frac{u\Delta t }{\Delta{x}} + \frac{v\Delta t }{\Delta{y}} + \frac{w\Delta t }{\Delta{z}} \right) 
\end{align}
is 0.8 and $\Delta{t}$~[\si{\second}] is the time step size. We simulate MD systems in counterflow operation and the inlet velocity of the simulations ranges from 0.001 to 0.7~\si{\meter\per\second}, with Reynolds number, $Re_{in}$, ranging between 5.0 and 3500 where $Re_{in}$ is defined as
\begin{align}
    &Re_{in} = \frac{u_{in}D_{h,in}}{\nu_0},
\end{align}
 with $D_{h,in}$~[\si{\meter}]  the hydraulic diameter of the inlet and $\nu_0=\num{1e-6}$~[\si{\meter\tothe{2}\per\second}]  the kinematic viscosity of pure water. 
The inlet temperatures of the feed and draw sides are 313.15~\si{\kelvin} and 293.15~\si{\kelvin}, which provides a temperature difference of 20~\si{\kelvin}. The inlet concentration of sodium chloride on the feed side is 35~\si{\kilogram\per\meter\tothe{3}} while that of the draw side is 0~\si{\kilogram\per\meter\tothe{3}}. All simulations are initiated with $\boldsymbol{u}(\boldsymbol{x}, 0) = 0$~\si{\meter\per\second}, $T(\boldsymbol{x}_f, 0) = 313.15$~\si{\kelvin}, $T(\boldsymbol{x}_d, 0) = 293.15$~\si{\kelvin}, $C(\boldsymbol{x}_f, 0) = 35$~\si{\kilogram\per\meter\tothe{3}} and $C(\boldsymbol{x}_d, 0) = 0$~\si{\kilogram\per\meter\tothe{3}} where $\boldsymbol{x}_f$ and $\boldsymbol{x}_d$ refer to locations within the feed and draw chamber, respectively. Table~\ref{tab:sim_params} summarizes the geometry and simulation parameters. The total simulation time $t_{f}$ is 100 $\tau$ to ensure the flow is sufficiently developed where 
\begin{align}
    \tau = \frac{L_f} {u_{in}}  \frac{\mathcal{A}_{c}}{\mathcal{A}_{in}},
\end{align}
represents the estimated  water retention time from inlet to outlet,  $\mathcal{A}_{in}$~[\si{\meter\tothe{2}}] $= L_I\times W_I$ is the area of the inlet, $\mathcal{A}_{c}$~[\si{\meter\tothe{2}}] $ = W_C\times H_C$ is the cross-sectional area of the main channel, and $L_f = 2H_I+2H_S+L_C = 0.06$~\si{\meter} is the approximate length of the systems.

\begin{table*}[hbtp!]
    \centering
    \caption{Summary of geometrical and simulation parameters.}
    \resizebox{!}{.35\paperheight}{%
    \resizebox{\linewidth}{!}{%
    \begin{tabular}{|M{0.3\textwidth} | M{0.15\textwidth}|M{0.15\textwidth}|M{0.15\textwidth}|M{0.15\textwidth}|}
    \hline \hline
        &   MD-5x10S    &   MD-5x10 &   MD-5x5  &   MD-3x3 \\ \hline 
    Inlet/Outlet ($L_I\times W_I \times H_I$~[10$^{-3}$\si{\meter]}) & 5$\times$10$\times$5 & 5$\times$10$\times$5 & 5$\times$5$\times$5 & 3$\times$3$\times$5 \\ \hline
    Side chamber ($L_S\times W_S \times H_S$~[10$^{-3}$\si{\meter]}) & 10$\times$10$\times$5 & \multicolumn{3}{c|}{5$\times$10$\times$10} \\ \hline
     Main channel ($L_C \times W_C \times H_C$~[10$^{-3}$\si{\meter}]) & 30$\times$10$\times$5 & \multicolumn{3}{c|}{30$\times$10$\times$5} \\ \hline
     Bends & No & \multicolumn{3}{c|}{Yes} \\  \hline
     Volumetric flow rate [10$^{-8}$\si{\meter\tothe{3}\per\second}] & 7.50~--~1500 & 7.50~--~1500 & 2.5~--~1750 & 3.0~--~600 \\\hline
     Inlet velocity ($u_{in}$~[\si{\meter\per\second}]) & 0.0015~--~0.3000 & 0.0015~--~0.3000 & 0.0010~--~0.7000 & 0.0033~--~0.6667 \\ \hline 
     Reynolds number ($Re_{in}$) & 10~--~2000 & 10~--~2000 & 5.0~--~3500 & 10~--~2000 \\ \hline 
     Feed inlet concentration ($C_{in,f}$~[\si{\kilogram\per\meter\tothe{3}}]) & \multicolumn{4}{c|}{35} \\ \hline
     Draw inlet concentration ($C_{in,d}$~[\si{\kilogram\per\meter\tothe{3}}]) & \multicolumn{4}{c|}{0} \\ \hline
     Feed inlet temperature ($T_{in,f}$~[\si{\kelvin}]) & \multicolumn{4}{c|}{313.15}\\ \hline
     Draw inlet temperature ($T_{in,d}$~[\si{\kelvin}]) & \multicolumn{4}{c|}{293.15} \\ \hline
    Maximum Counrant number $CFL_{\max}$ & \multicolumn{4}{c|}{0.8} \\ \hline 
    Number of simulations & 20 & 20 & 25 & 22\\ \hline \hline 
    \end{tabular}}}
    \label{tab:sim_params}
\end{table*}

\subsection{Grid convergence studies}
\label{sec:grid_conv}
\begin{figure*}
    \centering
    \begin{subfigure}[ht]{0.49\textwidth}
    \caption{}
    \centerline{
     {\includegraphics[width=\textwidth]{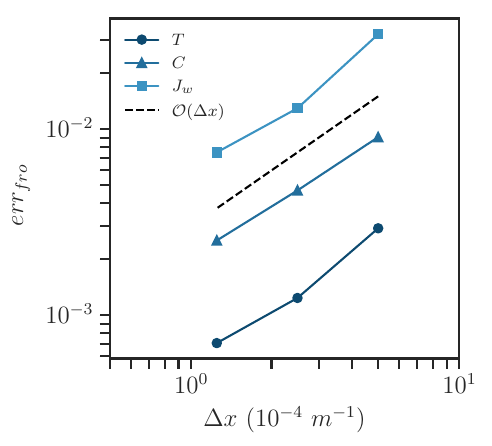}}}
    
    \label{fig:err_fro_mem}
    \end{subfigure}
    \begin{subfigure}[ht]{0.49\textwidth}
     \caption{}
    \centerline{
     {\includegraphics[width=\textwidth]{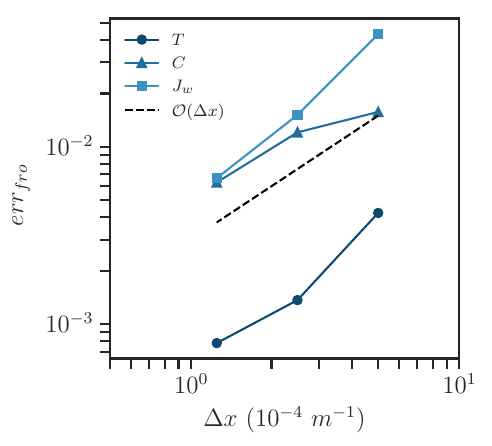}}}
    \label{fig:err_inf_mem}
     \end{subfigure}
     \caption{Grid convergence studies for temperature, concentration and permeate flux of two types of errors, (a) $err_{fro}$ and (b) $err_\infty$, respectively.}
     \label{fig:err_mem}
\end{figure*}

To understand the grid resolution required to obtain accurate simulation results, we perform grid convergence studies with geometry MD-5x5 of Figure~\ref{fig:schema_diag}(\subref{fig:5x5_diag}). The simulation is set up according to Section~\ref{sec:sim_setup} and the largest $Re_{in} = 2000$ is chosen as the inlet boundary condition. Table~\ref{tab:grid_conv} summarizes the resolution of the grid ($\Delta{x}$~[\si{\meter}]) and the total number of cells.  \revision{We define two types of error for each quantity of interest $QI$ on the membrane surface}, $err_{fro}$ and $err_{\infty}$, as
\begin{subequations}
\begin{align}
&err_{fro} = \frac{\left\Vert QI_{\Delta x} - QI_{\Delta x = 0.0625} \right\Vert_{fro}}{\enstmean{QI_{\Delta x = 0.0625}}}, \\
&err_{\infty} = \frac{\left\Vert QI_{\Delta x} - QI_{\Delta x = 0.0625} \right\Vert_{\infty}}{\enstmean{QI_{\Delta x = 0.0625}}},
\end{align}
\end{subequations}
where $QI$ is the quantity of interest, i.e.  $T$, $C$ and $J_w$, and $\enstmean{QI}$ is defined as the time- and spatial-average $QI$:
\begin{align}
\label{eq:t-x-avg-ops}
    \enstmean{QI} = \frac{1}{N_tN_xN_y} \sum_{n,i,j}^{N_t, N_x, N_y} QI_{i,j}^n,
\end{align}
where $QI_{i,j}^n=QI(x_i, y_j, t_n)$ is $QI$ at locations $x_i$ and $y_j$ and time $t_n$, $i=\{1,2, \cdots N_x\}$, $j=\{1,2, \cdots N_y\}$, $n=\{51\tau, 52\tau, \cdots 100\tau\}$, $N_x$ and $N_y$ are the mesh resolutions in the $x$ and $y$ directions, and $N_t=50$ is the number of time snapshots. The errors $err_{fro}$ and $err_{\infty}$ measure the total  and  maximum errors, respectively.  Figures~\ref{fig:err_mem}(\subref{fig:err_fro_mem}) and \ref{fig:err_mem}(\subref{fig:err_inf_mem}) show  $err_{fro}$ and $err_{\infty}$ as functions of $\Delta{x}$ for the feed-side membrane.  Overall, convergence in both errors has been observed. We choose $\Delta{x} = \SI{0.125e-3}{\meter}$ as the grid resolution for the subsequent simulations because both errors for all three quantities of interest are less than 1\%.
\begin{table*}[hbtp!]
    \centering
    \caption{Summary of grid resolutions and total number of cells used for grid convergence studies.}
    \begin{tabular}{|M{0.25\textwidth} | M{0.15\textwidth}|M{0.15\textwidth}|M{0.15\textwidth}|M{0.15\textwidth}|}
    \hline \hline
       $\Delta x$~[$10^{-3}$ \si{\meter}] &  0.5    &   0.25 &   0.125  &   0.0625 \\ \hline 
    No. of grid cells & 53,600 & 454,400 & 3,686,400  & 29,491,200  \\ \hline \hline 
    \end{tabular}
    \label{tab:grid_conv}
\end{table*}

\section{Results}
\label{sec:results}
\subsection{The effect of right-angled bend and inlets with sudden expansions on the formation of Dean vortices}
\label{sec:flow_char}
\begin{figure*}[hbtp!]
    \centering
    \begin{subfigure}[ht]{0.23\textwidth}
    \caption*{\centering MD-5x10S}
    \centerline{
     {\includegraphics[width=\textwidth, trim={0in 0in 0in 1.5in},clip]{5x10s_geom-eps-converted-to.pdf}}}
    \end{subfigure}
    \begin{subfigure}[ht]{0.23\textwidth}
    \caption*{\centering MD-5x10}
    \centerline{
     {\includegraphics[width=\textwidth, trim={0in 0in 0in 1.5in},clip]{5x10_geom-eps-converted-to.pdf}}}
    \end{subfigure}
    \begin{subfigure}[ht]{0.23\textwidth}
    \caption*{\centering MD-5x5}
    \centerline{
     {\includegraphics[width=\textwidth, trim={0in 0in 0in 1.5in},clip]{5x5_geom-eps-converted-to.pdf}}}
    \end{subfigure}
    \begin{subfigure}[ht]{0.23\textwidth}
     \caption*{\centering MD-3x3}
    \centerline{
     {\includegraphics[width=\textwidth, trim={0in 0in 0in 1.5in},clip]{3x3_geom-eps-converted-to.pdf}}}
     \end{subfigure}
    \begin{subfigure}[ht]{0.24\textwidth}
    \caption{}
    \centerline{
     {\includegraphics[width=\textwidth]{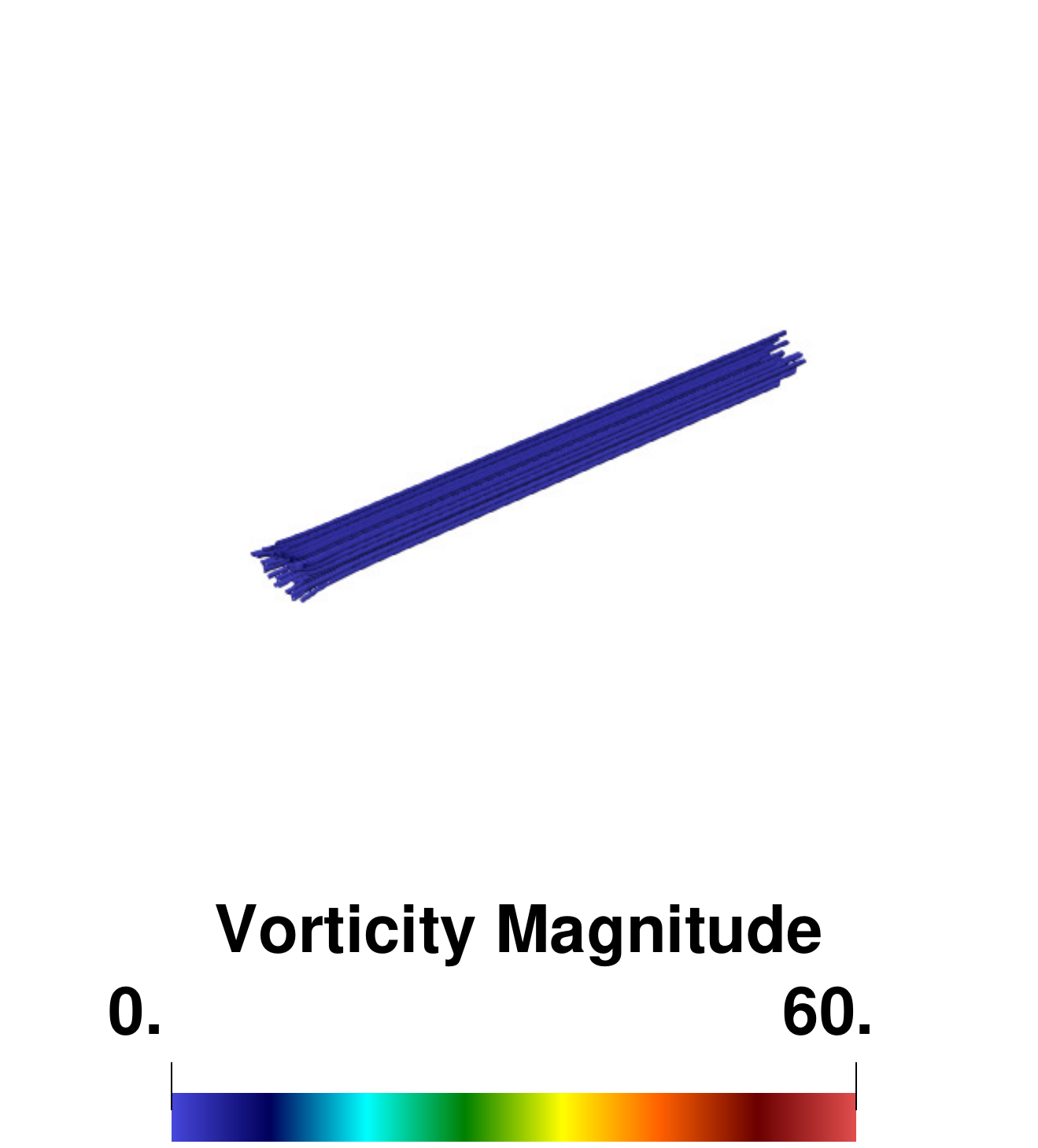}}}
    \label{fig:5x10s_0010_sl}
    \end{subfigure}
    \begin{subfigure}[ht]{0.24\textwidth}
    \caption{}
    \centerline{
     {\includegraphics[width=\textwidth]{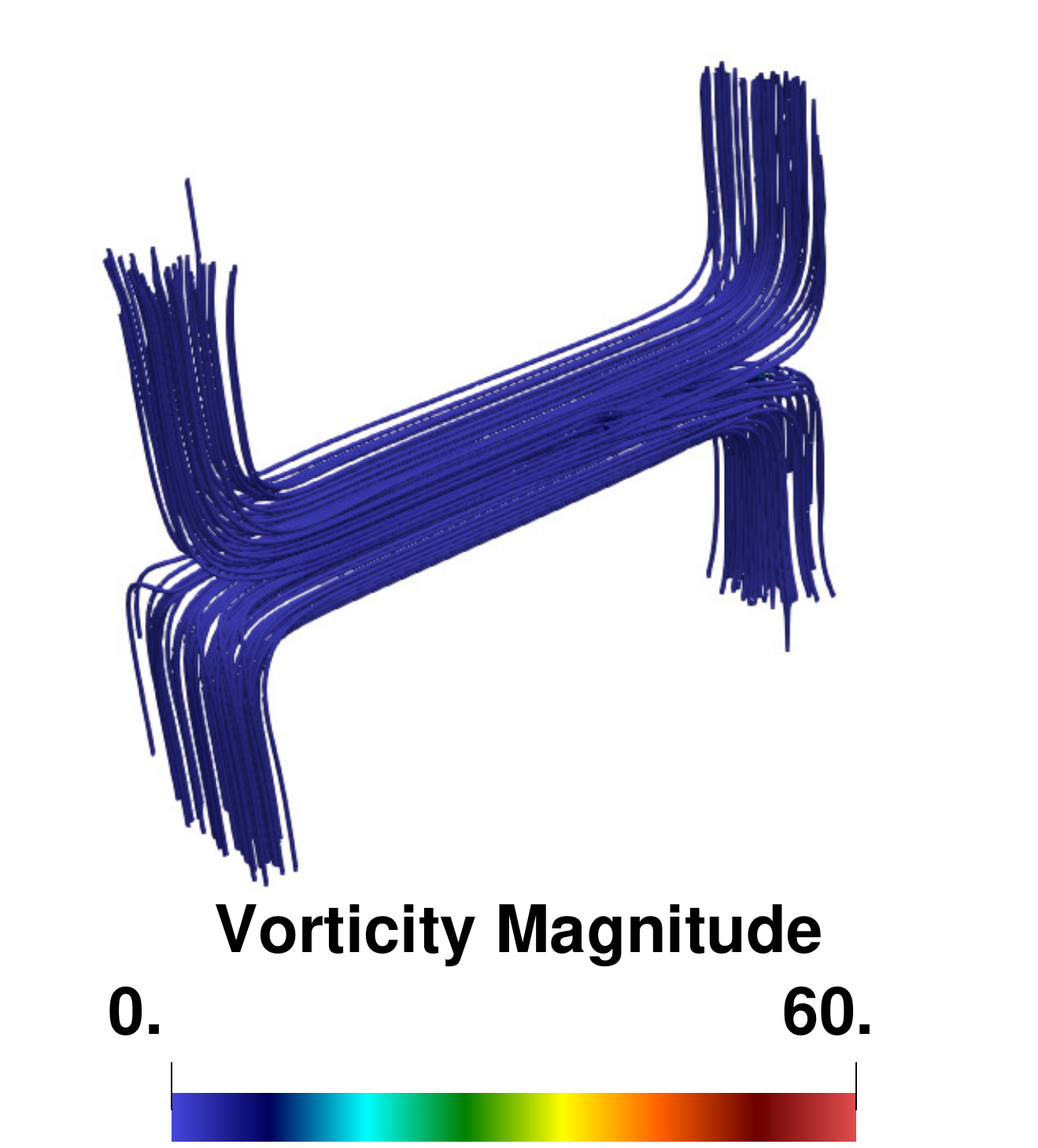}}}
    \label{fig:5x10_0010_sl}
    \end{subfigure}
    \begin{subfigure}[ht]{0.24\textwidth}
    \caption{}
    \centerline{
     {\includegraphics[width=\textwidth]{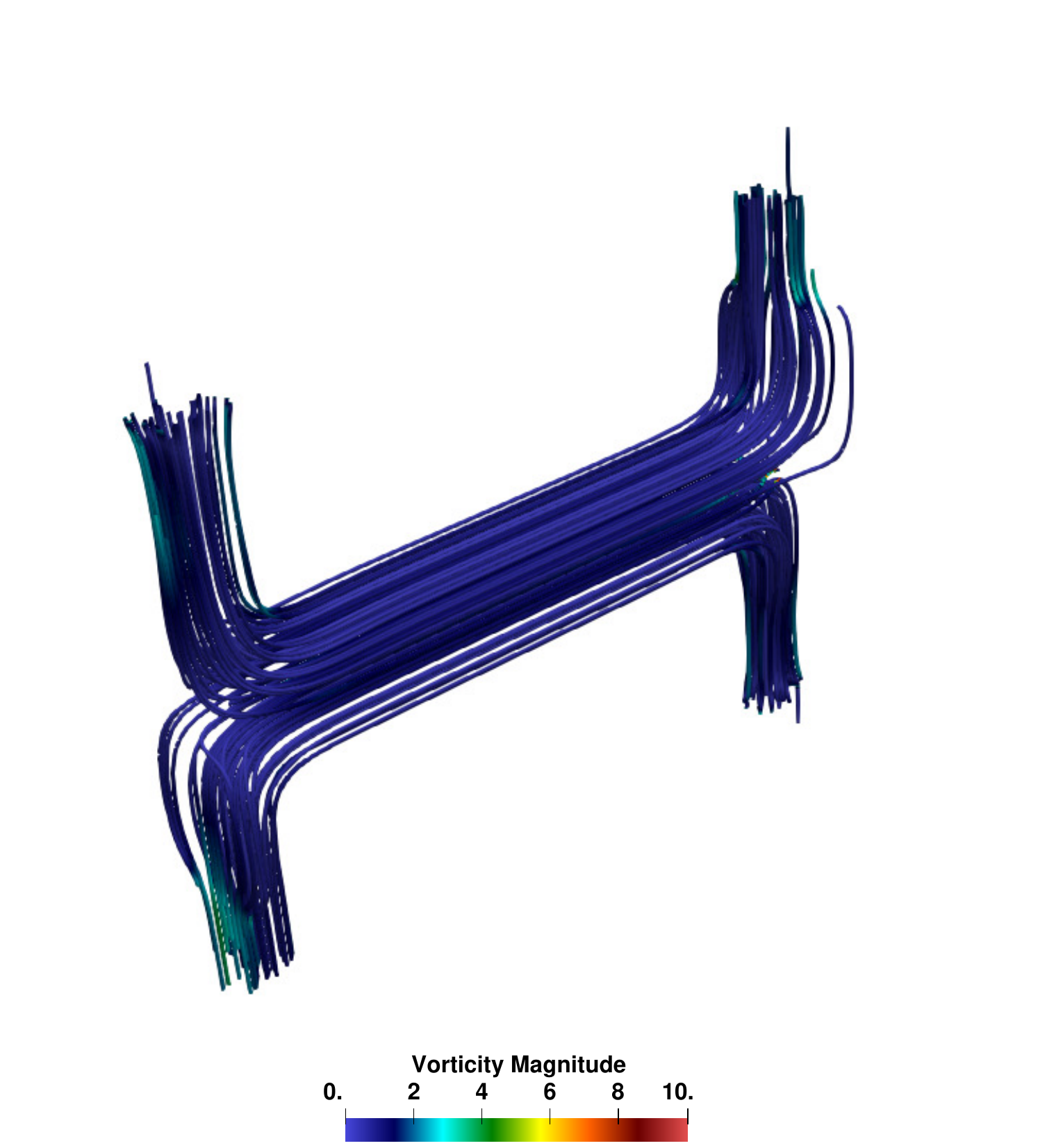}}}
    \label{fig:5x5_0010_sl}
    \end{subfigure}
    \begin{subfigure}[ht]{0.24\textwidth}
    \caption{}
    \centerline{
     {\includegraphics[width=\textwidth]{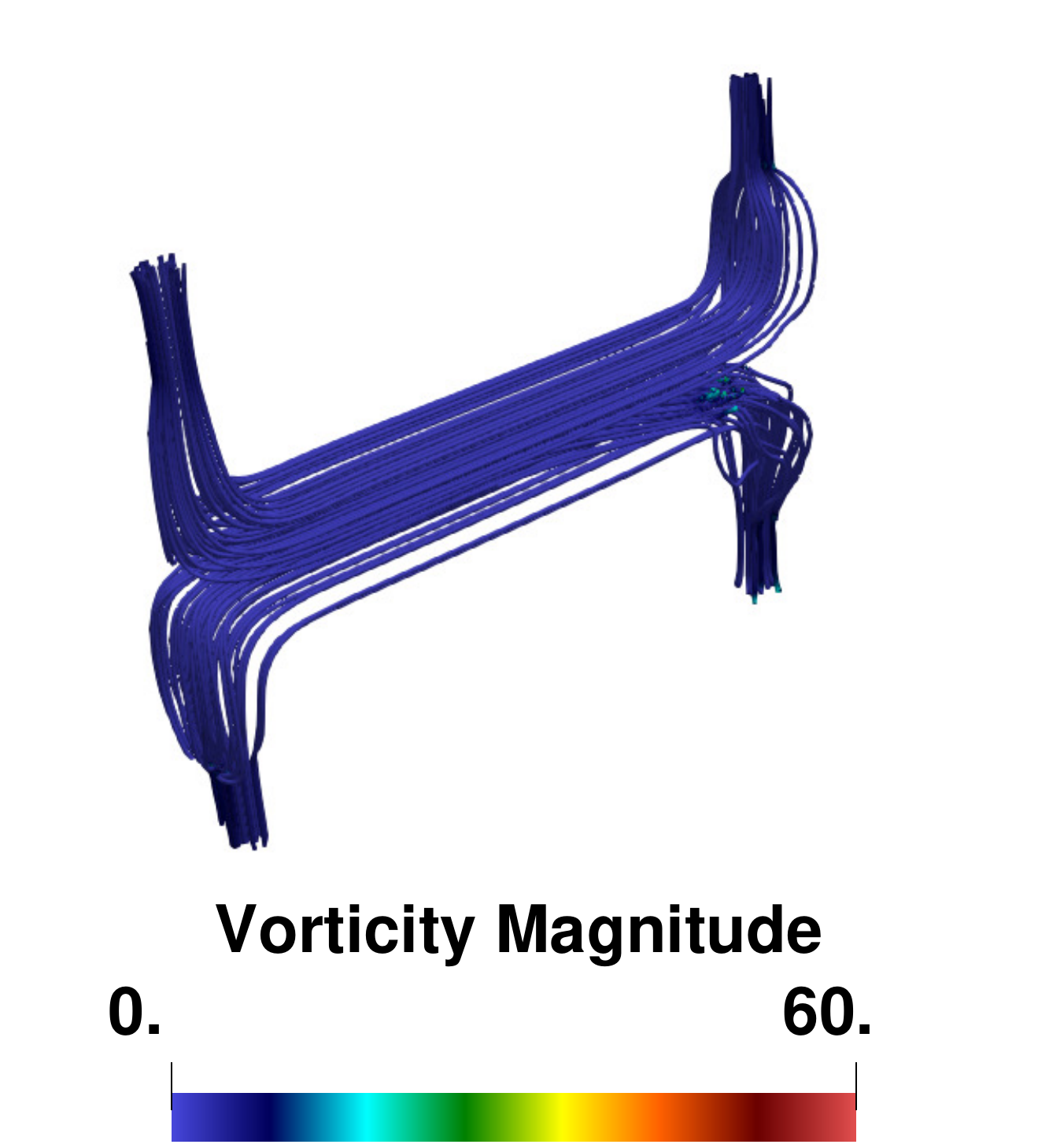}}}
    \label{fig:3x3_0010_sl}
    \end{subfigure}

    \medskip
    
    \begin{subfigure}[ht]{0.24\textwidth}
     \caption{}
    \centerline{
     {\includegraphics[width=\textwidth]{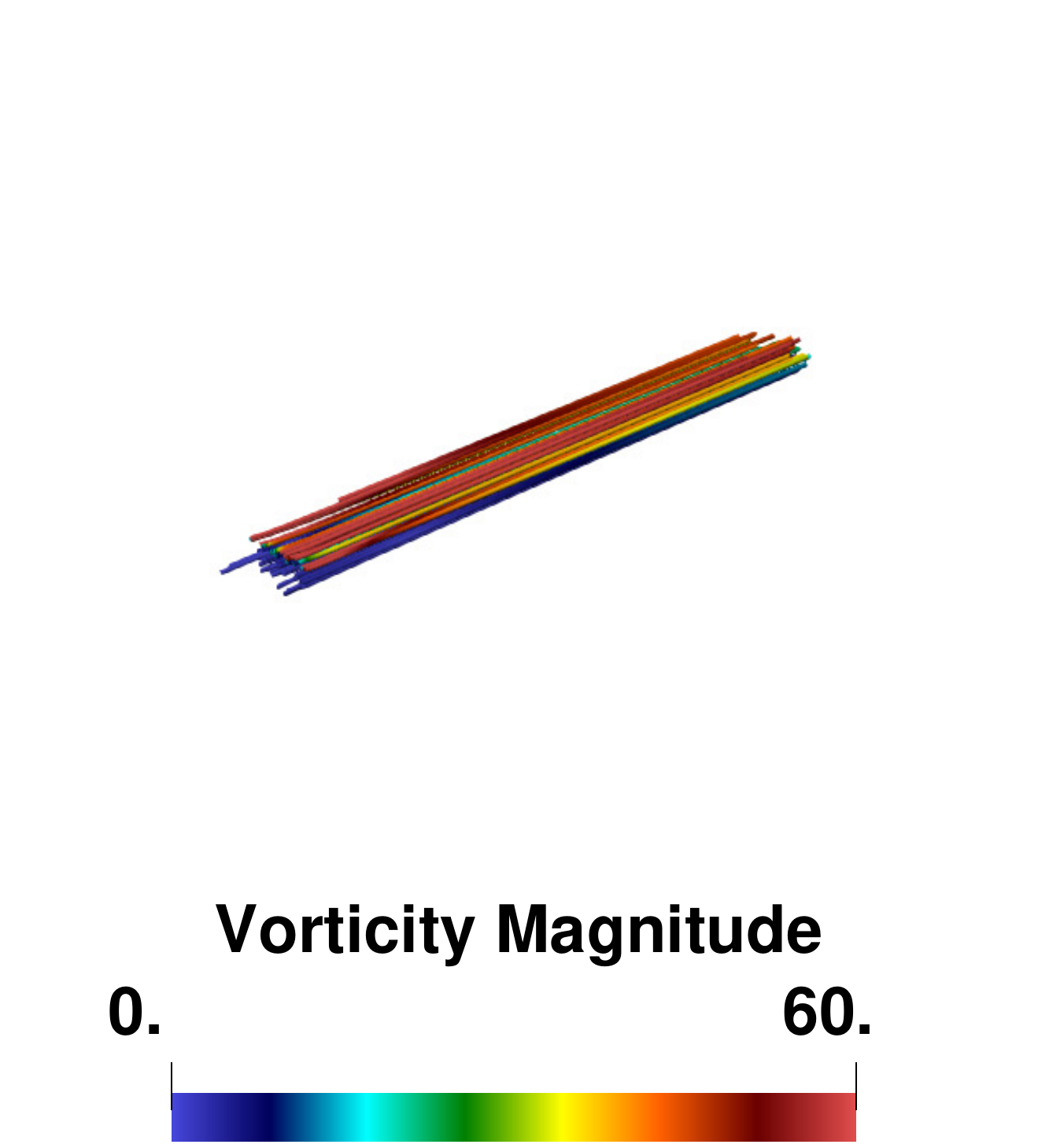}}}
    \label{fig:5x10s_0250_sl}
     \end{subfigure}
    \begin{subfigure}[ht]{0.24\textwidth}
     \caption{}
    \centerline{
     {\includegraphics[width=\textwidth]{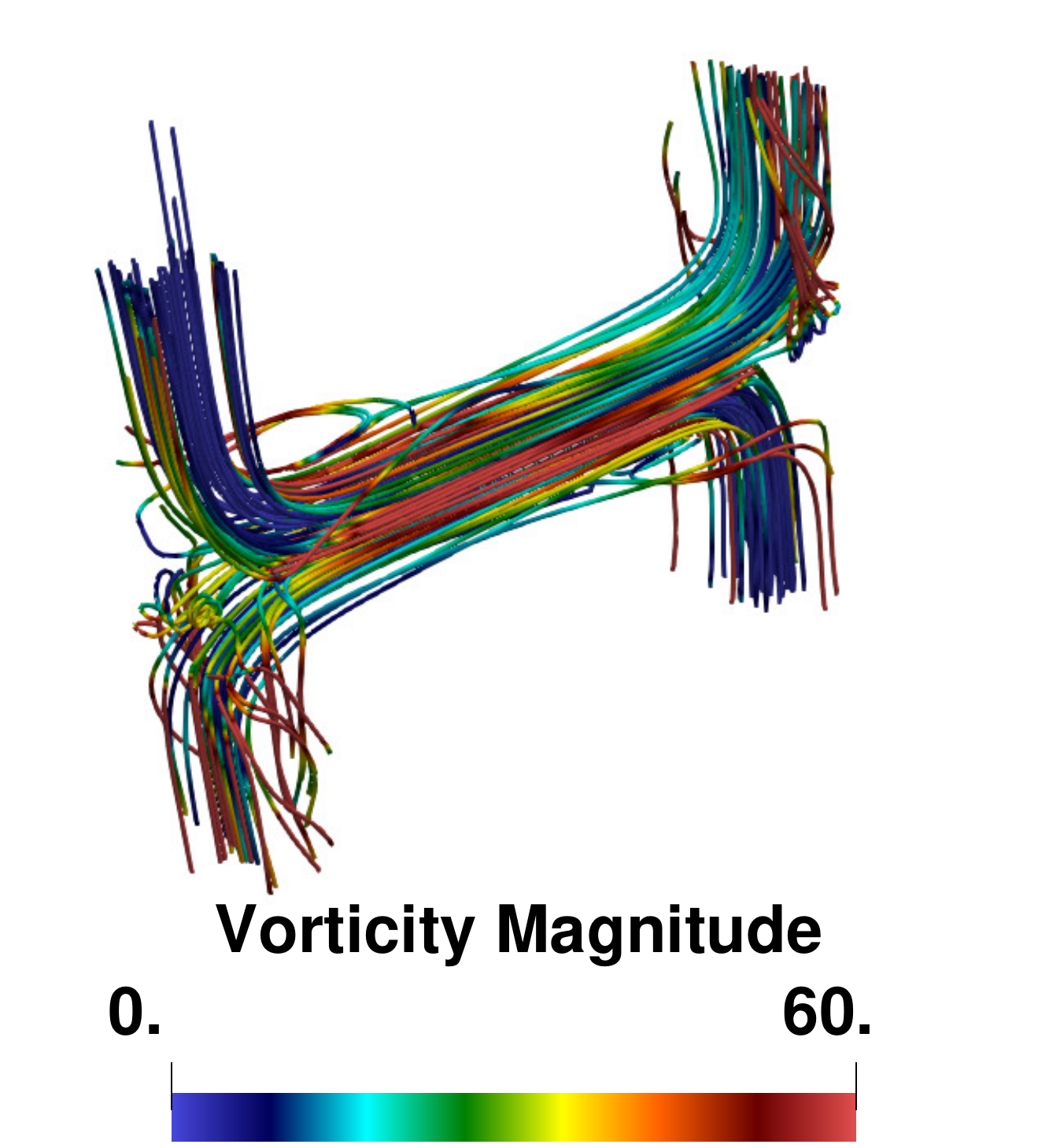}}}
    \label{fig:5x10_0250_sl}
     \end{subfigure}
    \begin{subfigure}[ht]{0.24\textwidth}
     \caption{}
    \centerline{
     {\includegraphics[width=\textwidth]{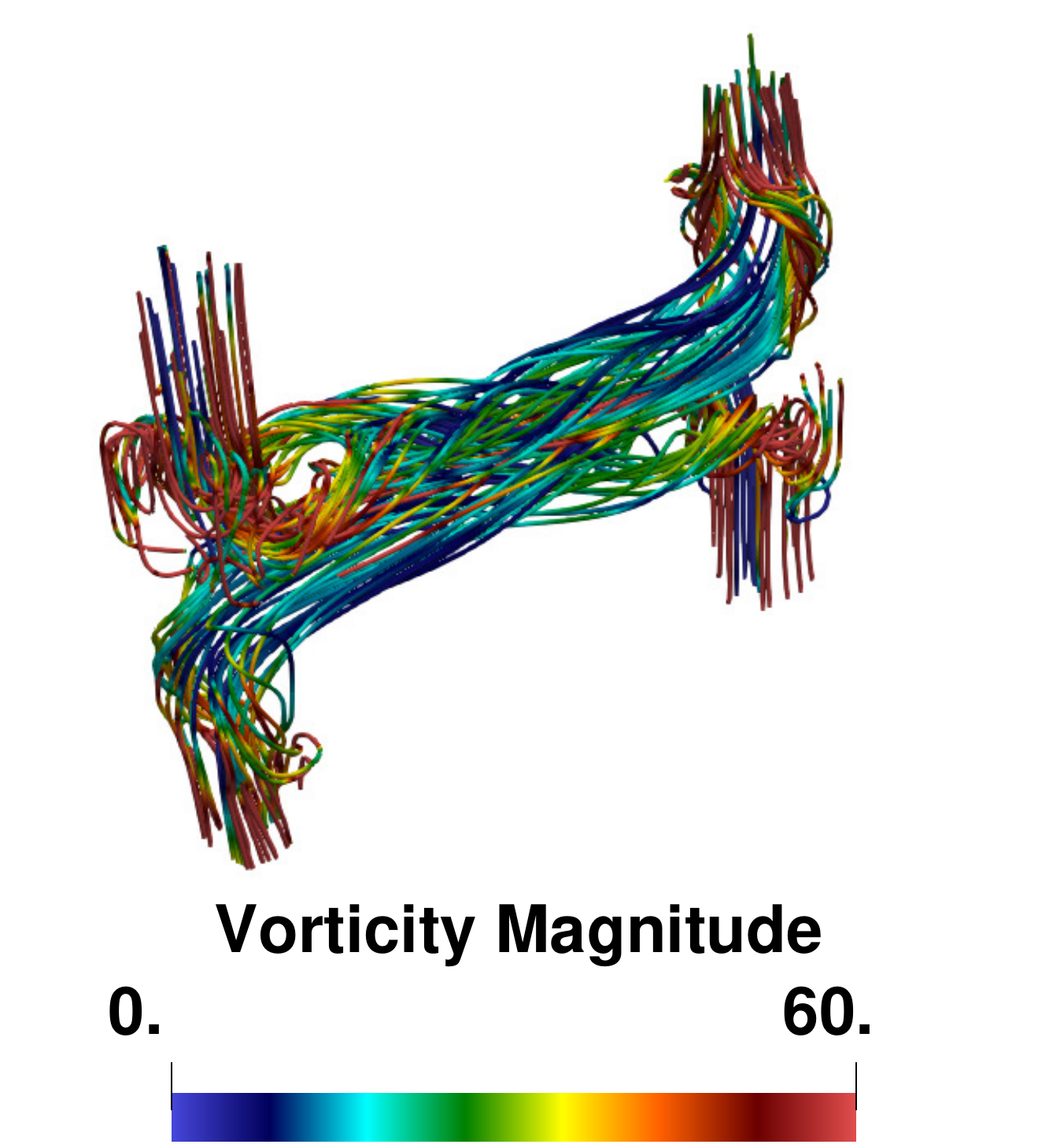}}}
    \label{fig:5x5_0250_sl}
     \end{subfigure}
    \begin{subfigure}[ht]{0.24\textwidth}
     \caption{}
    \centerline{
     {\includegraphics[width=\textwidth]{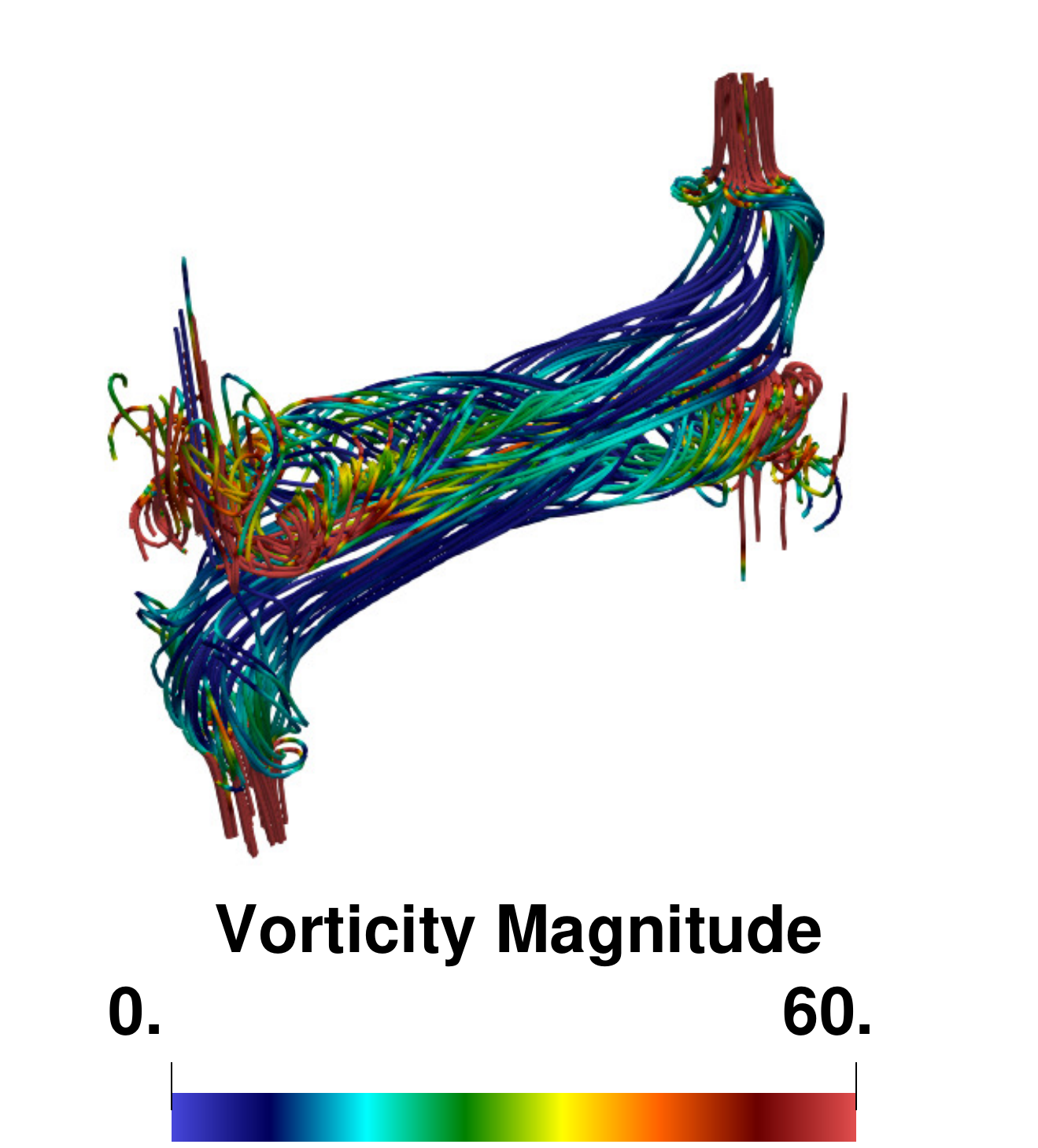}}}
    \label{fig:3x3_0250_sl}
     \end{subfigure}
     
    \medskip
    
     \begin{subfigure}[ht]{0.24\textwidth}
     \caption{}
    \centerline{
     {\includegraphics[width=\textwidth]{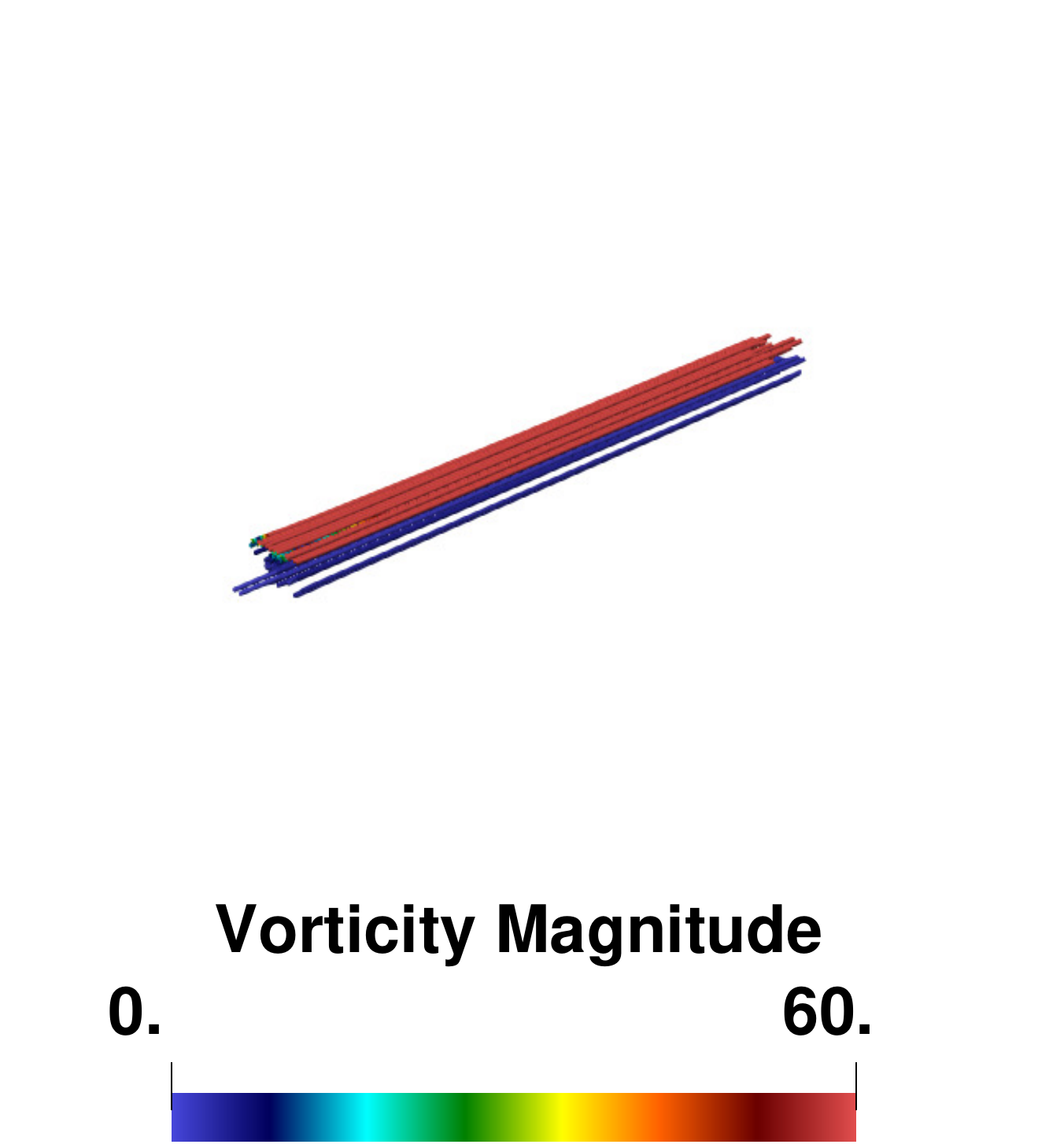}}}
    \label{fig:5x10s_2000_sl}
     \end{subfigure}
     \begin{subfigure}[ht]{0.24\textwidth}
     \caption{}
    \centerline{
     {\includegraphics[width=\textwidth]{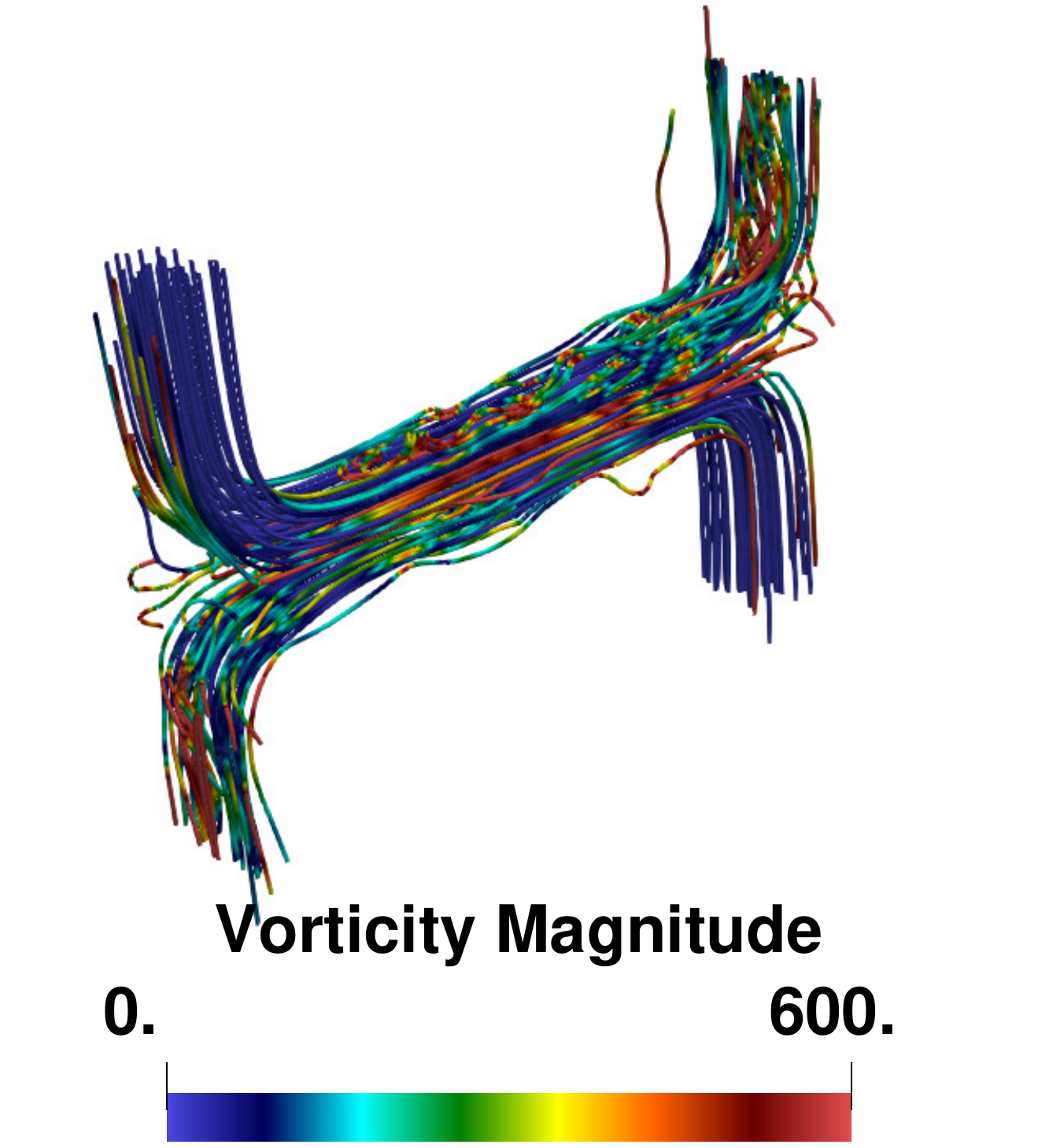}}}
    \label{fig:5x10_2000_sl}
     \end{subfigure}
     \begin{subfigure}[ht]{0.24\textwidth}
     \caption{}
    \centerline{
     {\includegraphics[width=\textwidth]{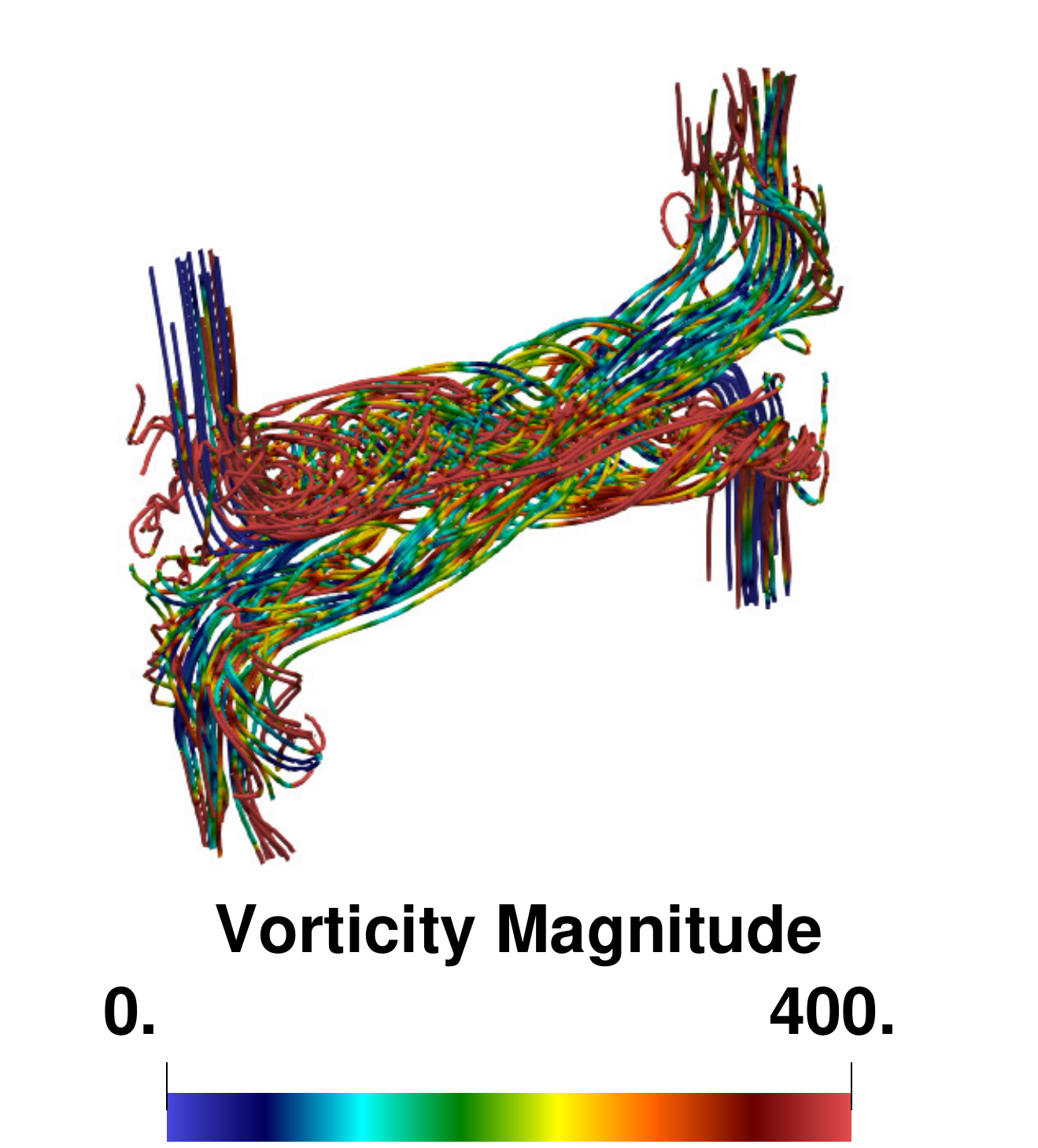}}}
    \label{fig:5x5_2000_sl}
     \end{subfigure}
     \begin{subfigure}[ht]{0.24\textwidth}
     \caption{}
    \centerline{
     {\includegraphics[width=\textwidth]{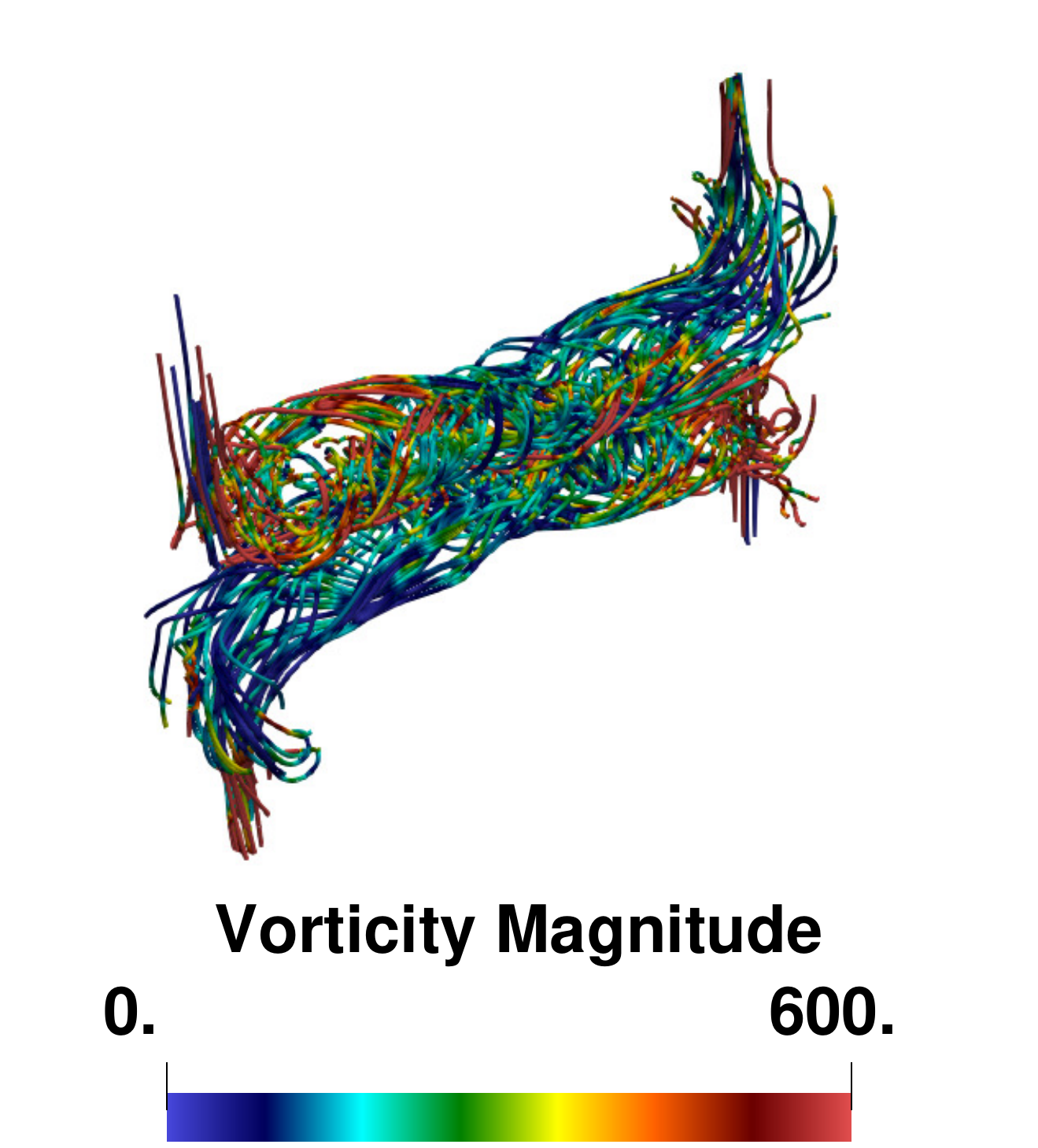}}}
    \label{fig:3x3_2000_sl}
     \end{subfigure}
     
     \caption{Instantenous streamlines colored by the magnitude of vorticity for $Re_{in}=10$ (first row) $Re_{in}=250$ (second row) and $Re_{in}=2000$ (third row) of MD-5x10S, MD-5x10, MD-5x5 and MD-3x3 (left to right).}
     \label{fig:cases_sl}
\end{figure*}

To understand the flow dynamics in MD systems, the instantaneous fluid streamlines for three representative Reynolds numbers $Re_{in}=10$, $250$ and $2000$ and all four geometries are calculated  and plotted in Figure~\ref{fig:cases_sl}. For a low Reynolds number ($Re_{in}=10$), the flow is laminar and no distortion or swirling is observed, as indicated by the smooth/straight streamlines in Figures~\ref{fig:cases_sl}(\subref{fig:5x10s_0010_sl})~--~(\subref{fig:3x3_0010_sl}). For a moderate Reynolds number ($Re_{in}=250$) (Figures~\ref{fig:cases_sl}(\subref{fig:5x10s_0250_sl})~--~(\subref{fig:3x3_0250_sl})), no swirling is observed for the straight channel systems (MD-5x10S). The extent of swirling gradually intensifies in the MD-5x10, MD-5x5 and MD-3x3 systems with the presence of right-angled bends and for inlets with sudden expansions. As the Reynolds number further increases ($Re_{in} = 2000$) (Figures~\ref{fig:cases_sl}(\subref{fig:5x10s_2000_sl})~--~(\subref{fig:3x3_2000_sl})), the streamlines in the MD-5x10, MD-5x5 and MD-3x3 systems become more chaotic while  those in the MD-5x10S geometry remain smooth. When comparing the moderate Reynolds number cases for designs MD-5x10 and MD-5x5 (Figures~\ref{fig:cases_sl}(\subref{fig:5x10_0250_sl}) vs~(\subref{fig:5x5_0250_sl})), the origination/appearance of the swirls differs. In MD-5x10, the swirling originates in the main channel after the right-angled bend, while in MD-5x5 design the onset of the swirling motion is  in the side chamber, i.e. after the inlet and before the right-angled bend. This suggests that  two  mechanisms may play a role in the formation of vortices. 

\begin{figure*}
    \centering
    \begin{subfigure}[ht]{0.23\textwidth}
    \caption*{\centering MD-5x10S}
    \centerline{
     {\includegraphics[width=\textwidth, trim={0in 0in 0in 1.5in},clip]{5x10s_geom-eps-converted-to.pdf}}}
    \end{subfigure}
    \begin{subfigure}[ht]{0.23\textwidth}
    \caption*{\centering MD-5x10}
    \centerline{
     {\includegraphics[width=\textwidth, trim={0in 0in 0in 1.5in},clip]{5x10_geom-eps-converted-to.pdf}}}
    \end{subfigure}
    \begin{subfigure}[ht]{0.23\textwidth}
    \caption*{\centering MD-5x5}
    \centerline{
     {\includegraphics[width=\textwidth, trim={0in 0in 0in 1.5in},clip]{5x5_geom-eps-converted-to.pdf}}}
    \end{subfigure}
    \begin{subfigure}[ht]{0.23\textwidth}
     \caption*{\centering MD-3x3}
    \centerline{
     {\includegraphics[width=\textwidth, trim={0in 0in 0in 1.5in},clip]{3x3_geom-eps-converted-to.pdf}}}
     \end{subfigure}
    \begin{subfigure}[ht]{0.24\textwidth}
    \caption{}
    \centerline{
     {\includegraphics[width=\textwidth]{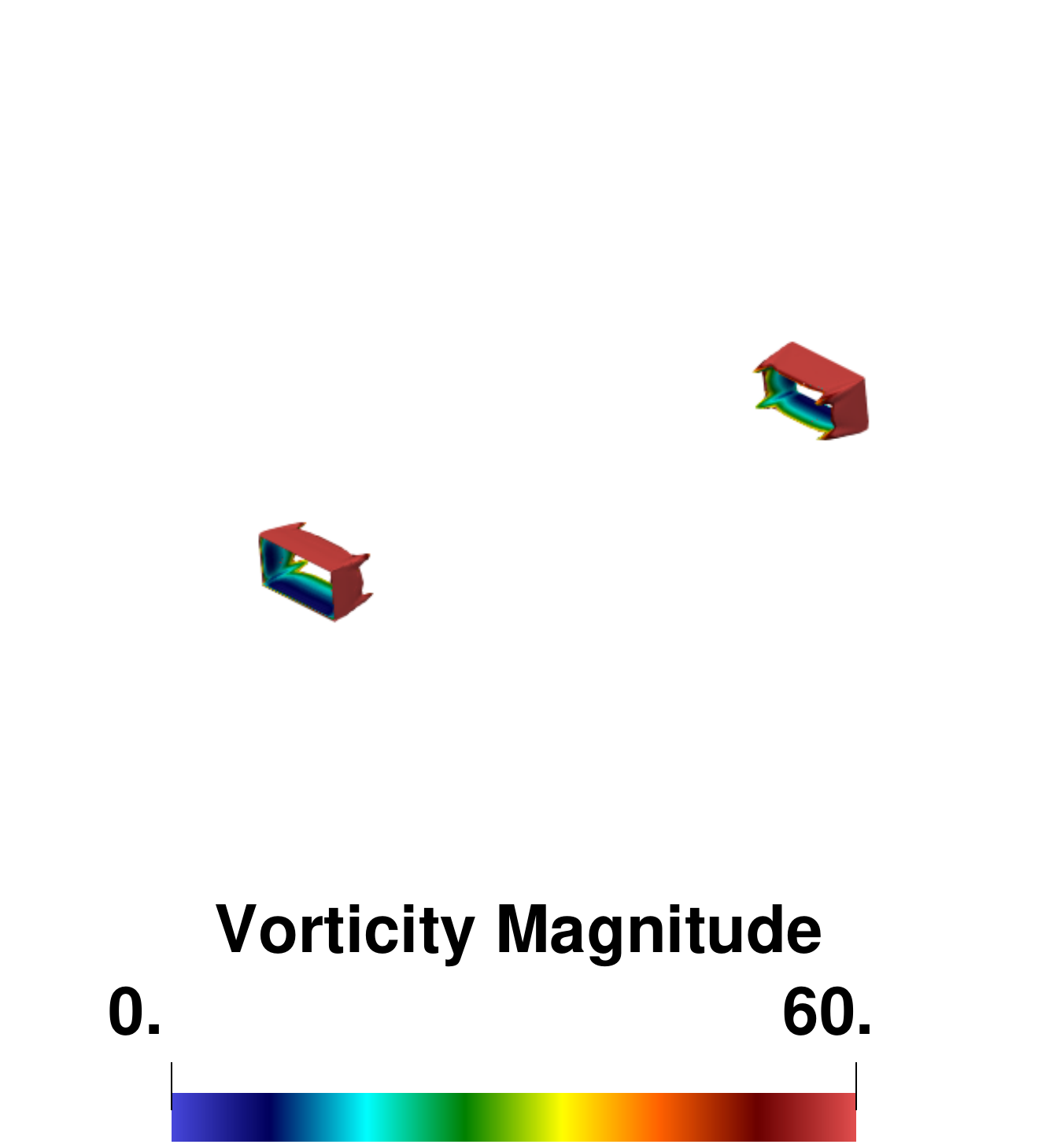}}}
    \label{fig:5x10s_0250_Q}
    \end{subfigure}
    \hfill
    \begin{subfigure}[ht]{0.24\textwidth}
    \caption{}
    \centerline{
     {\includegraphics[width=\textwidth]{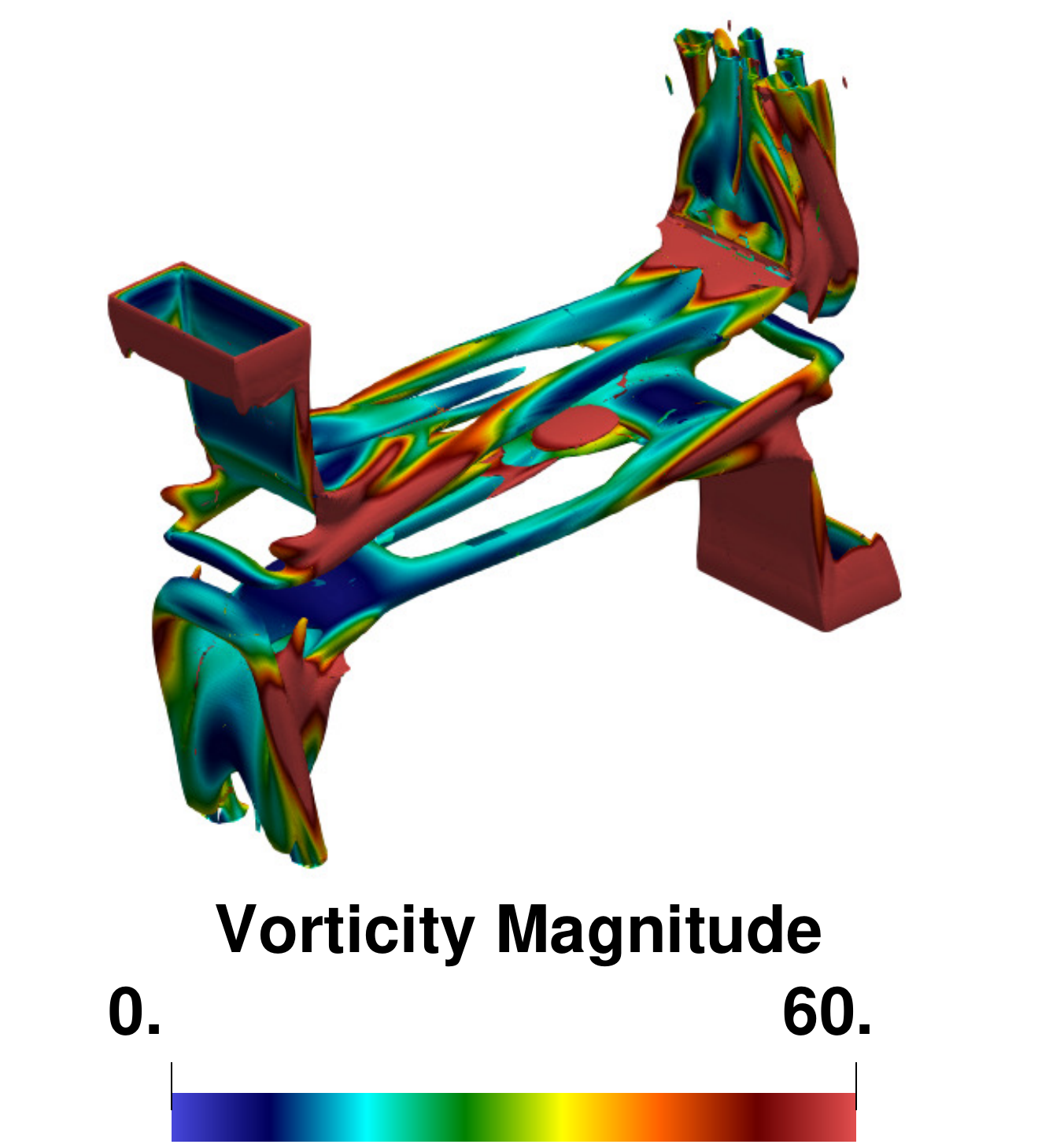}}}
    \label{fig:5x10_0250_Q}
    \end{subfigure}
    \hfill
    \begin{subfigure}[ht]{0.24\textwidth}
    \caption{}
    \centerline{
     {\includegraphics[width=\textwidth]{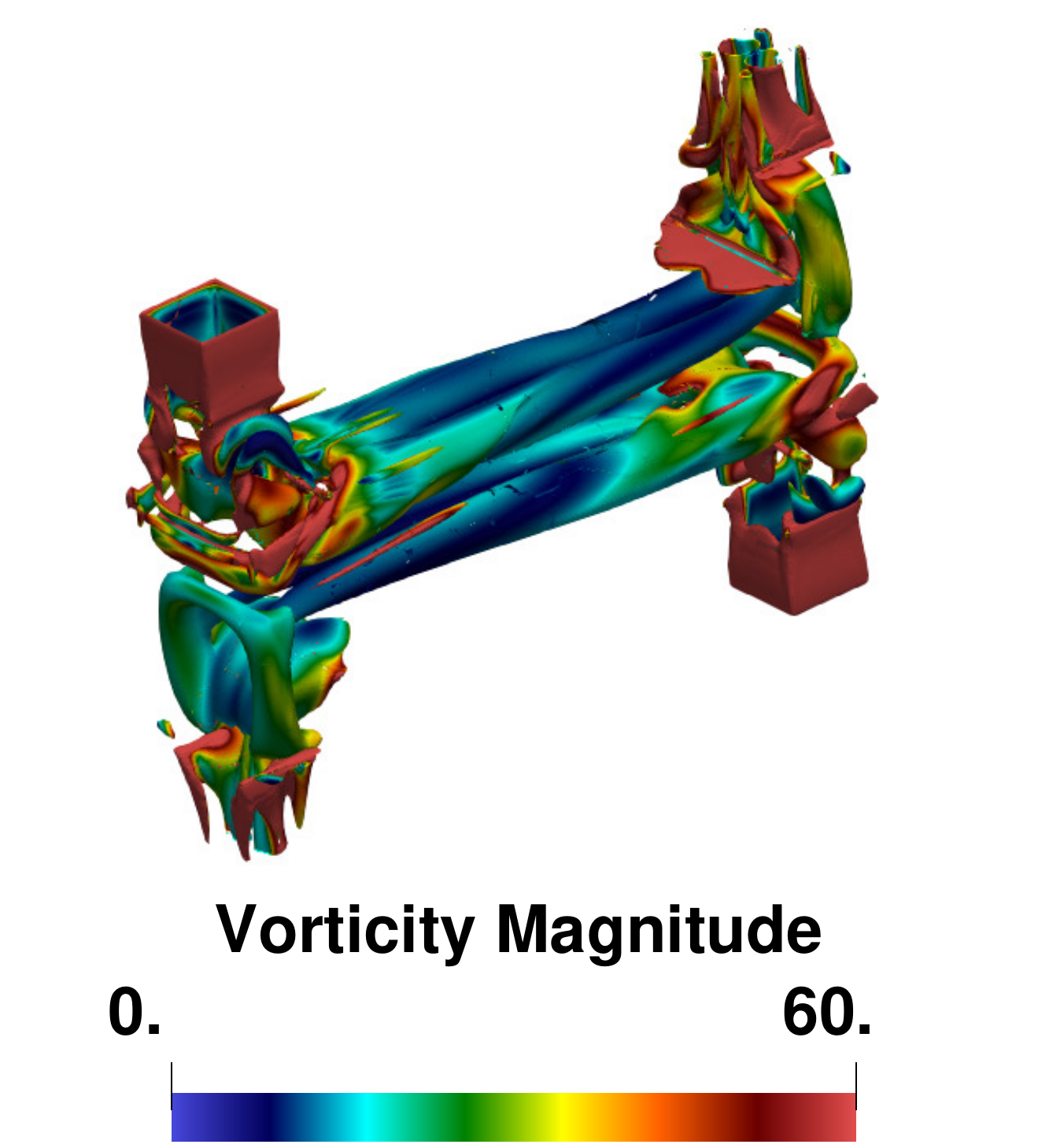}}}
    \label{fig:5x5_0250_Q}
    \end{subfigure}
    \hfill
    \begin{subfigure}[ht]{0.24\textwidth}
    \caption{}
    \centerline{
     {\includegraphics[width=\textwidth]{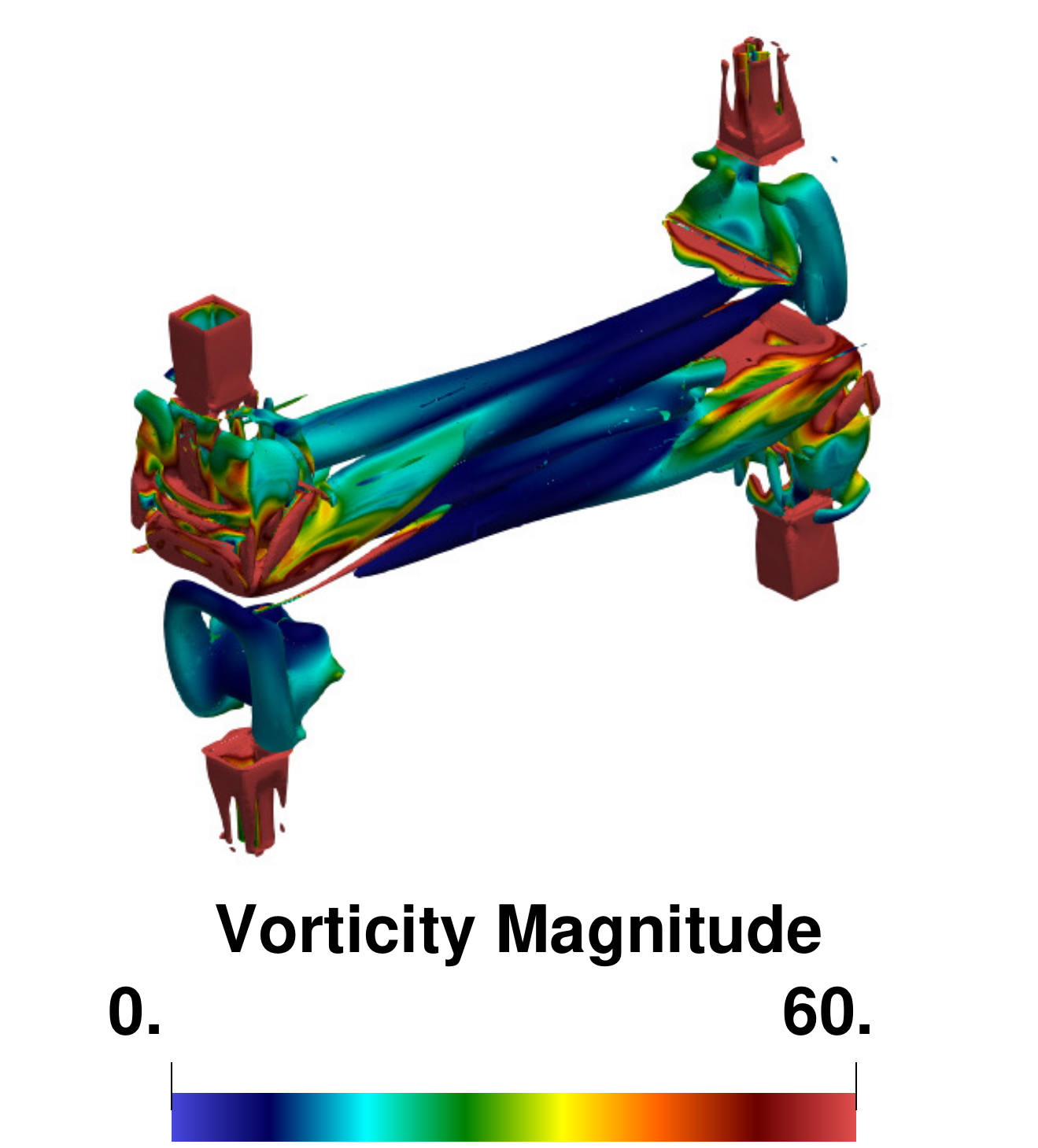}}}
    \label{fig:3x3_0250_Q}
    \end{subfigure}
    \medskip
    \begin{subfigure}[ht]{0.24\textwidth}
     \caption{}
    \centerline{
     {\includegraphics[width=\textwidth]{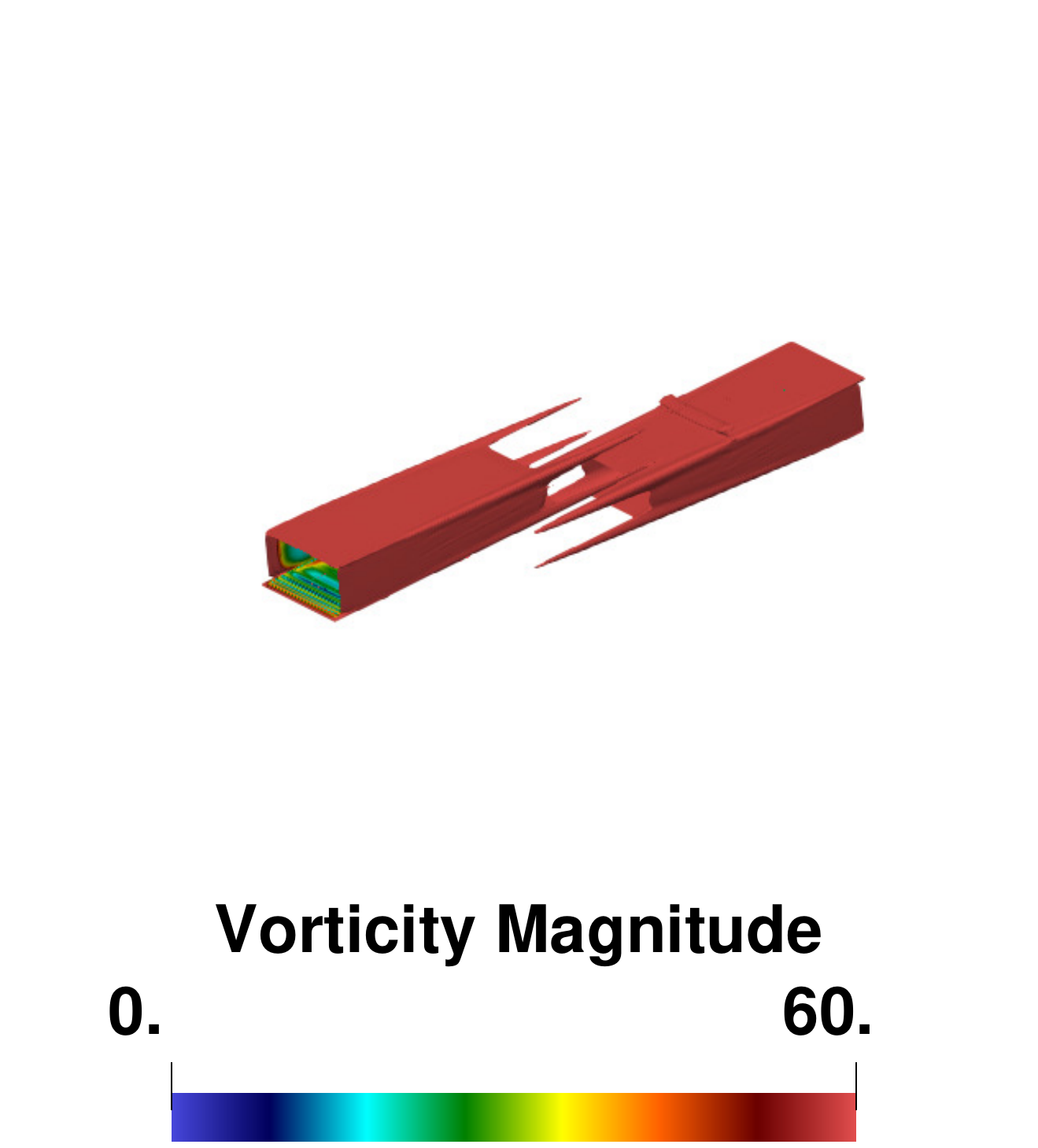}}}
    \label{fig:5x10s_2000_Q}
     \end{subfigure}
     \hfill
    \begin{subfigure}[ht]{0.24\textwidth}
     \caption{}
    \centerline{
     {\includegraphics[width=\textwidth]{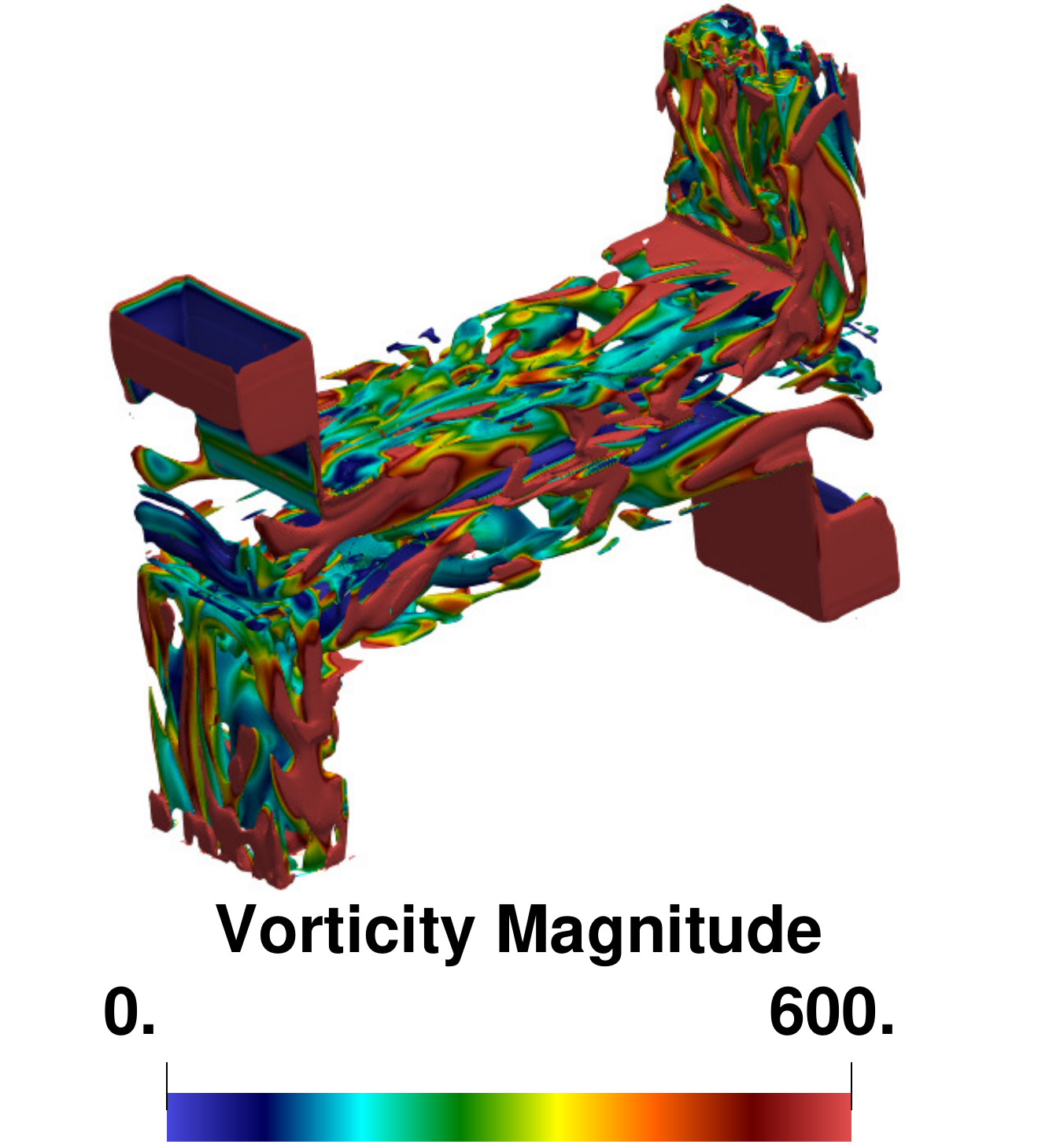}}}
    \label{fig:5x10_2000_Q}
     \end{subfigure}
     \hfill
    \begin{subfigure}[ht]{0.24\textwidth}
     \caption{}
    \centerline{
     {\includegraphics[width=\textwidth]{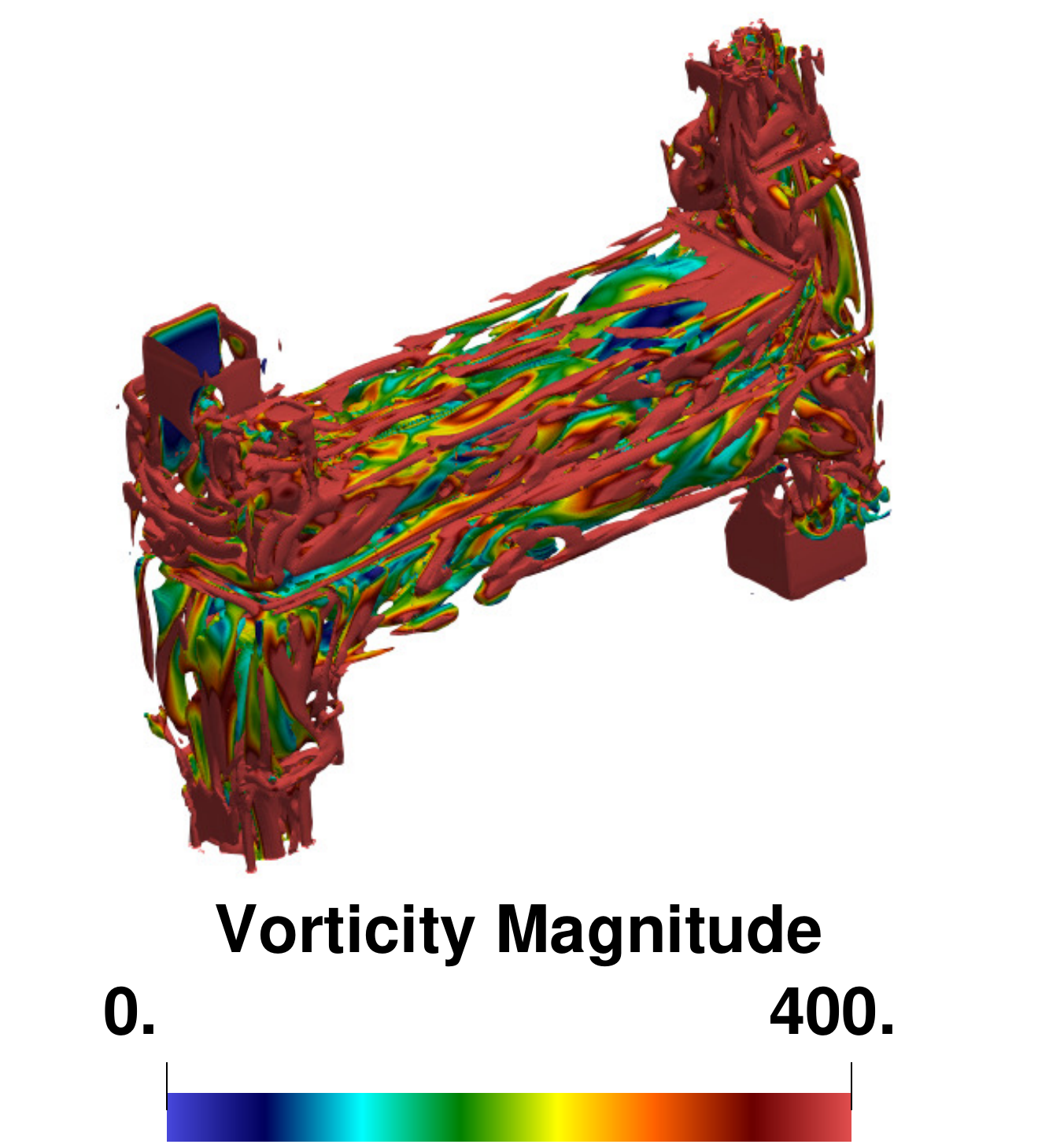}}}
    \label{fig:5x5_2000_Q}
     \end{subfigure}
     \hfill
    \begin{subfigure}[ht]{0.24\textwidth}
     \caption{}
    \centerline{
     {\includegraphics[width=\textwidth]{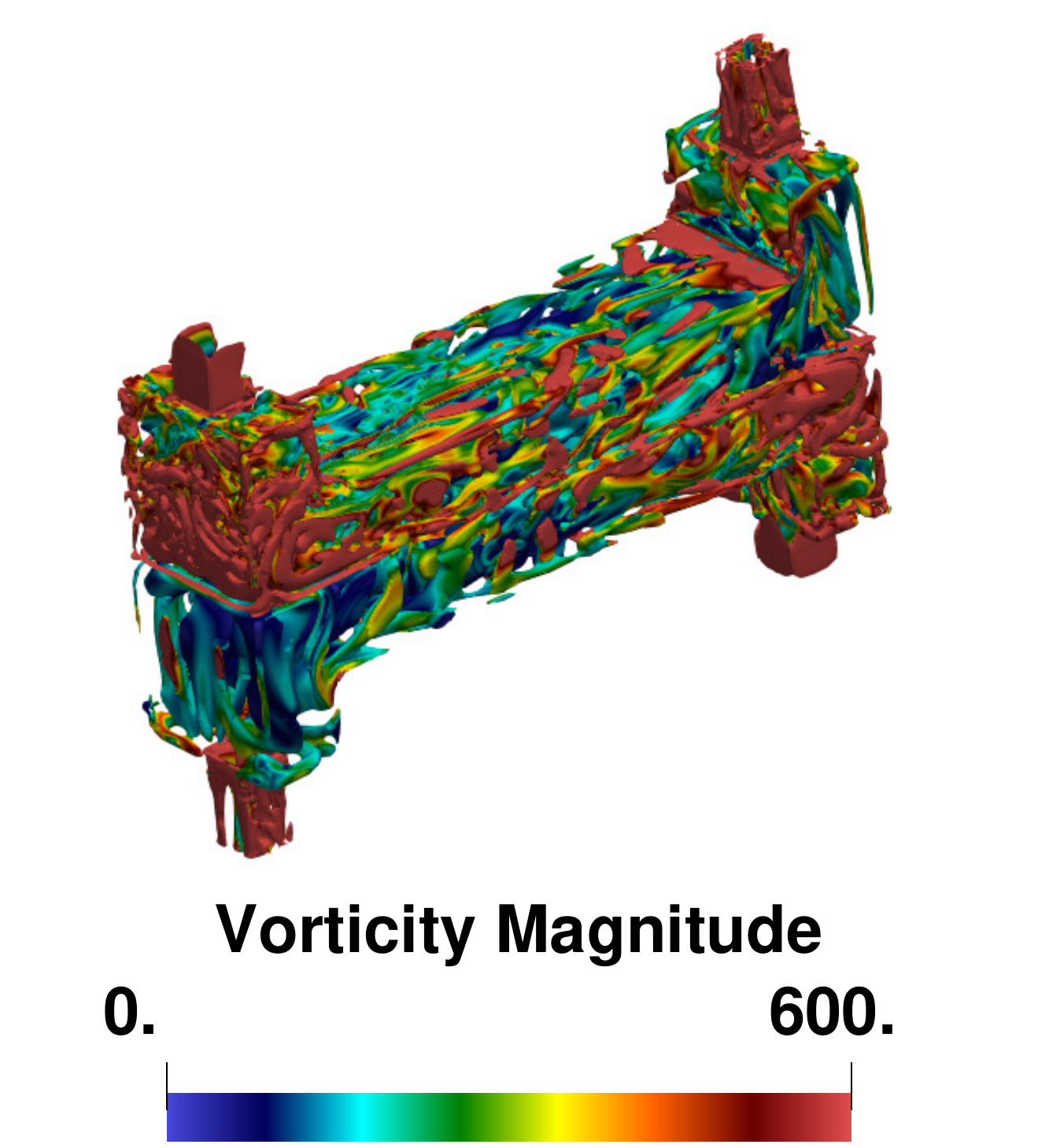}}}
    \label{fig:3x3_2000_Q}
     \end{subfigure}
     \caption{Isosurface of vortex tubes ($Q=50$) colored by the magnitude of vorticity at $Re_{in}=250$ for (a) MD-5x10S, (b) MD-5x10, (c) MD-5x5 and (d) MD-3x3, and $Re_{in}=2000$ for (e) MD-5x10S, (f) MD-5x10, (g) MD-5x5 and (h) MD-3x3.}
     \label{fig:cases_Q}
\end{figure*}

To visualize the swirling of the fluid in  MD systems, we calculate the $Q$-criterion ($Q$) of the velocity field as
\begin{align}
    Q &= \frac{1}{2}\left( \left\Vert \boldsymbol{\Omega} \right\Vert + \left\Vert \boldsymbol{\mathbf{S}} \right\Vert \right),
\end{align}
where 
\begin{subequations}
\begin{align}
   &\boldsymbol{\mathbf{S}} =  \frac{1}{2} \left( \grad{\boldsymbol{u}} + \left(\grad{\boldsymbol{u}}\right)^T \right), \\
   &\boldsymbol{\Omega} =  \frac{1}{2} \left( \grad{\boldsymbol{u}} - \left(\grad{\boldsymbol{u}}\right)^T \right),
\end{align}
\end{subequations}
are defined as the rate of strain and the vorticity tensor, respectively~\cite{Jeong1995-uo}. 
A positive $Q$-criterion ($Q>0$)  means that the magnitude of the vorticity is greater than the magnitude of the rate of strain, indicating the existence of vortices. Figure~\ref{fig:cases_Q} shows the isosurfaces of the $Q$-criterion ($Q=50$) for cases with different Reynolds numbers. For $Re_{in} =10$ (not shown), no pair of (Dean) vortex tubes or vortices are observed. For $Re_{in}=250$, the MD-5x10S design does not lead to the formation of visible Dean vortex tubes or vortex structures (Figure~\ref{fig:cases_Q}(\subref{fig:5x10s_0250_Q})). Fast-decaying Dean vortex tubes in the main channel can be observed for MD-5x10. No visible vortices were found in the side chamber below the inlet: this is consistent with the streamlines plot of Figure~\ref{fig:cases_sl}(\subref{fig:5x10s_0010_sl}), which suggests that the formation of vortices occurs in the main channel after the right-angled bend. Stable Dean vortex tubes are observed for both MD-5x5 and MD-3x3. This is likely due to the diverging flow or impinging jets when the inlet has a smaller opening than the side chamber, in presence of a wall where the flow can rebound. Significant vortex structures are observed in the side chamber for both MD-5x5 (Figure~\ref{fig:cases_Q}(\subref{fig:5x5_0250_Q})) and MD-3x3 (Figure~\ref{fig:cases_Q}(\subref{fig:3x3_0250_Q})). Therefore, it is likely that the formation of stable Dean vortex tubes in MD-5x5 and MD-3x3 is due both to the effects of right-angled bends and inlets with sudden expansions. For $Re_{in}=2000$, the number of fine-scale vortex structures increases drastically for all cases except MD-5x10S due to the flow transition to  chaotic regimes. Furthermore, Dean vortex tubes can only be observed for MD-5x5 and MD-3x3 with jet-like inlets. A possible explanation is that inlets with sudden expansions result in strong vortex structures that are less likely to be broken into smaller and irregular vortex structures by flow.

\begin{figure*}
    \centering
    \begin{subfigure}[ht]{0.23\textwidth}
    \caption*{\centering MD-5x10S}
    \centerline{
     {\includegraphics[width=\textwidth, trim={0in 0in 0in 1.5in},clip]{5x10s_geom-eps-converted-to.pdf}}}
    \end{subfigure}
    \begin{subfigure}[ht]{0.23\textwidth}
    \caption*{\centering MD-5x10}
    \centerline{
     {\includegraphics[width=\textwidth, trim={0in 0in 0in 1.5in},clip]{5x10_geom-eps-converted-to.pdf}}}
    \end{subfigure}
    \begin{subfigure}[ht]{0.23\textwidth}
    \caption*{\centering MD-5x5}
    \centerline{
     {\includegraphics[width=\textwidth, trim={0in 0in 0in 1.5in},clip]{5x5_geom-eps-converted-to.pdf}}}
    \end{subfigure}
    \begin{subfigure}[ht]{0.23\textwidth}
     \caption*{\centering MD-3x3}
    \centerline{
     {\includegraphics[width=\textwidth, trim={0in 0in 0in 1.5in},clip]{3x3_geom-eps-converted-to.pdf}}}
     \end{subfigure}
    \smallskip
    \begin{subfigure}[ht]{0.26\textwidth}
    \caption{}
    \centerline{
     {\includegraphics[width=\textwidth, trim={0in 0.09in 0in 0.09in},clip]{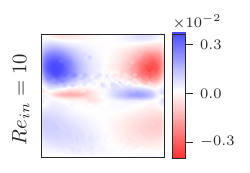}}}
    \label{fig:5x10s_0010_instan_vortex}
    \end{subfigure}
    \begin{subfigure}[ht]{0.23\textwidth}
     \caption{}
    \centerline{
     {\includegraphics[width=\textwidth, trim={0in 0.09in 0in 0.09in},clip]{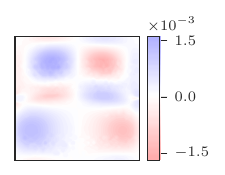}}}
    \label{fig:5x10_0010_instan_vortex}
     \end{subfigure}
     \begin{subfigure}[ht]{0.23\textwidth}
     \caption{}
    \centerline{
     {\includegraphics[width=\textwidth, trim={0in 0.09in 0in 0.09in},clip]{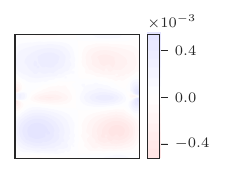}}}
    \label{fig:5x5_0010_instan_vortex}
     \end{subfigure}
     \begin{subfigure}[ht]{0.23\textwidth}
     \caption{}
    \centerline{
     {\includegraphics[width=\textwidth, trim={0in 0.09in 0in 0.09in},clip]{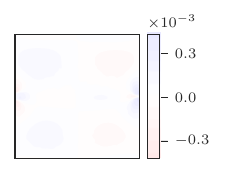}}}
    \label{fig:3x3_0010_instan_vortex}
     \end{subfigure}
    \smallskip
     \begin{subfigure}[ht]{0.26\textwidth}
     \caption{}
    \centerline{
     {\includegraphics[width=\textwidth, trim={0in 0.09in 0in 0.09in},clip]{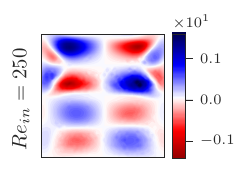}}}
    \label{fig:5x10s_0250_instan_vortex}
     \end{subfigure}
     \begin{subfigure}[ht]{0.23\textwidth}
     \caption{}
    \centerline{
     {\includegraphics[width=\textwidth, trim={0in 0.09in 0in 0.09in},clip]{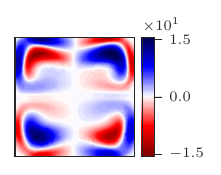}}}
    \label{fig:5x10_0250_instan_vortex}
     \end{subfigure}
     \begin{subfigure}[ht]{0.23\textwidth}
     \caption{}
    \centerline{
     {\includegraphics[width=\textwidth, trim={0in 0.09in 0in 0.09in},clip]{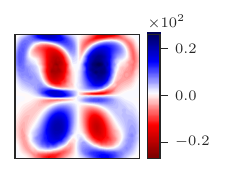}}}
    \label{fig:5x5_0250_instan_vortex}
     \end{subfigure}
     \begin{subfigure}[ht]{0.23\textwidth}
     \caption{}
    \centerline{
     {\includegraphics[width=\textwidth, trim={0in 0.09in 0in 0.09in},clip]{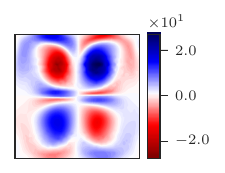}}}
    \label{fig:3x3_0250_instan_vortex}
     \end{subfigure}
    \smallskip
     \begin{subfigure}[ht]{0.26\textwidth}
     \caption{}
    \centerline{
     {\includegraphics[width=\textwidth]{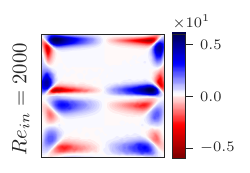}}}
    \label{fig:5x10s_2000_instan_vortex}
     \end{subfigure}
     \begin{subfigure}[ht]{0.23\textwidth}
     \caption{}
    \centerline{
     {\includegraphics[width=\textwidth]{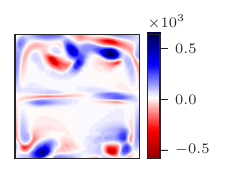}}}
    \label{fig:5x10_2000_instan_vortex}
     \end{subfigure}
     \begin{subfigure}[ht]{0.23\textwidth}
     \caption{}
    \centerline{
     {\includegraphics[width=\textwidth]{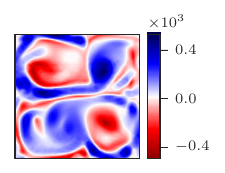}}}
    \label{fig:5x5_2000_instan_vortex}
     \end{subfigure}
     \begin{subfigure}[ht]{0.23\textwidth}
     \caption{}
    \centerline{
     {\includegraphics[width=\textwidth]{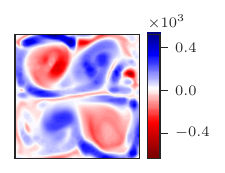}}}
    \label{fig:3x3_2000_instan_vortex}
     \end{subfigure}
     \caption{The instantaneous $x$-direction vorticity $\omega_x$ for $Re_{in}=10$ ((\subref{fig:5x10s_0010_instan_vortex})~--~(\subref{fig:3x3_0010_instan_vortex})) $Re_{in}=250$ ((\subref{fig:5x10s_0250_instan_vortex})~--~(\subref{fig:3x3_0250_instan_vortex})) and $Re_{in}=2000$ ((\subref{fig:5x10s_2000_instan_vortex})~--~(\subref{fig:3x3_2000_instan_vortex})) of MD-5x10S, MD-5x10, MD-5x5 and MD-3x3.}
     \label{fig:instan_vortex}
\end{figure*}

 Figures~\ref{fig:instan_vortex} shows the instantaneous vorticity in the $x$-direction, $\omega_x$, at the center of the channel ($x=0.02$~\si{\meter} for MD-5x10, MD-5x5 and MD-3x3, and $0.025$~\si{\meter} for MD-5x10S) with different Reynolds numbers where
\begin{align}
\label{eq:vort_x}
    \omega_x = \pdv{w}{y} - \pdv{v}{z}.
\end{align}
 At low Reynolds number ($Re_{in}=10$) (Figure~\ref{fig:instan_vortex}(\subref{fig:5x10s_0010_instan_vortex})~--~(\subref{fig:3x3_0010_instan_vortex})), no noticeable Dean vortices are observed, since the flow is laminar. The effects of right-angled bends and inlets with sudden expansions have no significant impact on flow characteristics. At moderate Reynolds number ($Re_{in}=250$) (Figure~\ref{fig:instan_vortex}(\subref{fig:5x10s_0250_instan_vortex})~--~(\subref{fig:3x3_0250_instan_vortex})), the right-angled bends and inlets with sudden expansions result in the formation of Dean vortices for MD-5x10, MD-5x5 and MD-3x3. Additionally,  inlets with sudden expansions (MD-5x5 and MD-3x3) lead to  Dean vortices of greater intensities and with more regular shapes. As the Reynolds number increases ($Re_{in}=2000$) (Figure~\ref{fig:instan_vortex}(\subref{fig:5x10s_2000_instan_vortex})~--~(\subref{fig:3x3_2000_instan_vortex})), the flow in MD-5x10 transitions into a chaotic flow with fine-scale vortex structures similar to those observed in turbulent flows. For MD-5x5 and MD-3x3, one can observe Dean-vortex-like structures with significant fluctuations. 

\begin{figure*}[hbpt!]
    \centering
    \begin{subfigure}[ht]{0.23\textwidth}
    \caption*{\centering MD-5x10S}
    \centerline{
     {\includegraphics[width=\textwidth, trim={0in 0in 0in 1.5in},clip]{5x10s_geom-eps-converted-to.pdf}}}
    \end{subfigure}
    \begin{subfigure}[ht]{0.23\textwidth}
    \caption*{\centering MD-5x10}
    \centerline{
     {\includegraphics[width=\textwidth, trim={0in 0in 0in 1.5in},clip]{5x10_geom-eps-converted-to.pdf}}}
    \end{subfigure}
    \begin{subfigure}[ht]{0.23\textwidth}
    \caption*{\centering MD-5x5}
    \centerline{
     {\includegraphics[width=\textwidth, trim={0in 0in 0in 1.5in},clip]{5x5_geom-eps-converted-to.pdf}}}
    \end{subfigure}
    \begin{subfigure}[ht]{0.23\textwidth}
     \caption*{\centering MD-3x3}
    \centerline{
     {\includegraphics[width=\textwidth, trim={0in 0in 0in 1.5in},clip]{3x3_geom-eps-converted-to.pdf}}}
     \end{subfigure}
    \smallskip
    \begin{subfigure}[ht]{0.26\textwidth}
    \caption{}
    \centerline{
     {\includegraphics[width=\textwidth, trim={0in 0.09in 0in 0.09in},clip]{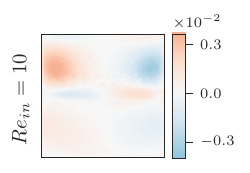}}}
    \label{fig:5x10s_0010_timeavg_vortex}
    \end{subfigure}
    \begin{subfigure}[ht]{0.23\textwidth}
     \caption{}
    \centerline{
     {\includegraphics[width=\textwidth, trim={0in 0.09in 0in 0.09in},clip]{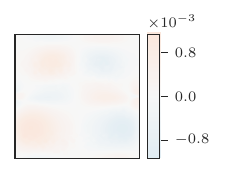}}}
    \label{fig:5x10_0010_timeavg_vortex}
     \end{subfigure}
    \begin{subfigure}[ht]{0.23\textwidth}
     \caption{}
    \centerline{
     {\includegraphics[width=\textwidth, trim={0in 0.09in 0in 0.09in},clip]{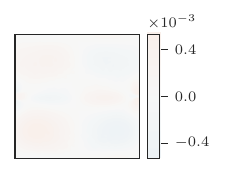}}}
    \label{fig:5x5_0010_timeavg_vortex}
     \end{subfigure}
    \begin{subfigure}[ht]{0.23\textwidth}
     \caption{}
    \centerline{
     {\includegraphics[width=\textwidth, trim={0in 0.09in 0in 0.09in},clip]{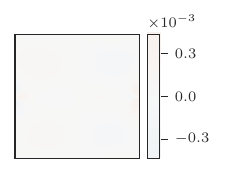}}}
    \label{fig:3x3_0010_timeavg_vortex}
     \end{subfigure}
    \smallskip
     \begin{subfigure}[ht]{0.26\textwidth}
     \caption{}
    \centerline{
     {\includegraphics[width=\textwidth, trim={0in 0.09in 0in 0.09in},clip]{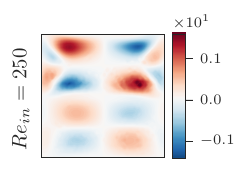}}}
    \label{fig:5x10s_0250_timeavg_vortex}
     \end{subfigure}
     \begin{subfigure}[ht]{0.23\textwidth}
     \caption{}
    \centerline{
     {\includegraphics[width=\textwidth, trim={0in 0.09in 0in 0.09in},clip]{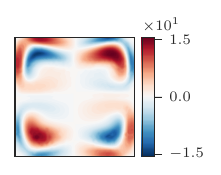}}}
    \label{fig:5x10_0250_timeavg_vortex}
     \end{subfigure}
     \begin{subfigure}[ht]{0.23\textwidth}
     \caption{}
    \centerline{
     {\includegraphics[width=\textwidth, trim={0in 0.09in 0in 0.09in},clip]{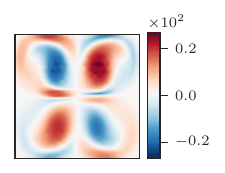}}}
    \label{fig:5x5_0250_timeavg_vortex}
     \end{subfigure}
     \begin{subfigure}[ht]{0.24\textwidth}
     \caption{}
    \centerline{
     {\includegraphics[width=\textwidth, trim={0in 0.09in 0in 0.09in},clip]{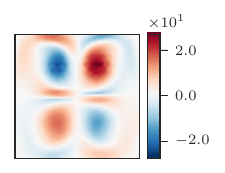}}}
    \label{fig:3x3_0250_timeavg_vortex}
     \end{subfigure}
    \smallskip
     \begin{subfigure}[ht]{0.26\textwidth}
     \caption{}
    \centerline{
     {\includegraphics[width=\textwidth, trim={0in 0.09in 0in 0.09in},clip]{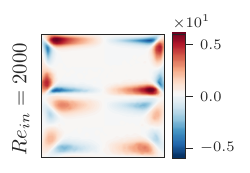}}}
    \label{fig:5x10s_2000_timeavg_vortex}
     \end{subfigure}
     \begin{subfigure}[ht]{0.23\textwidth}
     \caption{}
    \centerline{
     {\includegraphics[width=\textwidth, trim={0in 0.09in 0in 0.09in},clip]{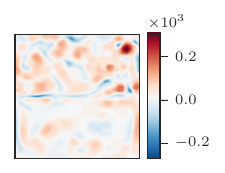}}}
    \label{fig:5x10_2000_timeavg_vortex}
     \end{subfigure}
     \begin{subfigure}[ht]{0.23\textwidth}
     \caption{}
    \centerline{
     {\includegraphics[width=\textwidth, trim={0in 0.09in 0in 0.09in},clip]{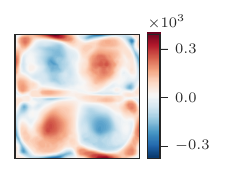}}}
    \label{fig:5x5_2000_timeavg_vortex}
     \end{subfigure}
     \begin{subfigure}[ht]{0.23\textwidth}
     \caption{}
    \centerline{
     {\includegraphics[width=\textwidth, trim={0in 0.09in 0in 0.09in},clip]{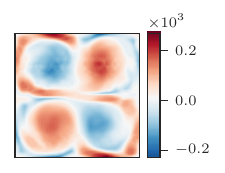}}}
    \label{fig:3x3_2000_timeavg_vortex}
     \end{subfigure}
     \caption{The time-average x-direction vorticity $\enstmean{\omega}_x$ for $Re_{in}=10$ ((\subref{fig:5x10s_0010_timeavg_vortex})~--~(\subref{fig:3x3_0010_timeavg_vortex})) $Re_{in}=250$ ((\subref{fig:5x10s_0250_timeavg_vortex})~--~(\subref{fig:3x3_0250_timeavg_vortex})) and $Re_{in}=2000$ ((\subref{fig:5x10s_2000_timeavg_vortex})~--~(\subref{fig:3x3_2000_timeavg_vortex})) of MD-5x10S, MD-5x10, MD-5x5 and MD-3x3.}
     \label{fig:timeavg_vortex}
\end{figure*}

To eliminate time fluctuations, we compute the time-average of $\omega_x$ as 
\begin{align}
    \tmean{\omega}_x = \frac{1}{N_t} \sum_{n} \omega_{x}^{n},
\end{align}
where $\omega_{x}^{n}=\omega_{x}(t_n)$ refers to the vorticity in $x$-direction at time $t_n$. Figures~\ref{fig:timeavg_vortex} shows the time-average vorticity $\tmean{\omega}_x$ for the respective cases at different Reynolds numbers.  For low and moderate Reynolds numbers,  no significant differences are observed between the instantaneous and time-averaged vorticity results: this is  expected since the flow is in the laminar regime. At  high Reynolds numbers ($Re_{in}=2000$), one can clearly identify the pair of Dean vortices for MD-5x5 and MD-3x3 (Figures~\ref{fig:timeavg_vortex}(\subref{fig:5x5_2000_timeavg_vortex}) and~(\subref{fig:3x3_2000_timeavg_vortex})) but not for MD-5x10 (Figures~\ref{fig:timeavg_vortex}(\subref{fig:5x10_2000_timeavg_vortex})):  this suggests that the inlet can stabilize the Dean vortices at high Reynolds numbers. Overall, instantaneous vorticity can differentiate between stable and unstable Dean vortices while time-average vorticity can identify Dean vortices in chaotic flows. 

\subsection{The effect of right-angled bend and inlet with sudden expansions on the temperature and concentration polarization}
\label{sec:polarization_effects}
One major concern in operating MD systems is temperature and concentration polarization, which will ultimately result in membrane fouling and an increase in the cost of operation. The ability to predict temperature and concentration polarization is critical to the development of efficient and low-cost MD systems. In this section, the effects of the right-angled bends and inlet design on  temperature and concentration polarization are studied.

\begin{figure*}
    \centering
    \begin{subfigure}[ht]{0.23\textwidth}
    \caption*{\centering MD-5x10S}
    \centerline{
    \vspace{-1.0\baselineskip}
     {\includegraphics[width=\textwidth, trim={0in 0in 0in 1.5in},clip]{5x10s_geom-eps-converted-to.pdf}}}
    \end{subfigure}
    \begin{subfigure}[ht]{0.23\textwidth}
    \caption*{\centering MD-5x10}
    \centerline{
    \vspace{-1.0\baselineskip}
     {\includegraphics[width=\textwidth, trim={0in 0in 0in 1.5in},clip]{5x10_geom-eps-converted-to.pdf}}}
    \end{subfigure}
    \begin{subfigure}[ht]{0.23\textwidth}
    \caption*{\centering MD-5x5}
    \centerline{
    \vspace{-1.0\baselineskip}
     {\includegraphics[width=\textwidth, trim={0in 0in 0in 1.5in},clip]{5x5_geom-eps-converted-to.pdf}}}
    \end{subfigure}
    \begin{subfigure}[ht]{0.23\textwidth}
     \caption*{\centering MD-3x3}
    \centerline{
    \vspace{-1.0\baselineskip}
     {\includegraphics[width=\textwidth, trim={0in 0in 0in 1.5in},clip]{3x3_geom-eps-converted-to.pdf}}}
     \end{subfigure}

    \begin{subfigure}[ht]{0.05\textwidth}
    \caption*{}
    \centerline{
    \vspace{-1.0\baselineskip}
     {\includegraphics[width=\textwidth, trim={0.0in 0.05in 0.4in 0.05in},clip]{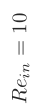}}}
     \end{subfigure}
    \begin{subfigure}[ht]{0.2275\textwidth}
    \caption{}
    \centerline{
    \vspace{-1.0\baselineskip}
     {\includegraphics[width=\textwidth, trim={0.1in 0.1in 0.1in 0.1in},clip]{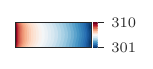}}}
    \label{fig:5x10s_0010_instan_Tmf}
    \end{subfigure}
    \begin{subfigure}[ht]{0.2275\textwidth}
     \caption{}
    \centerline{
    \vspace{-1.0\baselineskip}
     {\includegraphics[width=\textwidth, trim={0.1in 0.1in 0.1in 0.1in},clip]{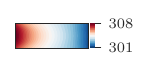}}}
    \label{fig:5x10_0010_instan_Tmf}
     \end{subfigure}
    \begin{subfigure}[ht]{0.2275\textwidth}
     \caption{}
    \centerline{
    \vspace{-1.0\baselineskip}
     {\includegraphics[width=\textwidth, trim={0.1in 0.1in 0.1in 0.1in},clip]{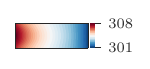}}}
    \label{fig:5x5_0010_instan_Tmf}
     \end{subfigure}
    \begin{subfigure}[ht]{0.2275\textwidth}
     \caption{}
    \centerline{
    \vspace{-1.0\baselineskip}
     {\includegraphics[width=\textwidth, trim={0.1in 0.1in 0.1in 0.1in},clip]{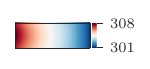}}}
    \label{fig:3x3_0010_instan_Tmf}
     \end{subfigure}

    \begin{subfigure}[ht]{0.05\textwidth}
    \caption*{}
    \centerline{
    \vspace{-1.0\baselineskip}
     {\includegraphics[width=\textwidth, trim={0.0in 0.05in 0.4in 0.05in},clip]{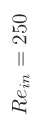}}}
     \end{subfigure}
     \begin{subfigure}[ht]{0.2275\textwidth}
     \caption{}
    \centerline{
    \vspace{-1.0\baselineskip}
     {\includegraphics[width=\textwidth, trim={0.1in 0.1in 0.1in 0.1in},clip]{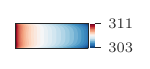}}}
    \label{fig:5x10s_0250_instan_Tmf}
     \end{subfigure}
     \begin{subfigure}[ht]{0.2275\textwidth}
     \caption{}
    \centerline{
    \vspace{-1.0\baselineskip}
     {\includegraphics[width=\textwidth, trim={0.1in 0.1in 0.1in 0.1in},clip]{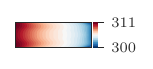}}}
    \label{fig:5x10_0250_instan_Tmf}
     \end{subfigure}
     \begin{subfigure}[ht]{0.2275\textwidth}
     \caption{}
    \centerline{
    \vspace{-1.0\baselineskip}
     {\includegraphics[width=\textwidth, trim={0.1in 0.1in 0.1in 0.1in},clip]{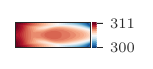}}}
    \label{fig:5x5_0250_instan_Tmf}
     \end{subfigure}
     \begin{subfigure}[ht]{0.2275\textwidth}
     \caption{}
    \centerline{
    \vspace{-1.0\baselineskip}
     {\includegraphics[width=\textwidth, trim={0.1in 0.1in 0.1in 0.1in},clip]{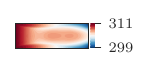}}}
    \label{fig:3x3_0250_instan_Tmf}
     \end{subfigure}
    
    \begin{subfigure}[ht]{0.05\textwidth}
    \caption*{}
    \centerline{
     {\includegraphics[width=\textwidth, trim={0.0in 0.05in 0.4in 0.05in},clip]{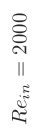}}}
     \end{subfigure}
     \begin{subfigure}[ht]{0.2275\textwidth}
     \caption{}
    \centerline{
     {\includegraphics[width=\textwidth, trim={0.1in 0.1in 0.1in 0.1in},clip]{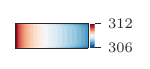}}}
    \label{fig:5x10s_2000_instan_Tmf}
     \end{subfigure}
     \begin{subfigure}[ht]{0.2275\textwidth}
     \caption{}
    \centerline{
     {\includegraphics[width=\textwidth, trim={0.1in 0.1in 0.1in 0.1in},clip]{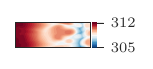}}}
    \label{fig:5x10_2000_instan_Tmf}
     \end{subfigure}
     \begin{subfigure}[ht]{0.2275\textwidth}
     \caption{}
    \centerline{
     {\includegraphics[width=\textwidth, trim={0.1in 0.1in 0.1in 0.1in},clip]{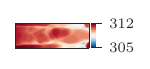}}}
    \label{fig:5x5_2000_instan_Tmf}
     \end{subfigure}
     \begin{subfigure}[ht]{0.2275\textwidth}
     \caption{}
    \centerline{
     {\includegraphics[width=\textwidth, trim={0.1in 0.1in 0.1in 0.1in},clip]{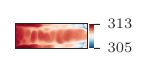}}}
    \label{fig:3x3_2000_instan_Tmf}
     \end{subfigure}
     \caption{The instantaneous temperature distribution for $Re_{in}=10$ (first row) $Re_{in}=250$ (second row) and $Re_{in}=2000$ (third row) of MD-5x10S, MD-5x10, MD-5x5 and MD-3x3 (left to right).}
     \label{fig:instan_Tmf}
\end{figure*}

\begin{landscape}
\begin{figure*}
    \centering
    \begin{subfigure}[ht]{2.4in}
    \caption{}
    \centerline{
     {\includegraphics[width=\textwidth]{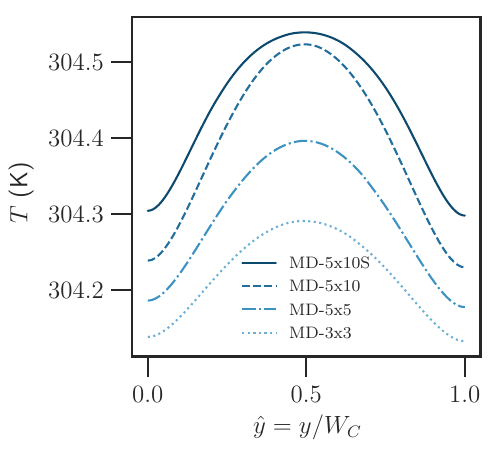}}}
    \label{fig:10_instan_Tmf_center}
    \end{subfigure}
    \begin{subfigure}[ht]{2.4in}
     \caption{}
    \centerline{
     {\includegraphics[width=\textwidth]{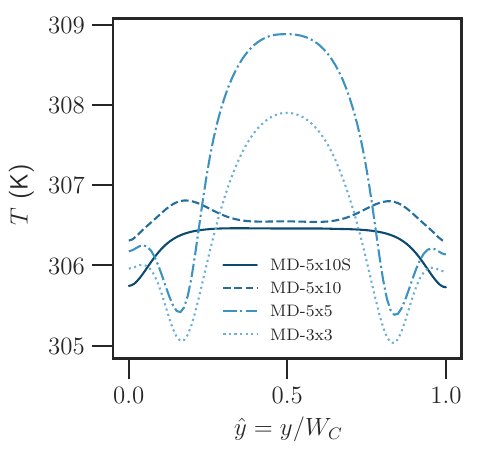}}}
    \label{fig:250_instan_Tmf_center}
     \end{subfigure}
    \begin{subfigure}[ht]{2.4in}
     \caption{}
    \centerline{
     {\includegraphics[width=\textwidth]{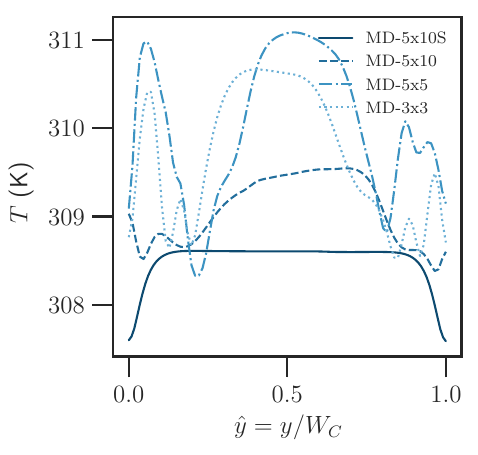}}}
    \label{fig:2000_instan_Tmf_center}
     \end{subfigure}
     \caption{The centerline instantaneous temperature in the $y$-direction on the feed-side membrane for (a) $Re_{in}=10$, (b) $Re_{in}=250$ and (c) $Re_{in}=2000$.}
     \label{fig:instan_Tmf_center}
\end{figure*}
\end{landscape}

Figure~\ref{fig:instan_Tmf} shows the instantaneous temperature distributions of the feed side membrane for three representative Reynolds numbers. At low Reynolds number ($Re_{in}=10$)(Figure~\ref{fig:instan_Tmf}(\subref{fig:5x10s_0010_instan_Tmf})~--~(\subref{fig:3x3_0010_instan_Tmf})), the temperature gradually decreases in the flow direction ($x$ direction) and is approximately homogeneous with insignificant boundary effects indicated by the lower temperature near the walls  in the perpendicular direction ($y$ direction). The difference in the temperature distribution for cases with right-angled bends (i.e. Figure~\ref{fig:instan_Tmf}(\subref{fig:5x10_0010_instan_Tmf})) and inlets with sudden expansions (i.e. Figures~\ref{fig:instan_Tmf}(\subref{fig:5x5_0010_instan_Tmf}) and~(\subref{fig:3x3_0010_instan_Tmf}))  is negligible because the flow is laminar. At moderate Reynolds numbers ($Re_{in}=250$)(Figure~\ref{fig:instan_Tmf}(\subref{fig:5x10s_0250_instan_Tmf})~--~(\subref{fig:3x3_0250_instan_Tmf})),  the temperature is approximately homogeneous in the perpendicular direction for MD-5x10S, while significant heterogeneity is observed for cases with right-angled bends  and  inlets with sudden expansions (MD-5x10, MD-5x5 and MD-3x3). For example, the temperature at the center in the flow direction is much higher than in the near-wall regions. As the Reynolds number further increases ($Re_{in}=2000$)(Figure~\ref{fig:instan_Tmf}(\subref{fig:5x10s_2000_instan_Tmf})~--~(\subref{fig:3x3_2000_instan_Tmf})), so do temperature  fluctuations. For MD-5x10 with only right-angled bends, the temperature distribution becomes more irregular. For MD-5x5 and MD-3x3, the high-temperature region appears to fluctuate. Figure~\ref{fig:instan_Tmf_center} shows the $y$-direction centerline of instantaneous temperature profiles. At a low Reynolds number (Figure~\ref{fig:instan_Tmf_center}(\subref{fig:10_instan_Tmf_center})), the temperature distributions are similar with a slight difference in magnitude between different designs. The magnitude difference between the low- and high-temperature regions at low Reynolds numbers is much smaller than that at moderate and high Reynolds numbers. At moderate Reynolds number (Figure~\ref{fig:instan_Tmf_center}(\subref{fig:250_instan_Tmf_center})), we can clearly observe three regions with alternating low and high temperatures for MD-5x5 and MD-3x3. As the Reynolds number increases, the instantaneous temperature for all cases except MD-5x10S fluctuates.

In MD systems, one of the challenges is to control membrane fouling due to both the temperature and concentration polarizations. 

We compute the time-average temperature polarization coefficient $\tmean{\mathbb{TPC}}$ as 
\begin{align}
    \label{eq:tpc_avg}
    \tmean{\mathbb{TPC}} = \frac{1}{N_t}\sum_n^{N_t} \frac{T^n_{m,f} - T^n_{m,d}}{T_{b,f} - T_{b,d}},
\end{align}
where $T^n_{m,i}=T_{m,i}(t_n)$ and $T_{b,i}$ are the temperature on the membrane surface ($m$) at $t_n$ and in the bulk fluid ($b$) for region $i$ that is equivalent to $T_{in,i}$, respectively, and $i=\{f,d\}$ refers to the feed and draw sides of the MD systems. Based on the definition in equation~\eqref{eq:tpc_avg}, $\tmean{\mathbb{TPC}}= 1$ indicates that there is no temperature polarization, while $\tmean{\mathbb{TPC}} < 1$ indicates the existence of temperature polarization. The concentration polarization coefficient $\tmean{\mathbb{CPC}}$ is calculated as
\begin{align}
    \tmean{\mathbb{CPC}} = \frac{1}{N_t}\sum_n^{N_t} \frac{C^n_{m,f}}{C_{b,f}},
\end{align}
where $C^n_{m,i}=C_{m,i}(t_n)$ and $C_{b,i}$ are the concentration on the membrane surface at $t_n$ and in the bulk fluid for region $i$ that is equivalent to $C_{in.i}$, respectively, and $\tmean{\mathbb{CPC}} > 1$ indicates concentration polarization.

    

\begin{landscape}
\begin{figure*}
    \centering
    \begin{subfigure}[ht]{2.4in}
    \caption{}
    \centerline{
     {\includegraphics[width=\textwidth]{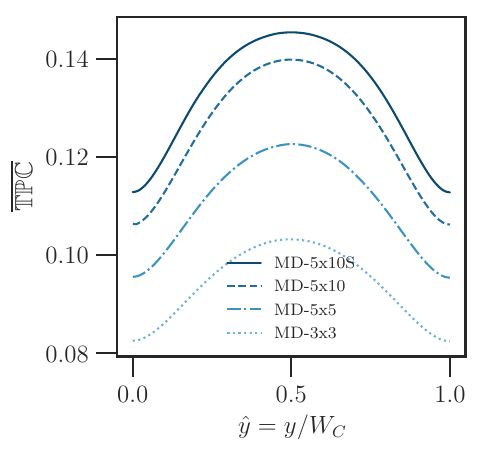}}}
    \label{fig:10_timeavg_TP_center}
    \end{subfigure}
    \begin{subfigure}[ht]{2.4in}
     \caption{}
    \centerline{
     {\includegraphics[width=\textwidth]{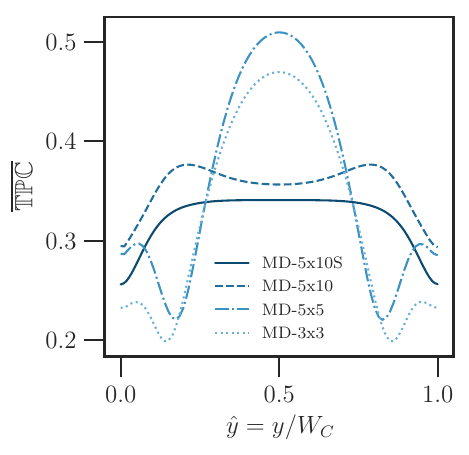}}}
    \label{fig:250_timeavg_TP_center}
     \end{subfigure}
    \begin{subfigure}[ht]{2.4in}
     \caption{}
    \centerline{
     {\includegraphics[width=\textwidth]{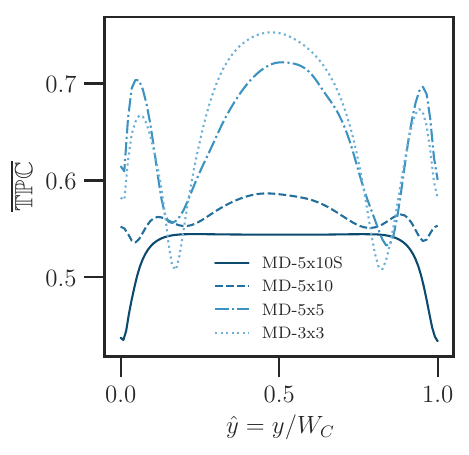}}}
    \label{fig:2000_timeavg_TP_center}
     \end{subfigure}
     \smallskip
    \begin{subfigure}[ht]{2.4in}
    \caption{}
    \centerline{
     {\includegraphics[width=\textwidth]{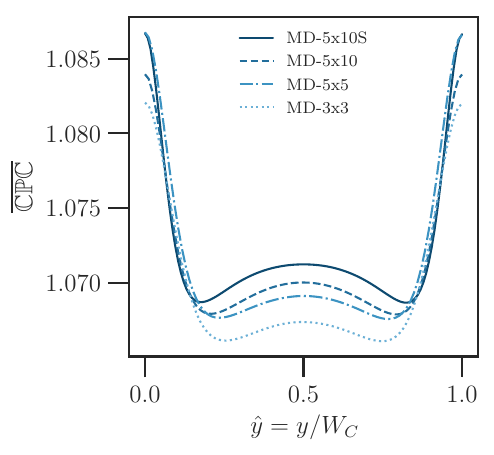}}}
    \label{fig:10_timeavg_CP_center}
    \end{subfigure}
    \begin{subfigure}[ht]{2.4in}
     \caption{}
    \centerline{
     {\includegraphics[width=\textwidth]{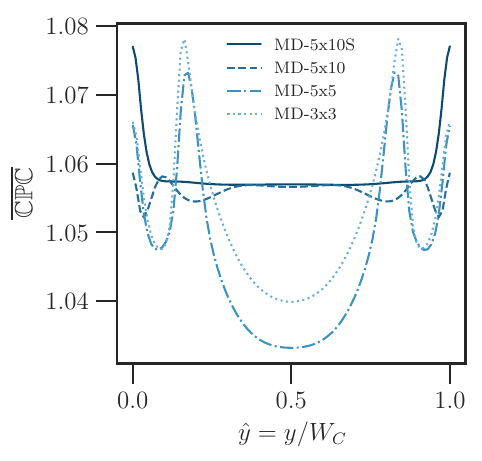}}}
    \label{fig:250_timeavg_CP_center}
     \end{subfigure}
    \begin{subfigure}[ht]{2.4in}
     \caption{}
    \centerline{
     {\includegraphics[width=\textwidth]{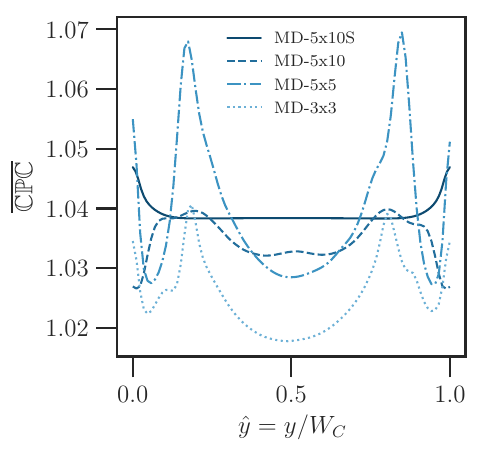}}}
    \label{fig:2000_timeavg_CP_center}
     \end{subfigure}
     \caption{The centerline time-average temperature polarization $\tmean{\mathbb{TPC}}$(\subref{fig:10_timeavg_TP_center}~--~\subref{fig:2000_timeavg_TP_center})) and concentration polarization $\tmean{\mathbb{CPC}}$(\subref{fig:10_timeavg_CP_center}~--~\subref{fig:2000_timeavg_CP_center})) in the $y$-direction on the feed-side membrane for (a, d) $Re_{in}=10$, (b, e) $Re_{in}=250$ and (c,f) $Re_{in}=2000$.}
     \label{fig:timeavg_TP_CP_center}
\end{figure*}
\end{landscape}
Figure~\ref{fig:timeavg_TP_CP_center}(\subref{fig:10_timeavg_TP_center})~--~(\subref{fig:2000_timeavg_TP_center}) show the centerline plots in the $y$-direction of the time average $\tmean{\mathbb{TPC}}$. At low Reynolds number (Figure~\ref{fig:timeavg_TP_CP_center}(\subref{fig:10_timeavg_TP_center})), all cases have similar distributions of $\tmean{\mathbb{TPC}}$ because the flow is approximately laminar. Interestingly, MD-5x10S has the least temperature polarization indicated by the largest $\tmean{\mathbb{TPC}}$. A possible explanation is that right-angled bends and inlets with sudden expansions result in more energy loss in the flow, but mixing enhancement mechanisms in presence of vortices are absent due to laminar flow. At moderate Reynolds number (Figure~\ref{fig:timeavg_TP_CP_center}(\subref{fig:250_timeavg_TP_center})), MD-5x10S has the most significant temperature polarization, indicated by small $\tmean{\mathbb{TPC}}$ while MD-5x5 and MD-3x3 have significantly less temperature polarization because of the enhanced mixing near the membrane surface due to strong Dean vortical structures. For MD-5x10, whose design includes only the effect of right-angled bends,  temperature polarization is lower than in MD-5x10S design but higher than in MD-5x5 and MD-3x3 designs. For high Reynolds number (Figure~\ref{fig:timeavg_TP_CP_center}(\subref{fig:2000_timeavg_TP_center})), the trends remain consistent while the difference in the extent of temperature polarization becomes more significant. 

    

Figure~\ref{fig:timeavg_TP_CP_center}(\subref{fig:10_timeavg_CP_center})~--~(\subref{fig:2000_timeavg_CP_center}) show the centerline plots in the $y$-direction of the time-average $\tmean{\mathbb{CPC}}$. Similar trends to time-average temperature polarization can be observed. At low Reynolds number (Figure~\ref{fig:timeavg_TP_CP_center}(\subref{fig:10_timeavg_CP_center})), all cases have a similar distribution, with MD-5x10S showing  the lower concentration polarization. At a moderate Reynolds number (Figure~\ref{fig:timeavg_TP_CP_center}(\subref{fig:250_timeavg_CP_center})), a slight improvement is observed for MD-5x10 with right-angled bends. For MD-5x5 and MD-3x3 designs with both right-angled bends and inlets with sudden expansions, a significant reduction in concentration polarization can be observed, indicated by the lower $\tmean{\mathbb{CPC}}$ at the center. As Reynolds number further increases (Figure~\ref{fig:timeavg_TP_CP_center}(\subref{fig:2000_timeavg_CP_center})), the trends remain similar to those for $Re_{in}=250$ with MD-3x3 showing the lowest polarization because of its strongest inlet effect.

\section{Discussion}
\label{sec:disscussion}
\subsection{Impact of vortices on predicted temperature and concentration polarization from Nusselt and Sherwood correlations}\label{Sec: sherwood-nusselt}

An established approach to predict temperature and concentration polarization in MD is to use the Nusselt and Sherwood correlation to predict temperature and concentration on the surface of the membrane~\cite{Hitsov2015-au}.  Dudchenko \emph{et al.}~\cite{Dudchenko2022-pz} have investigated the accuracy of different Nusselt correlations to predict temperature polarization. Once the thermophysical properties of the fluid mixture are defined,  membrane surface temperatures on both the feed and draw sides can be determined. However, as demonstrated in this work and other relevant studies~\cite{Lou2019-xa}, membrane temperature can be spatially highly heterogeneous. 
We first compute the time- and spatial-averaged temperature polarization coefficient $\enstmean{\mathbb{TPC}}$ through equation~\eqref{eq:t-x-avg-ops}. To compare the simulation results with predicted values, we follow the algorithm outlined in Hitsov \emph{et al.}~\cite{Hitsov2015-au}. Two Nusselt correlations by Stephan \emph{et al.}~\cite{Stephan1959-zq} and Gryta \emph{et al.}~\cite{Gryta1997-hx} are defined as
\begin{align}
    &Nu = 0.097 Re_{c}^{0.73} Pr_{b}^{0.13}\left(\frac{Pr_{b}}{Pr_{m}} \right)^{0.25}, \label{eq:nusselt_cor1} \\
    &Nu = 7.55 + \frac{0.024\left( \frac{Pr_m Re_c d_{ch}}{L_C} \right)}{1 + \frac{0.0358 Pr_m^{0.81}}{\left(\frac{L_C}{d_{ch} Re_c} \right)^{0.64}}} \left( \frac{\mu_b}{\mu_m} \right)^{0.14},
\end{align}
where $d_{ch}$~[\si{\meter}] is the hydraulic diameter of the channel, $\mu_b$~[\si{\kilogram\per\meter\per\second}] and $\mu_m$~[\si{\kilogram\per\meter\per\second}] are the dynamic viscosity of the fluid in the bulk and on the membrane, respectively, and $Pr$ is the Prandtl number defined as
\begin{align}
    Pr = \frac{c_p \mu}{k}.
\end{align}
Furthermore, we compute the concentration on the membrane surface with Sherwood correlation~\cite{Hitsov2015-au} such that
\begin{align}
    &C_m = C_b \exp(\frac{J_w}{\rho h_c}), \\
    &Sh = \frac{d_{ch} h_c}{D} = 1.86 \left(Re_{c} Sc \frac{d_{ch}}{L_C} \right)^{0.33}, \\
    & Sc = \frac{\mu}{\rho D} \label{eq:sc_cor}, 
\end{align}
where $L_C$~\si{\meter} is the length of the main channel, $C_m$~[\si{\density}] and $C_b$~[\si{\density}] are the salt concentration on the membrane surface and in the bulk fluid and $h_c$~[\si{\meter\per\second}] is the convective mass transfer coefficient.

\begin{figure*}
    \centering
    \begin{subfigure}[ht]{2.4in}
    \caption{}
    \centerline{
     {\includegraphics[width=\textwidth]{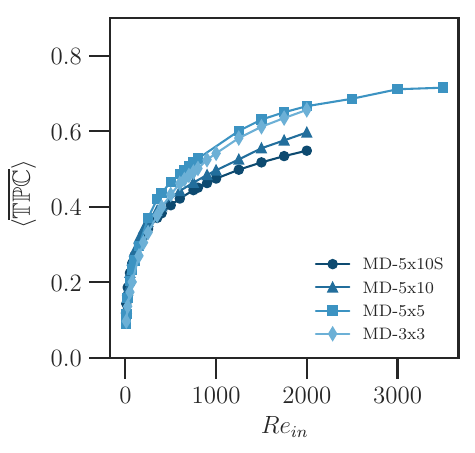}}}
    
    \label{fig:TPC_Rein}
    \end{subfigure}
    \begin{subfigure}[ht]{2.4in}
     \caption{}
    \centerline{
     {\includegraphics[width=\textwidth]{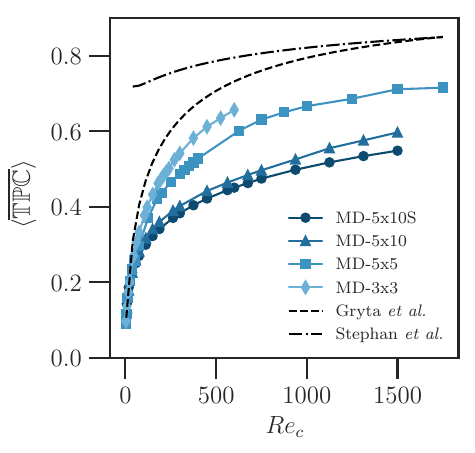}}}
    \label{fig:TPC_Rec}
     \end{subfigure}
     \caption{The time- and spatial-average temperature polarization coefficients $\enstmean{\mathbb{TPC}}$ as a function of (a) $Re_{in}$ and (b) $Re_{c}$. Model results are computed with models by Gryta \emph{et al.}~\cite{Gryta1997-hx} and Stephan \emph{et al.}~\cite{Stephan1959-zq}.} 
     \label{fig:TPC_re}
\end{figure*}

Figure~\ref{fig:TPC_re}(\subref{fig:TPC_Rein}) shows the time- and spatial-averaged temperature polarization $\enstmean{\mathbb{TPC}}$ as a function of Reynolds number $Re_{in}$. As $Re_{in}$ increases, $\enstmean{\mathbb{TPC}}$ increases, indicating weaker temperature polarization. At low $Re_{in}$,  all cases show similar values of $\enstmean{\mathbb{TPC}}$  since the flow is laminar and no  Dean vortices are present. As $Re_{in}$ increases, $\enstmean{\mathbb{TPC}}$ for MD-5x10, MD-5x5 and MD-3x3 is always higher than that of MD-5x10S, indicating the significant impacts of right-angled bends and jet-inducing inlets on the reduction of temperature polarization. However, $\enstmean{\mathbb{TPC}}$ for MD-3x3 with stronger impinging jets does not result in lower temperature polarization compared to MD-5x5. One possible explanation is that $Re_{in}$ does not account for the effect of the volumetric flow rate:  the same $Re_{in}$ does not necessarily correspond to the same volumetric flow rate because $Re_{in}$ is calculated based on the hydraulic diameter of the inlet. 



To understand the effect of the volumetric flow rate, we introduce the channel Reynolds number, $Re_{c}$ as  
\begin{align}
    &Re_{c} = \left(\frac{u_{in}H_C}{\nu_0}\right)\left(\frac{\mathcal{A}_{in}}{\mathcal{A}_{c}}\right),
\end{align}
where $H_C$~[\si{\meter}] is the height of the main channel. Figure~\ref{fig:TPC_re}(\subref{fig:TPC_Rec}) shows  $\enstmean{\mathbb{TPC}}$ as a function of channel Reynolds number $Re_{c}$ that includes the effect of volumetric flow rate. At low $Re_{c}$, $\enstmean{\mathbb{TPC}}$ for all cases are approximately the same, indicating negligible effects of right-angled bends and jet-inducing inlets. As $Re_{c}$ increases, we observe that $\enstmean{\mathbb{TPC}}$ for MD-3x3 is always the largest, followed by MD-5x5 and MD-5x10. This result suggests that the inlet with the strongest impinging jets results in the largest $\enstmean{\mathbb{TPC}}$ and the least temperature polarization for the same volumetric flow rate or $Re_{c}$. Comparing MD-5x10 with MD-5x10S, we have also demonstrated that right-angled bends results in less significant temperature polarization. As shown in Figure~\ref{fig:TPC_re}(\subref{fig:TPC_Rec}), at low $Re_{c}$, the Nusselt correlation of Gryta \emph{et al.}~\cite{Gryta1997-hx} successfully captures the behavior. For high $Re_{c}$, both models consistently overpredict the temperature polarization coefficient and underestimate the temperature polarization by more than 10\% for MD-3x3 and 40\% for MD-5x10S. Furthermore, current models do not capture the effects induced by right-angled bends and inlet.

\subsection{Relationship between the Dean vortices and polarization}
\label{sec:dean_polar}
\begin{figure*}
    \centering
    \begin{subfigure}[ht]{2.4in}
    \caption{}
    \centerline{
     {\includegraphics[width=\textwidth]{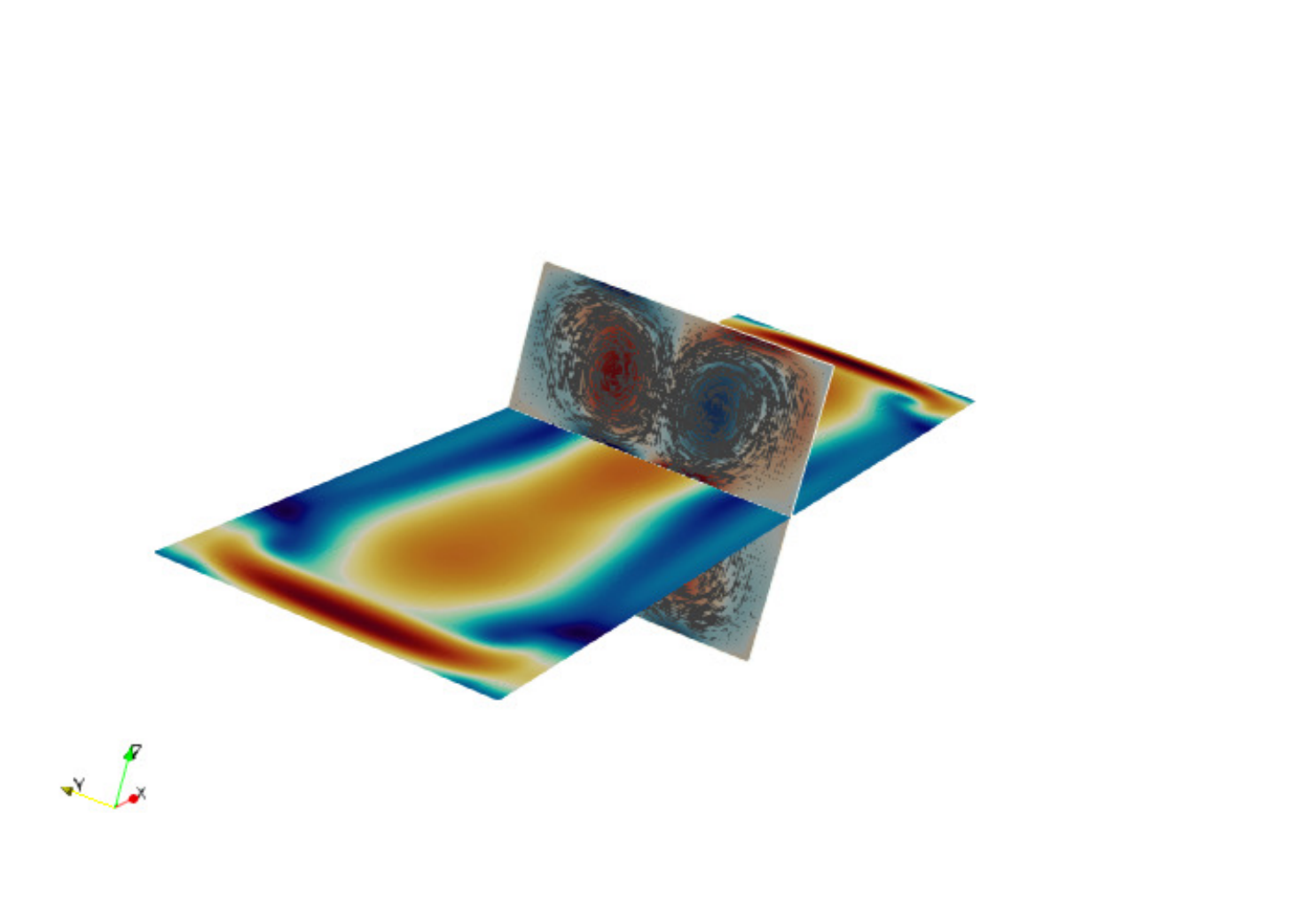}}}
    
    \label{fig:fig:5x5_250_intersect}
    \end{subfigure}
    \begin{subfigure}[ht]{2.4in}
     \caption{}
    \centerline{
     {\includegraphics[width=\textwidth]{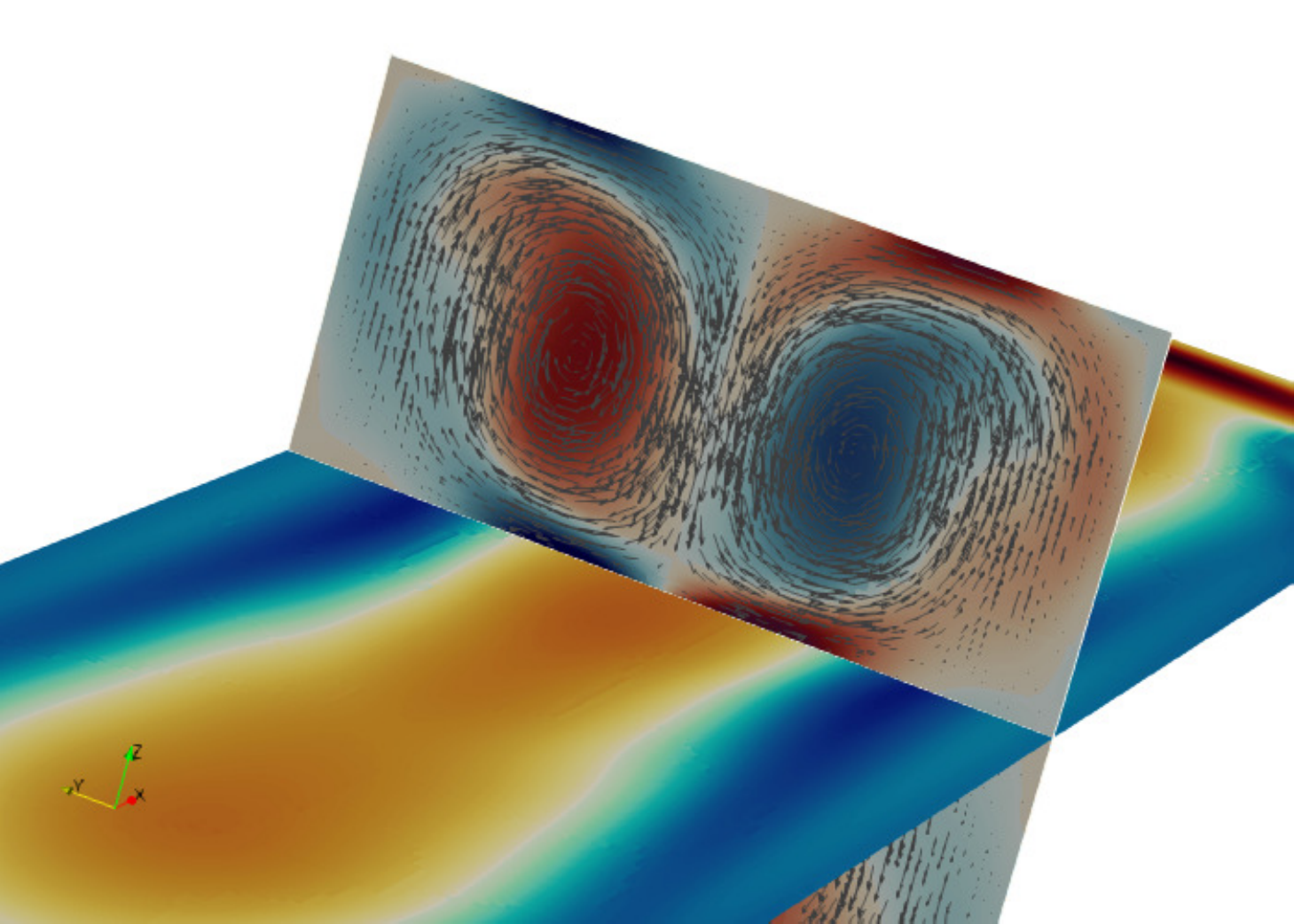}}}
    \label{fig:5x5_250_intersect_zoom}
     \end{subfigure}
     \caption{The intersection between instantaneous $x$-direction vorticity $\omega_x$ and temperature polarization on the membrane for MD-5x5 with $Re_{in}=250$: (a) standard and (b) zoomed-in. }
     \label{fig:fig:5x5_250_inter}
\end{figure*}

In previous sections, we discussed separately the effects of right-angled bends and inlet design on flow dynamics and temperature and concentration polarization. In this section, we focus on establishing the relationship between the formation of vortices and improvement in polarization effects. Figure~\ref{fig:fig:5x5_250_inter} shows a 3D visual of the flow field with Dean vortices and the underlying  temperature polarization on the membrane for MD-5x5 at $Re_{in}=250$. In the plot, we overlay the flow with the vortices, where the magnitude of the flow is proportional to the length of the arrows. As shown, the lowest temperature polarization, indicated by the red region, is observed at the location where two counter-rotating vortices touch each other, resulting in a strong flow perpendicular to the membrane surface, which leads to a  reduction of the temperature boundary layer. On the contrary, no significant flow nor vortices are observed near the wall corners, resulting in significant boundary layer effects and  stronger temperature polarization. Overall, the formation of Dean vortices enhances the flow perpendicular to the membrane, therefore reducing  boundary layer effects, and ultimately minimizing both temperature and concentration polarization.


\begin{figure*}
    \centering
    \begin{subfigure}[ht]{2.4in}
    \caption{}
    \centerline{
     {\includegraphics[width=\textwidth]{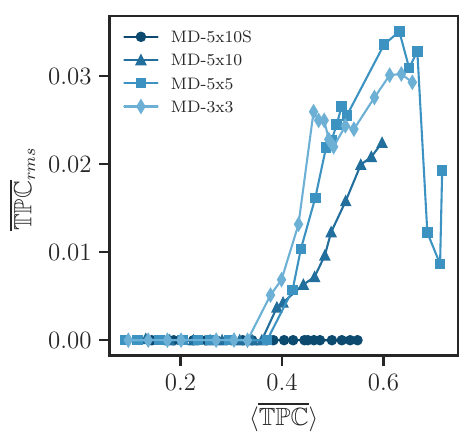}}}
    \label{fig:TPC_rms_TPC}
    \end{subfigure}
    \begin{subfigure}[ht]{2.4in}
     \caption{}
    \centerline{
     {\includegraphics[width=\textwidth]{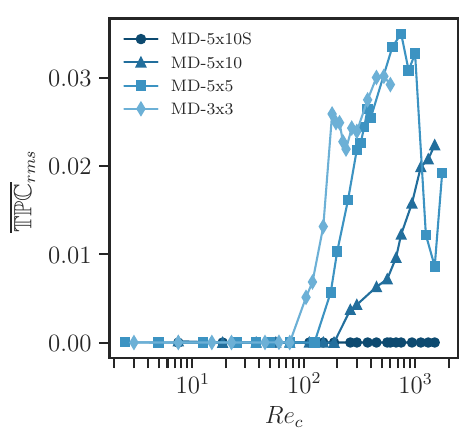}}}
    \label{fig:TPC_rms_t}
     \end{subfigure}
     \caption{The root mean square fluctuations of the temperature polarization $\tmean{\mathbb{TPC}}_{rms}$ as a function of (a) time- and spatial average temperature polarization coefficient $\enstmean{\mathbb{TPC}}$ and (b) Reynolds number $Re_{in}$.}
     \label{fig:TPC_rms}
\end{figure*}


To quantify the relationship, we introduce a metric to relate the effects of Dean vortices on temperature polarization. Theoretically, if there is no chaos in the flow, the distribution of temperature polarization on the membrane should remain constant over time (that is, $\mathbb{TPC}$ at a location should not change over time). Therefore, quantifying the fluctuation of $\mathbb{TPC}$ in time would indicate how chaotic the flow is. As demonstrated in Figure~\ref{fig:instan_vortex} and~\ref{fig:timeavg_vortex}, the chaos in the flow of MD-5x10, MD-5x5 and MD-3x3 is related to the formation of dean vortices, therefore, we could equate the fluctuation of $\mathbb{TPC}$ in time with the formation of dean vortices. The spatial average flucuation $\tmean{\mathbb{TPC}}_{rms}$ is defined as
\begin{align}
    \tmean{\mathbb{TPC}}_{rms} = \ensmean{\left(\mathbb{TPC}(\boldsymbol{x},t) - \tmean{\mathbb{TPC}}(\boldsymbol{x})\right)^2}.
\end{align}
Figure~\ref{fig:TPC_rms}(\subref{fig:TPC_rms_TPC}) shows the root mean square fluctuations of the temperature polarization $\tmean{\mathbb{TPC}}_{rms}$ as a function of  $\enstmean{\mathbb{TPC}}$. For MD-5x10S, $\tmean{\mathbb{TPC}}_{rms}$ remains negligible as $\enstmean{\mathbb{TPC}}$ increases, indicating that the temperature polarization coefficient does not change in time. This shows that the improvement in  temperature polarization is not due to the chaos in the flow. For other cases (MD-5x10, MD-5x5 and MD-3x3), two regimes are identified. Initially, $\enstmean{\mathbb{TPC}}$ increases when the fluctuations remain zero. As $\enstmean{\mathbb{TPC}}$ increases to a critical value, the fluctuations increase. This shows that the improvement in  temperature polarization after the critical value is correlated with the chaos in the flow.

Figure~\ref{fig:TPC_rms}(\subref{fig:TPC_rms_t}) shows the root mean square fluctuations of the temperature polarization $\tmean{\mathbb{TPC}}_{rms}$ as a function of $Re_{c}$. For MD-5x10S, $\tmean{\mathbb{TPC}}_{rms}$ remains negligible for all $Re_{c}$  as expected: since the MD-5x10S design is a straight channel of constant cross-section, Dean vortices are absent and mixing is negligible. For the other designs, at low Reynolds numbers, $\tmean{\mathbb{TPC}}_{rms}$ is approximately zero, indicating the absence of Dean vortices and mixing. As $Re_{c}$ increases beyond a threshold, we observe an increase in $\tmean{\mathbb{TPC}}_{rms}$. Combining the findings in Figure~\ref{fig:TPC_rms}, at low Reynolds numbers, the improvement in temperature polarization is due to the increase in volumetric flow rate, which reduces the thickness of the boundary layers. For cases with right-angled bends and jet-inducing inlets, vortices are generated when Reynolds number overcomes a threshold value: this causes an increase in the temporal fluctuations of the flow field which ultimately reduce temperature polarization.

\begin{figure*}
    \centering
     {\includegraphics[width=\textwidth]{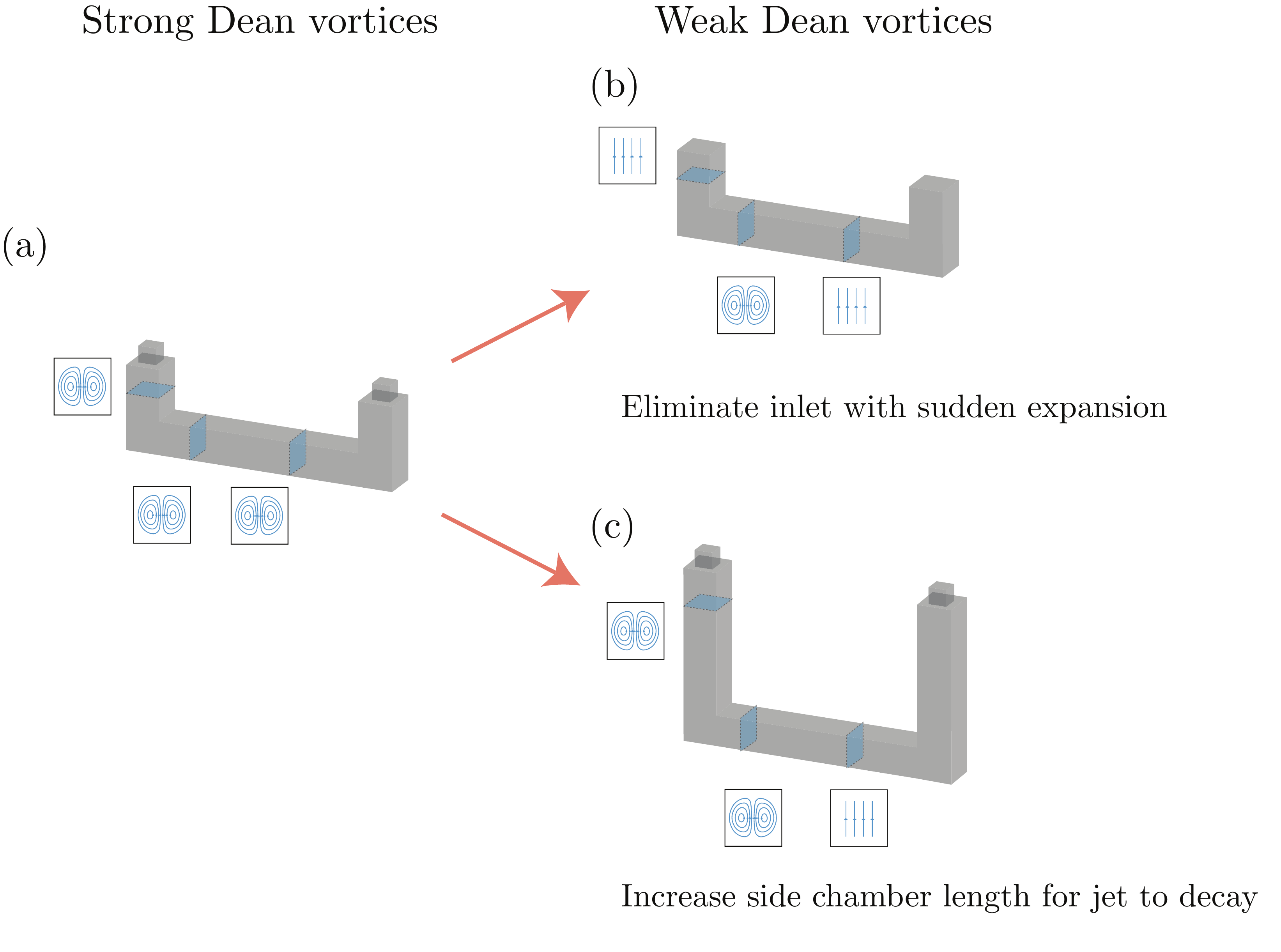}}
     \caption{\noah{Proposed design strategies to suppress and minimize Dean vortices.}}
     \label{fig:design}
\end{figure*}

\subsection{Implications on model development from lab-scale data}
In this study, we have demonstrated how  inlet designs of lab-scale systems can induce the formation of vortical structures in the flow field, which can greatly affect temperature and concentration polarization. Currently, experimental Sherwood and Nusselt correlations, calibrated  on data from lab-scale systems,  are  applied to predict the performance of full-scale systems. Yet, boundary effects related to the inlet design (both shape and bends) can have a large impact on  momentum, heat and mass transfer in  systems at the lab scale. As a result, concentration and temperature polarization estimated from models calibrated on bench scale systems may not be representative of the system performance at the full scale where these vortical structures are not persistent. Such models developed for lab-scale systems (with right-angled bends and jet-inducing inlets) would result in an underestimation of  temperature and concentration polarization in full-scale systems, where boundary effects  will likely be attenuated by the length of the systems. As a result, dynamic similarity between relevant momentum, mass and temperature transfer mechanisms needs to be assessed and established between lab-scale and full-scale systems, when Sherwood and Nusselt correlations are used for temperature and concentration polarization predictions  larger scales. \noah{To develop accurate correlations for full-scale systems from lab-scale data, the lab-scale system can be designed to suppress or minimize any entrance effect. One approach is to avoid sudden expansions of the inlet in combination with bends, which has been demonstrated to cause the formation of strong impinging jets and Dean vortices in the main channel (Figure~\ref{fig:design}(b)). An alternative method is to allow the use of inlets with sudden expansions but increase the length of the side chamber (Figure~\ref{fig:design}(c)): This will provide sufficient distance for the impinging jet to decay and will minimize the formation of Dean vortices in the main channel. Further studies are needed to understand the effects of the length of the side chamber on jet decaying.}

\section{Conclusions}\label{sec:conclusions}
Understanding and predicting membrane fouling in MD systems due to temperature and concentration polarization is critical to designing low-cost and efficient systems. In this study, we studied the effects of right-angled bends and inlet in  lab-scale MD systems on flow characteristics, temperature and concentration polarization. We conducted a total of 87 CFD simulations with OpenFOAM for four designs (MD-5x10S, MD-5x10, MD-5x5 and MD-3x3) with different volumetric flow rates. We discovered that both right-angled bends and inlet design are responsible for the formation of Dean vortices in the main channel of the MD systems, \noah{causing discrepancies in concentration and temperature polarization coefficients between different MD systems with same volumetric flow rate}. Designs with both right-angled bends and jet-forming inlets (MD-5x5 and MD-3x3) resulted in more stable Dean vortices at higher Reynolds numbers. In addition, at low Reynolds numbers, differences in temperature and concentration polarization are negligible across different designs. At moderate and high Reynolds numbers, MD systems with right-angled bends and inlets have more heterogeneous distributions of the temperature and concentration polarization coefficients. By calculating time- and spatial-average temperature and concentration polarization coefficients, we discovered that MD systems with right-angled bends and inlets of variable cross-sections present much lower temperature and concentration polarization. To assess the impact of vortex presence on  polarization, a new metric based on the spatial fluctuation of the temperature polarization was calculated. At low Reynolds numbers, fluctuations remain zero for all cases. At  moderate Reynolds numbers, the fluctuations increase significantly for cases with right-angled bends and inlet with varying cross-sections, while remaining zero for MD-5x10S. This shows that the improvement in  temperature and concentration polarizations at higher Reynolds numbers is due to the formation of vortical structures in the channel. Additionally, time- and spatial- average temperature polarization coefficients are compared with  values predicted from Sherwood and Nusselt correlations available in the literature:  the error between models and CFD simulation results can be as high as 40\%. \noah{These results  suggest that in order to develop models that can accurately predict temperature and concentration polarization, the formation of Dean vortices must be suppressed to ensure that the flow dynamics in the lab-scale systems is dynamically similar, i.e. comparable, to that in  full-scale systems.} 

\section*{Acknowledgments} This material is based upon work supported by the National Alliance for Water Innovation (grant number: 1242861-12-SDGBM), funded by the U.S. Department of Energy, Energy Efficiency and Renewable Energy Office, Advanced Manufacturing Office under Funding Opportunity Announcement DE-FOA-0001905.

\appendix
\section{Thermophysical properties of fluid}
\label{sec:thermo_prop}
The density of the fluid is calculated with the correlation proposed by Naftz \emph{et al.}~\cite{Naftz2011-ns} as 
\begin{subequations}
\begin{align}
    \label{eq:fluid_density}
    \rho = &\rho_0 + 184.01062 + 1.04708C - 1.21061T \\ \nonumber
    & + \num{3.14712e-4}C^2 + \num{0.00199}T^2 - \num{0.00112}CT, \\
    \rho_0 = &\left[\right. 999.83952 + 16.952577T_{c} - \num{7.9905127e-3}T_{c}^2 \\ \nonumber
    &- \num{4.6241757e-5}T_{c}^3 + \num{1.0574601e-7}T_{c}^4 \\ \nonumber
    &- \num{2.8103006e-10}T_{c}^5\left.\right] / \left[ 1 + \num{0.016887236}T_{c}\right].
\end{align}
\end{subequations}
where $\rho_0$~[\si{\density}] is the density of pure water and $T_c$~[\si{\celsius}] $= T - T_{stp}$ is the temperature in celsius. According to Naftz \emph{et al.}~\cite{Naftz2011-ns}, the correlation is valid for fluid density with a temperature range of 5~\si{\celsius} to 50~\si{\celsius} and a concentration range of 23~\si{\density} to 182~\si{\density}. 

We follow the models by Lou \emph{et al.}~\cite{Lou2019-xa} for the dynamic viscosity, specific heat capacity and latent heat of vaporization. The dynamic viscosity of the sodium chloride solution is calculated as 
\begin{align}
    \mu(T, C) = \left(\mathbb{P}_{\mu}\boldsymbol{T}_{\mu}\right) \cdot \boldsymbol{C}_\mu,
\end{align}
where $\boldsymbol{T}_\mu = \begin{pmatrix} T_c^0 & T_c^1 & T_c^2 & T_c^3 & T_c^4 \end{pmatrix}^T$ and $\boldsymbol{C}_\mu = \begin{pmatrix} C^0 & C^1 & C^2 & C^3 & C^4 \end{pmatrix}^T$ and $\mathbb{P}_{\mu}$ is the matrix of polynomial coefficients given by

\begin{align}
    \mathbb{P}_{\mu} = \begin{pmatrix}
    \num{3.3922e-11} & \num{-8.6874e-9} & \num{9.0999e-7} & \num{-5.1893e-5} & \num{1.7415e-3} \\
    \num{1.0124e-6} & \num{4.1167e-8} & \num{-1.5232e-9} & \num{1.8820e-11} & \num{-8.3450e-14} \\
    \num{1.0827e-8} & \num{-6.9728e-10} & \num{1.5945e-11} & \num{-1.6487e-13} & \num{6.6306e-16} \\
    \num{4.7647e-15} & \num{1.2275e-12} & \num{-3.1776e-14} & \num{3.1028e-16} & \num{-1.1551e-18} \\
    \num{2.5617e-14} & \num{-3.3743e-15} & \num{8.9878e-17} & \num{-9.8135e-19} & \num{4.0912e-21} 
    \end{pmatrix}.
\end{align}


The specific heat capacity of sodium chloride solution is modeled as
\begin{align}
    c_p(T, C) = \left(\mathbb{P}_{c_p}\boldsymbol{T}_{c_p}\right) \cdot \boldsymbol{C}_{c_p},
\end{align}
where $\boldsymbol{T}_{c_p} = \begin{pmatrix} T_c^0 & T_c^1 & T_c^2  \end{pmatrix}^T$, $\boldsymbol{C}_{c_p} = \begin{pmatrix} C^0 & C^1 & C^2 \end{pmatrix}^T$ and $\mathbb{P}_{c_p}$ is the matrix of polynomial coefficients given by
\begin{align}
    \mathbb{P}_{c_p} = \begin{pmatrix}
    \num{4129.8} & \num{0.75986} & \num{-6.1128e-4}  \\
    \num{-4.6391} & \num{-2.2851e-3} & \num{2.2508e-5}  \\
    \num{3.2167e-8} & \num{-9.1455e-6} & \num{2.4487e-3} 
    \end{pmatrix}.
\end{align}
The thermal conductivity of sodium chloride solution is calculated with the model by Ramires \emph{et al.}~\cite{Ramires1994-ng} as
\begin{align}
    k(T, C) = \left(\mathbb{P}_{k}\boldsymbol{T}_{k}\right) \cdot \boldsymbol{C}_{k},
\end{align}
where $\boldsymbol{T}_{k} = \begin{pmatrix} T_c^0 & T_c^1 & T_c^2  \end{pmatrix}^T$, $\boldsymbol{C}_{k} = 0.01716\begin{pmatrix} C^0 & C^1 & C^2 \end{pmatrix}^T$ and $\mathbb{P}_{k}$ is the matrix of polynomial coefficients given by
\begin{align}
    \mathbb{P}_k = \begin{pmatrix}
    \num{0.5621} & \num{0.00199} & \num{-8.6e-6}  \\
    \num{-0.01394} & \num{0.000294} & \num{-2.3e-6}  \\
    \num{0.00177} & \num{-6.3e-5} & \num{4.5e-7} 
    \end{pmatrix}.
\end{align}
The latent heat of water is evaluated as
\begin{align}
    \lambda(T) = -2438.18T_c + 2502800,
\end{align}
while the mass diffusivity of sodium chloride solution is calculated using the correlation by Harned and Hildreth~\cite{Harned1951-ec}
\begin{gather}
    D(T) = \num{17.872e-14}T \frac{\lambda_{Na,25}\lambda_{Cl,25} (1+\alpha(T_c-25))^2}{(\lambda_{Na,25}+\lambda_{Cl,25})(1+\alpha(T_c-25))},
\end{gather}
where $\lambda_{Na,25}=50.09$ and $\lambda_{Cl,25}=76.23$.

\bibliographystyle{elsarticle-num}

\end{document}